\global\def\draftcontrol{0}

   \def\versionno{ quenches code}

\catcode`\@=11

\expandafter\ifx\csname draftcontrol\endcsname\relax\global\def\draftcontrol{0}
\fi

{\count255=\time\divide\count255 by 60
\xdef\hourmin{\number\count255}
\multiply\count255 by-60\advance\count255 by\time
\xdef\hourmin{\hourmin:\ifnum\count255<10 0\fi\the\count255}}
\def\draftdate{\number\month/\number\day/\number\year\ \ \ \hourmin }

\newcommand\makepapertitle{\par
  \begingroup
    \renewcommand\thefootnote{\@fnsymbol\c@footnote}%
    \def\@makefnmark{\rlap{\@textsuperscript{\normalfont\@thefnmark}}}%
    \long\def\@makefntext##1{\parindent 1em\noindent
            \hb@xt@1.8em{%
                \hss\@textsuperscript{\normalfont\@thefnmark}}##1}%
     \newpage
     \global\@topnum\z@   
     \@makepapertitle
     \thispagestyle{empty}\@thanks
  \endgroup
  \setcounter{footnote}{0}%
  \global\let\thanks\relax
  \global\let\makepapertitle\relax
  \global\let\@makepapertitle\relax
  \global\let\@thanks\@empty
  \global\let\@author\@empty
  \global\let\@date\@empty
  \global\let\@title\@empty
  \global\let\title\relax
  \global\let\author\relax
  \global\let\date\relax
  \global\let\and\relax
  \def\version{\let\version\@version\@gobble}
}
\def\@makepapertitle{%
  \newpage
   \ifnum\draftcontrol=1 {}
   \version\versionno
   \vskip 3em%
   \else
   \hfill\hbox to 3cm {\parbox{4cm}{\@pubnum}\hss}%
   \vskip 3em%
   \fi
   \begin{center}%
   \let \footnote \thanks
     {\LARGE {\@title}}%
     \vskip 1.5em%
     {\normalsize
       \lineskip .5em%
       \begin{tabular}[t]{c}%
         \@author
       \end{tabular}\par}%
     \vskip 1.5em%
     {\@bstract}%
     \end{center}%
     \vskip 1.5em
     \@date%
   \par
}

\gdef\@pubnum{}
\def\pubnum#1{%
  \gdef\@pubnum{#1}}

\gdef\@bstract{}
\def\Abstract#1{%
  \gdef\@bstract{%
   \parbox{\textwidth-0pc}{%
   \centerline{\bf Abstract}\penalty1000%
\kern.2cm%
\noindent
\renewcommand\baselinestretch{1.0}%
{#1}}}
}

\def\ps@paper{\let\@mkboth\@gobbletwo%
     \ifnum\draftcontrol=1
    \def\@oddfoot{\hbox to \textwidth{\tiny \versionno \hfil\tiny\draftdate}%
    \hskip -\textwidth \hbox to \textwidth{\hfil\rm\thepage\hfil}}%
     \else\def\@oddfoot{\hbox to \textwidth{\hfil\rm\thepage\hfil}}
     \fi
     \let\@evenfoot\@oddfoot
}

\def\body{\clearpage
          \pagestyle{paper}
    }

\def\@version#1{\ifnum\draftcontrol=1
\typeout{}\typeout{#1}\typeout{}
\vskip3mm\centerline{\hbox{\fbox{\normalsize{\tt DRAFT -- #1 -- }
                   {\draftdate}}}}\vskip3mm
\fi}
\let\version\@version
\long\def\eqlabel#1{\ifnum\draftcontrol=1
                    \tag@false  
                    \tag*{(\theequation) \hbox to -0.2cm{\hspace{0cm}\small{#1}\hss}}
                    \refstepcounter{equation}
                    \edef\@currentlabel{\theequation}
                    \ltx@label{#1}          
                    \else
                    \label{#1}
                    \fi
                    }
\let\st@bibitem\@bibitem
\let\st@lbibitem\@lbibitem
\ifnum\draftcontrol=1
  \def\@bibitem#1{%
    \st@bibitem{#1}\a@@label{#1}\ignorespaces}
  \def\@lbibitem[#1]#2{%
    \st@lbibitem[#1]{#2}\a@@label{#2}\ignorespaces}
  \def\a@@label#1{%
    \gdef\a@lab{\smash{\normalfont\small#1}}
    \ifvmode
      \if@inlabel
        \global\setbox\@labels\hbox{%
          \llap{\a@lab\let\a@lab\relax
                \kern\@totalleftmargin\kern\marginparsep}%
          \box\@labels}%
      \fi
    \fi}
\fi

\documentclass[12pt,letterpaper]{article}

\usepackage{graphicx}
\usepackage{amsmath,amssymb,array,calc,epsfig,rotating,bm}
\usepackage[sort]{cite}
\usepackage{psfrag,verbatim}
\usepackage{mathtools}

\ifnum\draftcontrol=1
\fi

\renewcommand\baselinestretch{1.25}
\renewcommand\section{\@startsection {section}{1}{\z@}%
                                   {-3.5ex \@plus -1ex \@minus -.2ex}%
                                   {2.3ex \@plus.2ex}%
                                   {\normalfont\large\bfseries}}
\renewcommand\subsection{\@startsection{subsection}{2}{\z@}%
                                   {-3.25ex\@plus -1ex \@minus -.2ex}%
                                   {1.5ex \@plus .2ex}%
                                   {\normalfont\normalsize\bfseries}}
\renewcommand\subsubsection{\@startsection{subsubsection}{3}{\z@}%
                                   {-3.25ex\@plus -1ex \@minus -.2ex}%
                                   {1.5ex \@plus .2ex}%
                                   {\normalfont\normalsize\it}}
\renewcommand\paragraph{\@startsection{paragraph}{4}{\z@}%
                                   {-3.25ex\@plus -1ex \@minus -.2ex}%
                                   {1.5ex \@plus .2ex}%
                                   {\normalfont\normalsize\bf}}

\numberwithin{equation}{section}

\def\revise#1       {\raisebox{-0em}{\rule{3pt}{1em}}%
                     \marginpar{\raisebox{.5em}{\vrule width3pt\
                     \vrule width0pt height 0pt depth0.5em
                     \hbox to 0cm{\hspace{0cm}{%
                     \parbox[t]{4em}{\raggedright\footnotesize{#1}}}\hss}}}}

\newcommand\nxt[1]  {\\\fnxt#1}
\newcommand{\ie}{{\it i.e.,}\ }
\newcommand{\eg}{{\it e.g.,}\ }

\def\cala         {{\cal A}}

\def\calb         {{\cal B}}
\def\calc         {{\cal C}}

\def\calf         {{\cal F}}

\def\caln         {{\cal N}}
\def\calo         {{\cal O}}
\def\calp         {{\cal P}}

\def\del          {\partial}

\def\sqr#1#2{{\vcenter{\vbox{\hrule height.#2pt
 \hbox{\vrule width.#2pt height#1pt \kern#1pt
 \vrule width.#2pt}\hrule height.#2pt}}}}

\newcommand{\ft}[2]{{\textstyle{\frac{#1}{#2}}}}

\def\a{\alpha}
\def\b{\beta}

\def\r{\rho}
\def\dd{\delta}

\def\aa1{\phi}
\def\cc1{\psi}

\def\D{\Delta}

\def\l{\lambda}

\def\ha{\hat{a}}

\def\t{\tau}

\def\hp{\hat{\phi}}

\catcode`\@=12

\begin{document}


\title{\bf   Nonlocal probes of  thermalization in holographic quenches with spectral methods}

\date\today

\author{
Alex Buchel,$^{1,2}$  Robert C. Myers,$^1$ and Anton van Niekerk$^{1,3}$\\[0.4cm]
\it $^1$\,Perimeter Institute for Theoretical Physics\\
\it Waterloo, Ontario N2J 2W9, Canada\\[0.2cm]
\it $^2$\,Department of Applied Mathematics,
University of Western Ontario\\
\it London, Ontario N6A 5B7, Canada\\[0.2cm]
\it $^3$\,Department of Physics \& Astronomy and Guelph-Waterloo Physics Institute\\
\it University of Waterloo, Waterloo, Ontario N2L 3G1, Canada\\[0.2cm]
}

\Abstract{We describe the application of pseudo-spectral methods
to problems of holographic thermal quenches of relevant couplings in strongly
coupled  gauge theories. We focus on  quenches of a fermionic mass term in
a strongly coupled $\caln=4$ supersymmetric Yang-Mills plasma, and the subsequent equilibration
of the system. From the dual gravitational perspective, we study the gravitational collapse
of a massive scalar field in asymptotically anti-de Sitter geometry with a prescribed boundary
condition for its non-normalizable mode. Access to the full background geometry
of the gravitational collapse allows for the study of nonlocal probes of the thermalization
process. We discuss the evolution of the apparent and the event horizons, the two-point
correlation functions of operators of large conformal dimensions, and the evolution of the
entanglement entropy of the system. We compare the thermalization process from the viewpoint of
local (the one-point) correlation functions and these nonlocal probes, finding that the
thermalization time as measured by the probes is length dependent, and approaches the thermalization time of the 
one-point function for longer probes. We further discuss how
the different energy scales of the problem contribute to its thermalization.
}

\makepapertitle

\body

\version\versionno
\tableofcontents

\section{Introduction}

Quantum quenches are processes where an isolated system is driven to a
far-from-equilibrium state by rapidly varying some control parameters.
It has been possible to produce and study such processes in laboratory
experiments in recent years, in particular, with ultra-cold atomic gases
\cite{exp,more}.
This experimental progress has provided a great impetus to improve our theoretical description of quenched systems. Certainly theoretical progress made with
investigations within a variety of different frameworks,
\eg
two-dimensional conformal field theories \cite{2dcft},
(nearly) free field theories \cite{cc0,cc1} and integrable models \cite{cc0,integrable}, as well as some results applying to weakly interacting relativistic quantum field theories in higher dimensions \cite{higherd}.
It remains a challenge to find broadly applicable and efficient techniques, as well as extracting insights into general organizing principles for the behaviour
of far-from-equilibrium systems.

A new theoretical tool allows for the investigation of quenches for (certain) strongly coupled field theories is gauge/gravity duality \cite{m1}.  Assuming the robustness of this holographic duality
in non-equilibrium situations, as studied in \eg \cite{early,minwalla}, it is
possible to study the behaviour of the boundary field theory, either when it is perturbed, or far from
equilibrium.  Initial work by \cite{cy,initial} has led to a large body of work in the field of
quantum quenches of field theories at strong coupling, including
\cite{adsnumerics,blm,blmv,bmv,bdyterm,newer,periodic,evennewer}.  In the gravity dual of the quantum field theory, the quench usually has a simple
geometric interpretation and is introduced, \eg in the form of
a gravitational shock wave collapsing into a black hole and a collapsing shell of matter
described by the Vaidya metric, collapsing into a black hole \cite{minwalla,therm1,vaidya,hong}.  Applications of holographic quenches include quenches across phase transition points in the field theory \cite{critical}, and to model hadron collisions in particle accelerators such as RHIC \cite{hadron}.

In an ongoing research program including \cite{blm}, \cite{blmv} and \cite{bmv}, we study the response of a strongly coupled $\caln=4$ supersymmetric Yang-Mills thermal plasma, quenched by a relevant
operator, using the holographic duality.  Having
previously studied such quenches, we now apply more powerful numerical techniques to find the full time-dependent profiles of the perturbations of the metric and scalar field in the dual AdS spacetime.  This allows us to utilize nonlocal
probes such as two-point functions and entanglement entropy to better understand thermalization at various distance scales.

The quench that we study here for the Yang-Mills plasma in four spacetime dimensions, is that of switching on a fermionic operator in a smooth manner, by giving it a time-dependent mass.  This is dual to a radially collapsing scalar field in an AdS black brane geometry in five dimensions. Since the mass is turned on in a homogeneous manner in the boundary theory, we are studying a global quench. Further, as implied in our description of the gravitational dual, we are studying a thermal quench where the theory begins in a thermal state, rather than in the vacuum --- the latter simplifies the analysis, as we will describe below.

It is straightforward to find the solution for a static scalar field on an AdS background containing a planar black hole --- see \cite{scalarsol1} for the solution in $\mathcal{N}=2^\ast$ theory, and \cite{scalarsol2} for general operators satisfying the constraint $2\leq\Delta<4$. However, once the scalar field is given a time-dependent source, the nonlinear Einstein and Klein-Gordon equations become highly nontrivial to solve.  Treating the scalar field as a perturbation backreacting on the metric only at second-order, the problem becomes more tractable, since the scalar field and metric components decouple at leading order in the Klein-Gordon equation.  The solution to the metric then becomes that of the static background, plus a
time-dependent contribution which is second-order in the amplitude of the scalar.  Despite this
simplification, in the asymptotic series for the scalar field, one (time-dependent) coefficient remains undetermined, and can only be solved by evolving the scalar field forward in time from a known initial configuration.  In \cite{blm} and \cite{blmv}, a finite difference method was employed, which is computationally quite costly.  These studies were limited to first-order in the amplitude of
the scalar, meaning that the normalizable coefficient in the asymptotic expansion of the scalar
field was calculated, as well as some terms in the metric's asymptotic expansion which could be
directly calculated from this normalizable coefficient.  However, the full second-order profile of the
metric could not be determined in this way, making the calculation of nonlocal probes in the
geometry impossible.

Chebyshev spectral methods are powerful methods for solving systems of differential equations \cite{chebyshev}.  Representing the solution to the equations by a series of Chebyshev polynomials, we can approximate the full
radial profile of the solutions to a high degree of accuracy.  In the present paper we will apply these methods to the problem of solving for a massive scalar field in a five-dimensional AdS spacetime, as well as the
time-dependent profiles of the metric perturbations.

The organization of the rest of our paper is as follows:  we first introduce the physical setup of the scalar field on the AdS-black brane spacetime in sections $2$ and $3$,
as well as the coordinate system used.  We then go on in section $4$ to show the calculation of the thermalization of the
system, by studying different nonlocal quantities that can be calculated in this spacetime.  First we examine the
evolution of the apparent and event horizons.  We then go on to calculate the two-point correlation
functions.  Finally we calculate
the entanglement entropy of a strip on the boundary.  These calculations require knowing the full time-dependent geometry, \ie the full profile of the second-order metric components, and therefore rely on our numerical simulations of the evolution.  We find in particular
that the apparent and event horizons thermalize much sooner than the local one-point functions of the quenching operator,
and that wider separations in the two-point correlator and entanglement entropy thermalize later than for
narrower regions.  Furthermore, wider two-point functions and entanglement entropies  approach the thermalization time of the one-point function, and we expect wide enough surfaces to thermalize even later.

In section $5$ we investigate the thermalization behaviour of the previous section in closer detail.
The entanglement entropy and two-point functions are dual to minimal surfaces and geodesics extending
into the geometry of the spacetime, respectively.  Since the radial direction in the AdS geometry is related to
the energy scales in the field theory,
we can see the thermalization as happening due to the interaction of a range of energy scales,
rather than a scalar quantity equilibrating over time.  We end this section by discussing how
the different scales of the problem contribute to the thermalization.

In appendix \ref{a1} we discuss a method for solving the perturbative problem using interpolating Chebyshev polynomials, which is far less costly computationally than the finite difference methods
used in \cite{blm,blmv}.  This
method leads to equivalent results compared to the finite difference method used in the above-mentioned papers.  We also discuss convergence properties of this method.

In the present paper we will focus on a bulk spacetime of dimension $d+1=5$, with a scalar field of
mass\footnote{We set the AdS radius to $1$.} $m^2=\D\left(\D-d\right)=-3$.  We emphasize that we specialize to these cases by way of example, and the methods and algorithms described in this paper can easily be adapted for different $d$ and $\D$ (with the restriction $\frac{d}{2}\leq\D< d$).

This will serve as a prelude to the new solution of the full non-perturbative backreaction of the scalar field on the AdS-black brane geometry.

\section{The physical setup}

The physical system we would like to study is that of a scalar field $\phi$ on an AdS-black brane spacetime.
The evolution equations of the metric and scalar can be found by varying the five-dimensional Einstein-Hilbert
action
\begin{equation}
S_5=\frac{1}{16\pi G_5}\int d^5\xi \sqrt{-g} \left(R+12-\frac 12(\del\phi)^2
-\frac 12 m^2\phi^2+\calo(\phi^4)\right).
\eqlabel{action5}
\end{equation}
Of particular interest to us is the case $m^2=-3$, since the scalar field is then dual to a fermionic
mass operator with $\D=3$ in a thermal $\mathcal{N}=2^*$ gauge theory living in four flat spacetime dimensions \cite{pw,bpp,j}. 
We use the background ansatz of an infalling Eddington-Finkelstein metric and a scalar field which depend only on the radial and time directions in the spacetime, while being isotropic in the four transverse directions:
\begin{equation}
ds_5^2=-A(v,r)\ dv^2+\Sigma(v,r)^2\ (d\vec{y})^2+2dr dv,\qquad \phi=\phi(v,r).
\eqlabel{ansatz}
\end{equation}
In \eqref{ansatz} $r$ is the light-like radial coordinate of the spacetime, $v$ is the time coordinate and $\vec{y}$ are the coordinates corresponding to the spatial directions on the conformal boundary.
We would like to send in a scalar field $\phi\left(v,r\right)$ from the boundary of this spacetime at
$r=\infty$.  Varying the metric and scalar field in \eqref{action5} leads to the equations of motion
\cite{blm}
\begin{equation}
\begin{split}
0=&\Sigma\, \del_r\!(\dot{\Sigma})+2\dot{\Sigma}\,\del_r\!\Sigma -2\Sigma^2+\frac{1}{12}m^2\phi^2\Sigma^2,\\
0=&\del_r^2\!A-\frac{12}{\Sigma^2}\dot{\Sigma}\,\del_r\!\Sigma+4+\dot{\phi}\,\del_r\phi-\frac 16 m^2\phi^2,\\
0=&\frac 2A\ \del_r\!(\dot{\phi})+\frac{3\,\del_r\!\Sigma}{\Sigma A}\ \dot{\phi}+\frac{3\,\del_r\phi}{\Sigma A}\
\dot{\Sigma}-\frac{m^2}{A}\ \phi,\\
0=&\ddot{\Sigma}-\frac 12 \del_r\!A\,\dot{\Sigma}+\frac 16 \Sigma\, (\dot{\phi})^2,\\
0=&\del_r^2\,\Sigma+\frac 16\Sigma\, (\del_r\phi)^2,
\end{split}
\eqlabel{eoms}
\end{equation}
where
\begin{equation}
\dot{h}\equiv \del_v h+\frac 12 A\, \del_r h,
\eqlabel{der}
\end{equation}
for any $h$.

Setting the scalar field to zero, there is no longer a source for dynamics in the spacetime.  We are
then left with a static, planar black hole metric which can be described by the line-element \cite{blm,blmv}
\begin{equation} \eqlabel{statmetric}
ds^2=-(r^2-\frac{\mu^4}{r^2})\,dv^2+r^2(d\vec{y})^2+2dr dv,
\end{equation}
where $r=\mu$ is the position of the event horizon. Of course, this parameter also sets the temperature of the corresponding plasma in the boundary theory, \ie
$T=\mu/\pi$. Since the scalar field is initially zero when we turn on the quench in the asymptotic past, the
above static spacetime is the initial equilibrium configuration of our system and $\mu$ sets the initial temperature in our thermal quenches.  As the mass coupling of the fermionic operator is switched on, the changing boundary conditions excite the scalar field in the AdS-black brane background, collapsing into the black hole.  The scalar field excitations evolve
and backreact on the metric, and the bulk fields reach a different equilibrium configuration in the asymptotic
future.  In this final static configuration, the metric will be modified from its initial form in
\eqref{statmetric}. In particular, the black hole would have grown due to the energy it absorbed from the
infalling scalar field excitations.  The scalar field will also have a nonzero static profile because of the new boundary conditions imposed at asymptotic infinity.

Beginning with an initial state allows us to simplify the analysis of the quenches. In particular, the initial state is provides an energy scale, \ie the initial temperature $T_i$, and we will only study quenches where the final mass $m_f$ of the fermionic operator is small compare to that scale. That is, we only consider quenches where $m_f/T_i\ll1$, following \cite{blm,blmv}. In the dual gravitational description, this choice corresponds to
treating the scalar as a perturbation on the background geometry. In other words, we assume that the black hole is
very large so that it is possible to perform an expansion in the amplitude of the scalar  field
in equations \eqref{eoms}, as in \cite{blm,blmv}.
To leading order in its amplitude the scalar field equation becomes the
equation of a scalar field on the static background metric \eqref{statmetric}.  In references \cite{scalarsol1,scalarsol2}, 
the analytic profile of the static perturbative scalar field was found for both particular and general values of $m$, respectively.  In the special case where $m^2=-3$ that we will be considering, the scalar field
was found to have the profile \cite{blm}
\begin{equation}
\phi(r) = \ell\pi^{-1/2}\Gamma\left(\frac{3}{4}\right)^2\,\left(\frac{\mu}{r}\right)^3
\,_2F_1\left(\frac{3}{4},\frac{3}{4},1,1-\frac{\mu^4}{r^4}\right),
\end{equation}
where $\ell$ parameterizes the amplitude of the bulk scalar and also the value of the mass coupling in the boundary theory. Regardless of the dynamics during the evolution of the quench, the scalar will relax to profile of the above form in the final equilibrium configuration. In \cite{blm,blmv}, it was still necessary to
know the full evolution of the bulk scalar in order to calculate late-time quantities such as the change
in the stress-energy tensor of the dual field theory and the change in temperature, which were determined in
terms of integrals of the normalizable mode over time.

\section{Dimensionless coordinates}

We introduce the dimensionless coordinates by scaling out a factor of the black hole horizon position
$\mu$ (which has units of energy)
\begin{equation}
\begin{split}
&\r=\frac\mu r,\qquad \t=\mu v.
\end{split}
\eqlabel{dimensionless}
\end{equation}
After scaling out the appropriate factor of $\mu$, the warp factors become
\begin{equation}
\begin{split}
A=\mu^2 a,\qquad \Sigma=\mu s.
\end{split}
\eqlabel{dimensionless2}
\end{equation}
Despite having dimensions of length, a more careful analysis shows that the scalar field should not be rescaled
by a factor of $\mu^{-1}$.
The new radial coordinate $\rho$ is particularly useful, since now the conformal boundary of
the spacetime is located at $\rho=0$.  As noted above the dimensionful constant
$\mu$ can be interpreted as
the initial position of the black brane horizon.  Therefore, the initial horizon is now situated at
$\rho=1$.
With these redefinitions, the field equations \eqref{eoms} become the dimensionless equations
\begin{align}
0=&\del_{\t}\del_{\r}\phi-\frac12 \r^2 a \del_{\r}^2\phi+\biggl(-\frac12 \r^2 \del_\r a-\r a-\frac 32 a \r^2 \del_\r \ln s +\frac 32
\del_\t \ln s\biggr) \del_\r \phi+\frac 32 \del_\r \ln s \del_\t \phi\nonumber\\
&+ \frac{m^2 \phi}{2\r^2},
\eqlabel{eq1}\\
0=&\del_{\t}\del_{\r}s-\frac 12 \r^2 a \del_{\r}^2s-\frac{\r^2 a}{ s} (\del_\r s)^2+\frac 2s  \del_\t  s \del_\r  s
+\biggl(-\frac12 \r^2 \del_\r a-\r a\biggr) \del_\r s\nonumber\\
&+\frac{1}{12} \frac{s (24-m^2 \phi^2)}{\r^2},
\eqlabel{eq2}\\
0=&\del_{\r}^2a+\frac 2\r \del_\r a+\biggl(-6 (\del_\r \ln s)^2+\frac12 (\del_\r \phi)^2\biggr) a
+\frac{12}{\r^2} \del_\r \ln s \del_\t \ln s+\frac{4}{\r^4}-\frac {1}{\r^2}\del_\r \phi \del_\t \phi\nonumber\\
&-\frac{m^2 \phi^2}{6\r^4}.
\eqlabel{eq3}
\end{align}
These are the evolution equations for the scalar field and the two warp factors in the metric.  Along with these, the Einstein equations provide two constraints, namely
\begin{align}
0=&\frac 16 (\del_\r \phi)^2 s+\del_{\r}^2s+\frac 2\r \del_\r s,
\eqlabel{eq4}\\
0=&\del_{\t}^2s-\frac 12\r^2  \del_\t a \del_\r s -a\r^2 \del_{\t}\del_{\r}s +\frac14 a^2 \r^4 \del_{\r}^2s
+\frac 12 a^2 \r^3 \del_\r s
+\frac12 \r^2 \del_\r a \del_\t s+\frac16 s (\del_\t \phi)^2\nonumber\\
&-\frac16 s a \r^2 \del_\t \phi  \del_\r \phi
+\frac{s a^2}{24\r^4} (\del_\r \phi)^2.
\eqlabel{eq5}
\end{align}
Used in combination, \eqref{eq4} and \eqref{eq5} determine the response of the warp factor $a$ up to an arbitrary integration constant \cite{blm}.

\subsection{$m^2=-3$}
Specializing equations \eqref{eq1} -- \eqref{eq5} to a scalar with mass $m^2=-3$, we find an asymptotic solution to the scalar and warp factors as $\r\to 0$ of \cite{blm}
\begin{equation}
\begin{split}
\phi=&p_0\ \r + p_0'\ \r^2 +\r^3\biggl(p_2+\ln\r\left(\frac 12 p_0''+\frac 16 p_0^3\right)\biggr)+\calo(\r^4\ln\r),\\
a=&\frac{1}{\r^2}-\frac 16 p_0^2+\r^2\left(a_2+\ln\r\biggl(\frac 16 (p_0')^2-\frac 16 p_0 p_0''-\frac{1}{36}p_0^4\biggr)\right)
+\calo(\r^3\ln\r),\\
s=&\frac{1}{\rho}-\frac{1}{12}\rho p_0^2-\frac19\rho^2 p_0p_0'+\calo(\r^3\ln\r),
\end{split}
\eqlabel{m3a}
\end{equation}
where $p_0$, $p_2$ and $a_2$ are functions of $\t$, a prime here denotes a derivative with respect to $\t$.  The coefficient $p_0$ is the so-called `non-normalizable mode' or the source coefficient \cite{adsreview}.
Here we will choose this coefficient to have a time-dependent profile, implying that the scalar field is sourced at the conformal boundary of the AdS spacetime and excitations are sent into the bulk geometry in a time-dependent manner.  The coefficient $p_2$ is the so-called `normalizable mode', or the response coefficient.  This is the coefficient which is to be determined given a source $p_0$.  While analytic solutions of $p_2$ are known when $p_0$ varies very slowly from time $\t=-\infty$ to $\t=+\infty$ \cite{blm, blmv}, as well as for $p_0$ made time-dependent abruptly
and over a very short period of time \cite{bmv}, no analytic solutions for the normalizable mode are currently known for a source with general time-dependence.

The solutions presented in \eqref{m3a} are for the full nonlinear equations.  There is an additional constraint on $a_2$ coming from eqs.~\eqref{eq4} and \eqref{eq5} \cite{blm}:
\begin{equation}
0=-\frac 16 p_0' p_2+\frac{1}{36} p_0' p_0''-\frac{5}{108} p_0^3 p_0'+\frac 12
a_2'-\frac 19 p_0 p_0'''+\frac16 p_0 p_2'.
\eqlabel{const}
\end{equation}
The full warp factors can in principle therefore be determined completely given $p_0$ and $p_2$.

The nonlinearities in the equation determining the scalar field make it challenging to extract the response coefficient $p_2$.  For this reason, in \cite{blm,blmv}, the scalar field was treated as a perturbation on the spacetime, linearizing the Klein-Gordon equation \eqref{eq1}.  It then becomes a simple procedure to numerically determine the response.  In the following, we also carry this amplitude expansion to
second-order in the metric coefficients in order to determine the leading-order backreaction of the
scalar field on the background.

In the following subsection we will describe the asymptotic solution of the scalar field and warp factors in this perturbative regime.  In the appendix we show how to solve the system using Chebyshev interpolation methods, which allow us to find the full profile of these metric perturbations, rather than single terms in
the asymptotic expansion.

\subsection{Leading-order backreaction}

Since in our analysis, the scalar field backreacts only
perturbatively on the spacetime, we are implicitly probing the limit
of a very large black brane in the AdS spacetime.  This is the
gravitational dual of switching on an operator in a thermal plasma at a
very high temperature.  In the boundary theory, the dual of the expansion in the amplitude of the scalar field
is the expansion in $\frac{m_f}{T_1}\ll1$, where $m_f$ indicates the mass of the fermionic operator and $T_i$ is the temperature of the initial thermal state, as described above. As discussed in detail in \cite{blm, bmv}, the
long-time evolution of the warp factors found in the perturbative
regime is questionable in the very short quench limit.  In this limit
we find that the change in radius of the black brane becomes
significant.  Nonetheless, we were able to find fully general results for certain questions,
namely the scaling of the response (at early times) and the energy injected into the system,
as described in \cite{bmv}.  This justifies the perturbative approach at
all quenching rates in \cite{blm, blmv}.

Expanding the scalar field in a small parameter $\ell$, it backreacts only at order $\ell^2$ on the metric, because the stress tensor of the gravity theory is quadratic (and higher) in $\phi$.
Hence the metric has a static part and a dynamical part as
\begin{equation}
\begin{split}
\phi(\t,\r)=&\ell\ \hp(\t,\r)+\calo(\ell^3),\\
a(\t,\r)=&\frac{1}{\r^2}-\r^2+\ell^2\ \ha(\t,\r)+\calo(\ell^4), \\
s(\t,\r)=& \frac 1\r e^{\ell^2 b(\t,\r)}+\calo(\ell^4). \\
\end{split}
\eqlabel{fdef}
\end{equation}
Given equation \eqref{m3a}, the leading perturbative part of the scalar field and metric are then
\begin{equation}
\hp(\t,\r)=\r\left(p_0+\r\ p_0'+\r^2\ \biggl(p_2+\frac 12 p_0''\ln\r\biggr) +\dots\right),\label{hpl2}
\end{equation}
\begin{equation}
\hat{a}(\t,\r)=-\frac 16 p_0^2+\r^2\left(a_{2,2}
+\ln\r\biggl(\frac 16 (p_0')^2-\frac 16 p_0 p_0''\biggr)\right)+\dots,\label{al2}
\end{equation}
\begin{equation}
b(\t,\r)=-\frac{1}{12}\rho^2 p_0^2-\frac19\rho^3 p_0p_0'+\dots,\label{bl2}
\end{equation}
and the equations of motion \eqref{eq1}-\eqref{eq3} take form:
\begin{equation}
\begin{split}
0=&\del_{\t}\del_{\r}\hp -\frac {1-\r^4}{2}\del_{\r}^2\hp+\frac{3+\r^4}{2\r}\del_\r\hp-\frac{3}{2\r}\del_\t\hp-\frac{3}{2\r^2}\hp,
\end{split}
\eqlabel{neq1}
\end{equation}
\begin{equation}
\begin{split}
0=&\del^2_{\r}\ha+\frac 2\r \del_\r \ha-\frac{6}{\r^2}\ha+\frac{12(1-\r^4)}{\r^3}\del_\r b-\frac{12}{\r^3}\del_\t b
+\frac{1-\r^4}{2\r^2}(\del_\r\hp)^2-\frac{1}{\r^2}\del_\r\hp\del_\t\hp+\frac{1}{2\r^4}\hp^2,
\end{split}
\eqlabel{neq2}
\end{equation}
\begin{equation}
\begin{split}
0=&\del_{\t}\del_{\r}b- \frac{1-\rho^4}{2}\del^2_{\r}b+\frac{3-\rho^4}{\r} \del_\r b-\frac{3}{\r} \del_\t b
+\frac\r2 \del_\r \ha-\ha+\frac {1}{4\r^2} \hp^2,
\end{split}
\eqlabel{neq3}
\end{equation}
while the constraints \eqref{eq4}-\eqref{eq5} take form:
\begin{equation}
\begin{split}
0=&\del^2_{\r} b+\frac 16 (\del_\r\hp)^2,
\end{split}
\eqlabel{neq4}
\end{equation}
\begin{equation}
\begin{split}
0=&\del^2_{\t}b+\left(\r^7-4\r^3+\frac{3}{\r}\right)\del_\r b+\frac{\r^4-3}{\r} \del_\t b
+(1-\r^4) \left(\frac 12 \r \del_\r \ha-\ha\right)
+\frac 12 \r \del_\t \ha\\
&+\frac16 (\del_\t \hp)^2-\frac{1-\r^4}{6} \del_\r\hp\del_\t \hp
+\frac{(1-\r^4)^2}{12} (\del_\r \hp)^2+\frac{1-\r^4}{4\r^2} \hp^2.
\end{split}
\eqlabel{neq5}
\end{equation}

\subsection{Rescaling the parameters} \label{rescaling}

In this paper we give the scalar field source $p_0$ the time-dependent profile
\begin{equation}
p_0(\t)=\frac{1}{2}\left(1+\tanh\left(\frac{\t}{\a}\right)\right),
\end{equation}
where $\a$ is the characteristic timescale on which the quench takes place.  It will be useful to rescale our coordinates
and fields such that we can compare different quenches on the same time and length scale, which make it
easier to see how quantities behave in the fast quench limit.

As discussed in \cite{blm}, the required rescaling is $\r\to\a\r$,
$\t\to\a\t$ and $\vec{y}\to\a\vec{y}$.  The fields rescale
as\footnote{More correct is to say that this is a leading order fast-quench rescaling.
The rescaling to second-order is
$a\to\frac{1}{\a^2}\left(\frac{1}{\r^2}-\a^4\r^2\right)+\ell^2\hat{a}$.
In other words $\hat{a}$ is not rescaled} $a\to a/\a^2$ , $s\to s/\a$ and $\phi\to\a\phi$.
In these rescaled coordinates the horizon
of the black hole will be located at $\r=1/\a$, and the source of the scalar field would be
\begin{equation}
p_0(\t)=\frac{1}{2}\left(1+\tanh\left(\t\right)\right).
\eqlabel{fastres1}
\end{equation}
The expression for the metric remains unchanged:
\begin{equation}
ds_5^2=-a\ d\t^2+s^2\ (d\vec{y})^2-2\frac{d\r d\t}{\r^2}.
\eqlabel{fastres2}
\end{equation}

\section{Probes of thermalization}

Knowing the profiles of the metric coefficients and the response coefficient in the asymptotic expansion
of the scalar field, we would like to obtain a meaningful measure of the thermalization time of the
field theory following the quench.  Since the geometry fluctuates, but then returns to a static configuration after some
time, we can conclude that the gauge theory plasma does (effectively) thermalize.
An interesting question
to ask then is whether the broken conformality of the theory introduces different scales for which
thermalization occurs at different rates.  Indeed, in \cite{therm1,vaidya,hong}
it was observed for Vaidya-type metrics that the theory thermalizes at the UV (short distance) range before thermalizing in the IR (large distance) range.

The Vaidya approach \cite{therm1,vaidya,hong} considers a thin planar collapsing shell of
null dust in AdS spacetime (an expanding shell in its original construction \cite{vaidya43}), which produces a
metric outside the shell equal to that of an AdS-black brane, and leaves the inside of the shell
to be that of empty AdS spacetime.  While this may seem an exotic form of bulk matter to consider,
 these constructions can be related to a collapsing thin shell of a massless scalar field in AdS
\cite{minwalla}. In any event, the gravitational picture suggests that the dual field
theory thermalizes instantaneously at the higher energy scales, while working its way down
from the UV to the IR scales.  This type of setup certainly describes exceptional quenches
of the dual field theory, and it is not clear to what extent the lessons learned from these studies extend to general quenches.  Hence in the present case, as in \cite{blm,blmv}, we are considering rapid but smooth
quenches, where in the gravitational dual the bulk scalar field evolves smoothly in space and time throughout
the background geometry.

In \cite{blm, blmv}, the thermalization time was approximated by observing when the response
coefficient of the scalar field was within $5\%$ of its final equilibrium value.\footnote{In this paper,
 we will use a stricter $2\%$ criterion --- see below.}  Of course, one
limitation of this method is that it is essentially measuring the thermalization time using the
one-point correlator of the quenching operator $\langle\mathcal{O}_{\D}\rangle$.
While in principle, the response of the one-point function depends on the whole range of energies from the IR to the UV, it cannot be used to distinguish between the different
contributions from the different scales.
It would therefore be interesting to employ nonlocal probes, as in \cite{therm1,vaidya},
to study the thermalization process more carefully.

An important step in this direction was made by \cite{periodic}, in which the authors probed the
thermalization of a periodically driven quench using holographic two-point functions and entanglement
entropy.  We will now extend their results for a non-periodic quench.  An important difference between
the current paper and \cite{periodic} is that the source we use is not periodic, and can be tuned to be a
step function in the case of an instantaneous quench.  Periodic quenches are not truly realizable in
the perturbative regime we are considering, since after a finite time the full nonlinear backreaction of the
scalar field on the background must be considered.

In this section we first describe the analytic and numerical methods used to calculate the perturbation of the
apparent and event horizons of the black brane.  We then go on to discuss the calculation of the two-point
function and the entanglement entropy in the field theory using holographic methods.  We then show that our
results agree with the results in the literature \cite{therm1,vaidya}, namely that wider probes have longer
thermalization times than narrower ones.

\subsection{Evolution of the apparent and event horizons}

As the scalar excitations are sent into the bulk geometry, and fall onto the black hole,
the black hole will necessarily grow.
While one would need a fully nonlinear evolution of the spacetime to see the full reaction of the geometry, it is still possible
to probe the growth of the black hole horizon in the perturbative regime, as per \cite{periodic}.

When one speaks of the black hole horizon, it can mean either the apparent or the event horizon.  In the case of a static black
hole, the two horizons necessarily coincide.  In the dynamical case, they can evolve at different rates, with the condition that
they coincide again once equilibrium is reached.

The apparent horizon is located at the radius where an outward pointing null geodesic stays at constant radius at that moment in time, \ie
it is the trapped surface of null geodesics.  The event horizon is the surface outside which a light ray must be in order to escape to infinity.
Intuitively a light ray may be able to move toward the outside of the black hole, but if it is inside the event horizon, then the apparent horizon, which
is also growing outward, will eventually catch up with the light ray and cause it to fall into the black hole.

Locating the apparent and event horizons is useful, because it gives a nonlocal measure for the
thermalization of the quenched system.  It is further  a good consistency check of our numerics, since
these properties of a spacetime are well understood.  If our numerical methods correctly evolve the
metric, we would expect the apparent horizon to always be located inside the event horizon.  We would
also always expect the area of the event horizon to grow monotonically as we pump energy into the
black hole.

The apparent horizon is located at the radius where the expansion $\theta$ of a congruence of outward pointing
null vectors vanishes (\ie it stops expanding outwards).  Working in the coordinates of equation
\eqref{ansatz}, we we characterize such a congruence with
the null vector $k=\partial_v+\frac{A}{2}\partial_r$.
The null vector $k$ points toward the boundary of the spacetime outside of the initial stationary black hole,
and points inward inside the initial horizon.

Following \cite{poisson}, the expansion of a congruence of affine parameterized null vectors $n$ is given by
\begin{equation}
\theta=\nabla_{\a}n^{\a}.
\end{equation}
However, it turns out that $k^{\b}\nabla_{\b}k^{\a}=\frac{1}{2}A'k^{\a}$, \ie $k$ is not affine (the prime meaning the derivative with respect to $r$).
To remedy this, we rescale $k$ by $\exp\{-\int \left(\frac{1}{2}A'\right)d\l\}$, where $\l$ is the parameter along
which the congruence $k$ evolves.  This ensures that the rescaled null vector satisfies the geodesic equation with $\lambda$ as an affine parameter.
Reference \cite{poisson} then gives the expansion of $k$ to be
\begin{equation} \label{expansion}
\theta = \exp\left[-\int \left(\frac{1}{2}\dot{a}\right)d\l\right]\left(\nabla_{\a}k^{\a}-\frac{1}{2}a'\right).
\end{equation}
Substituting in for $\nabla_{\a}k^{\a}$, we see that $\theta=0$, when
\begin{equation} \eqlabel{apparentr}
A \Sigma'+2\dot{\Sigma}=0,
\end{equation}
where the prime represents a derivative with respect to $r$, and the dot represents a derivative with respect to $v$.

In order to solve the equation, we change coordinates to the rescaled coordinates $\r$ and $\t$, in which the
unperturbed event horizon is located at $\r=\frac 1\a$, $\a$ being the quenching rate.  Equation \eqref{apparentr} then gets modified to be
\begin{equation} \eqlabel{apparentr2}
\a^2\r^2 a\,s'-2\dot{s}=0,
\end{equation}
where the equation is now in terms of the new radial and time coordinates $\r$ and $\t$.  Expanding $a$ and $s$ in terms of the perturbation
parameter $\ell$, and using the ansatz that the time-dependent position of the apparent horizon is $\frac 1\a+\ell^{2}\r_{a}(\tau)$, we see
that to zeroth-order in $\ell$, \eqref{apparentr2} is trivially satisfied. However, at order $\ell^{2}$, \eqref{apparentr2} gives an
expression for $\r_{a}$, namely
\begin{equation} \eqlabel{ra}
\r_{a}=\left[\frac{1}{4\a} \hat{a}+\frac{\dot{b}}{2\a^2 }\right]_{\r=\frac{1}{\a}}.
\end{equation}

The entropy of a black hole at equilibrium is related to its horizon
surface area by the Bekenstein-Hawking entropy formula \cite{ghent}
\begin{equation}
S = \frac{A_{hor}}{4\,G}, \label{yoza}
\end{equation}
$G$ being Newton's constant. Of course, for the planar black hole under consideration, the area of the
event horizon is infinite and hence we consider instead the {\it area density} of the horizon.
That is, we can calculate the measure which would be integrated over the (spatial) gauge theory directions to
evaluate the total area of the horizon. In a static configuration, \ie at equilibrium, this area density can
be related as above in (\ref{yoza}) to an entropy density
\begin{equation}
\mathcal{S} = \frac{V_{hor}}{4\,G},
\end{equation}
which is dual to the thermal entropy density of the corresponding plasma in the dual field theory.
It was proposed, \eg in \cite{edensity}, that this entropy density of the apparent horizon should have the same
interpretation as the dual entropy density in the boundary theory even in dynamical situations.

As above, we use $V$ to denote the area density or volume element of the black hole horizon.
The full (dynamical) area density is then given by
\begin{eqnarray}
V_{a} &=& s(\t,\frac 1\a+\ell^2 \r_{a})^3\equiv 1+\ell^2\ \dd V_a \nonumber\\
&=& 1+\ell^2\ \left[-3\a\r_{a}+3 b\right]_{\r=\frac{1}{\a}}.\eqlabel{vafull}
\end{eqnarray}
Therefore the perturbation of the area density of the apparent horizon is given by
\begin{equation}
\dd V_{a} = \left[3 b-\frac{3}{4} \hat{a}-\frac{3}{2\a}\dot{b}\right]_{\r=\frac{1}{\a}}.
\eqlabel{aaold}
\end{equation}
Furthermore, using the constraint \eqref{neq5} evaluated at the horizon, $\r=\frac 1\a$, 
we find 
\begin{equation}
\frac{d}{d\t}\dd V_{a}=\biggl[\frac{1}{4\a} \left(\del_\t \hat{\phi}\right)^2\biggr]_{\r=\frac 1\a}\qquad 
\Longrightarrow\qquad \dd V_{a}=\int_{-\infty}^\tau dt \biggl[\frac{1}{4\a} \left(\del_t \hat{\phi}\right)^2\biggr]_{\r=\frac 1\a}\,,
\eqlabel{dvdt}
\end{equation} 
implying that the area of the apparent horizon {\it monotonically increases} with time,
in agreement with  the area theorems of \cite{Hayward:1993mw} (see also \cite{booth})\footnote{We would like 
to thank Mukund Rangamani and Moshe Rozali for valuable discussions regarding the area theorems.}.
Note that at late times, after the system has equilibrated, $\dot{b}=0$, and therefore
\begin{equation}
\dd V_{a} = \left[3 b-\frac{3}{4} \hat{a}\right]_{\r=\frac{1}{\a}}
\end{equation}
at equilibrium. 
In what follows we compare the computation of  the perturbation of the area density of the apparent horizon
as given by  \eqref{aaold} and \eqref{dvdt}. An agreement represents a useful test of our numerics, in particular 
the constraint equation \eqref{neq5}.

We would also like to calculate the location and area of the event horizon of the black hole.  This is
a more involved calculation, since it is a global property of the spacetime, and therefore cannot
be read off from the fields at any one moment in time.

The equation satisfied by the event horizon can be obtained from the line-element.  The position of the
event horizon is the outermost radius at a point in time from which a null ray cannot escape to infinity.
Since the event horizon is an expanding null surface, an outward-pointing null ray lying on the event horizon will move
outward with the event horizon and will stay at the same radius as the event horizon throughout the
evolution, until it becomes stationary again when equilibrium is reached.  In other words, the event
horizon follows a null trajectory.

Working in the rescaled coordinate system, and working at fixed position in the transverse
directions, we can therefore set the proper-time along a radial geodesic situated at the event
horizon to zero:
\begin{equation} \eqlabel{lineel}
0=-a d\t^2 - 2\frac{d\r\, d\t}{\r^2}.
\end{equation}
Substituting for $a=\frac{1}{\a^2\r^2}-\a^2\r^2+\ell^2\hat{a}$, and $\r=\frac 1\a+\ell^2\r_e$, and dividing by $d\t^2$,
equation \eqref{lineel} simplifies to \cite{periodic}
\begin{equation} \eqlabel{reeq}
\frac{d\r_e}{d\t} = 2\a\r_e - \frac{1}{2}\hat{a}.
\end{equation}
Notice that at late times when $\frac{d\r_e}{d\t}=0$, the equation has the solution $\r_e=\frac{1}{4\a}\hat{a}$,
therefore coinciding with the radius of the apparent horizon \eqref{ra} at late times.  The differential
equation \eqref{reeq} has a general solution
\begin{equation} \eqlabel{resol}
\r_e(\t) = e^{2\a\t}\left(\r_i e^{-2\a\t_i} - \frac{1}{2}
\int^{\t}_{\t_i}e^{-2\a t}\hat{a}\left(t,\frac 1\a\right)dt\right),
\end{equation}
where $\r_i$ is the radius of the event horizon at initial time $\t_i$. Taking the limit for late
times $\t\rightarrow\infty$ in equation \eqref{resol}, we see that the prefactor $\exp{(2\a\t)}$ diverges.
In order that the null ray (lying on the event horizon) does not shoot off to plus or minus infinity, and
for the solution to make physical sense, we need
\begin{equation}
\biggl[\r_i e^{-2\a\t_i} - \frac{1}{2}
\int^{\infty}_{\t_i}e^{-2\a t}\hat{a}\left(t,\frac 1\a\right)dt\biggr]\to0\quad\text{as}\quad\t\to\infty.
\end{equation}
This gives the position of the event horizon at time $\t_i$ as
\begin{equation}
\r_e(\t_i)=\frac{1}{2}e^{2\a\t_i}\int^{\infty}_{\t_i}e^{-2\a t}\hat{a}\left(t,\frac 1\a\right)dt.
\end{equation}
To calculate this quantity numerically, it is somewhat easier to integrate the interval
$-\infty<\t\leq\t_i$ than $\t_i\leq\t<\infty$, since one only needs to know the past evolution of
the system (as well as the overall evolution).  The position of the event horizon can then be expressed as
\begin{equation}
\r_e(\t_i)=\frac{1}{2}e^{2\a\t_i}\left(\int^{\infty}_{-\infty}-\int^{\t_i}_{-\infty}\right)
e^{-2\a t}\hat{a}\left(t,\frac 1\a\right)dt.
\end{equation}
To implement the evolution of the event horizon, we calculate the function
\begin{equation}
\r_{temp}(\t)=\int^{\t}_{-\infty}e^{-2\a t}\hat{a}\left(t,\frac 1\a\right)dt,
\end{equation}
at each timestep. At the end of the numerical evolution of the spacetime, we  can
calculate $\r_e$ as
\begin{equation}
\r_e(\t)=\frac{1}{2}e^{2\a\t}\left(\r_{temp}(\infty)-\r_{temp}(\t)\right).
\end{equation}
The value of $\r_{temp}(\infty)$ is determined by numerically taking the limit
$\lim_{\t\rightarrow\infty}\r_{temp}(\t)$.  The result is accurate, because we calculate the function
up until late times, after the evolution has reached equilibrium.

Similar to the case above, the area density of the event horizon is given by
\begin{equation} \eqlabel{vefull}
V_{e} \equiv 1+\ell^2\ \dd V_e = 1+\ell^2\left[-3\a\r_{e}+3 b\right]_{\r=\frac{1}{\a}}.
\end{equation}

\begin{figure}
\begin{center}
\psfrag{t}[Br][tl]{{$\scriptstyle{\t}$}}
\psfrag{v}[c]{{$\scriptstyle{\dd V_{e/a}}$}}
\psfrag{a}[cc][][1.5]{{$\scriptstyle{\a=1}$}}
\psfrag{b}[cc][][1.5]{{$\scriptstyle{\a=\frac{1}{2}}$}}
\psfrag{c}[cc][][1.5]{{$\scriptstyle{\a=\frac{1}{4}}$}}
\psfrag{d}[cc][][1.5]{{$\scriptstyle{\a=\frac{1}{8}}$}}
  \includegraphics[width=3in]{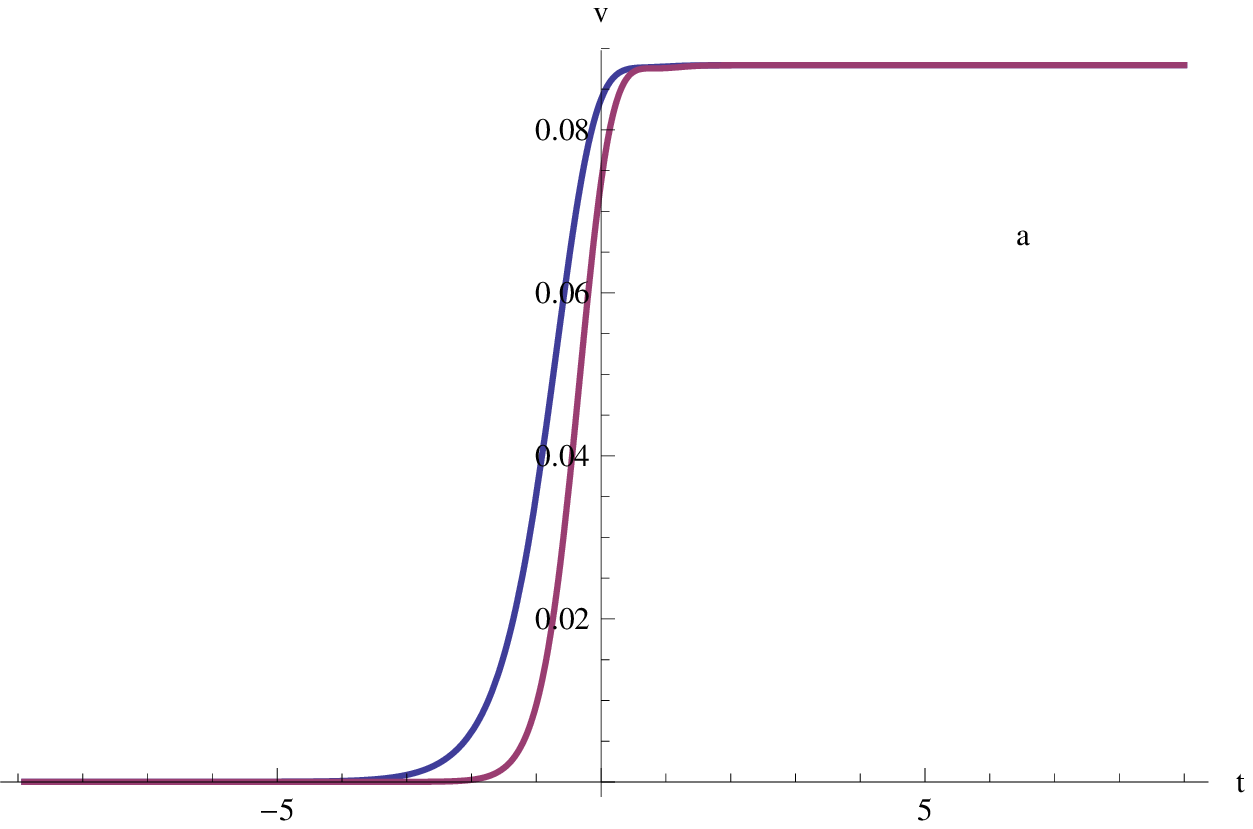}
    \includegraphics[width=3in]{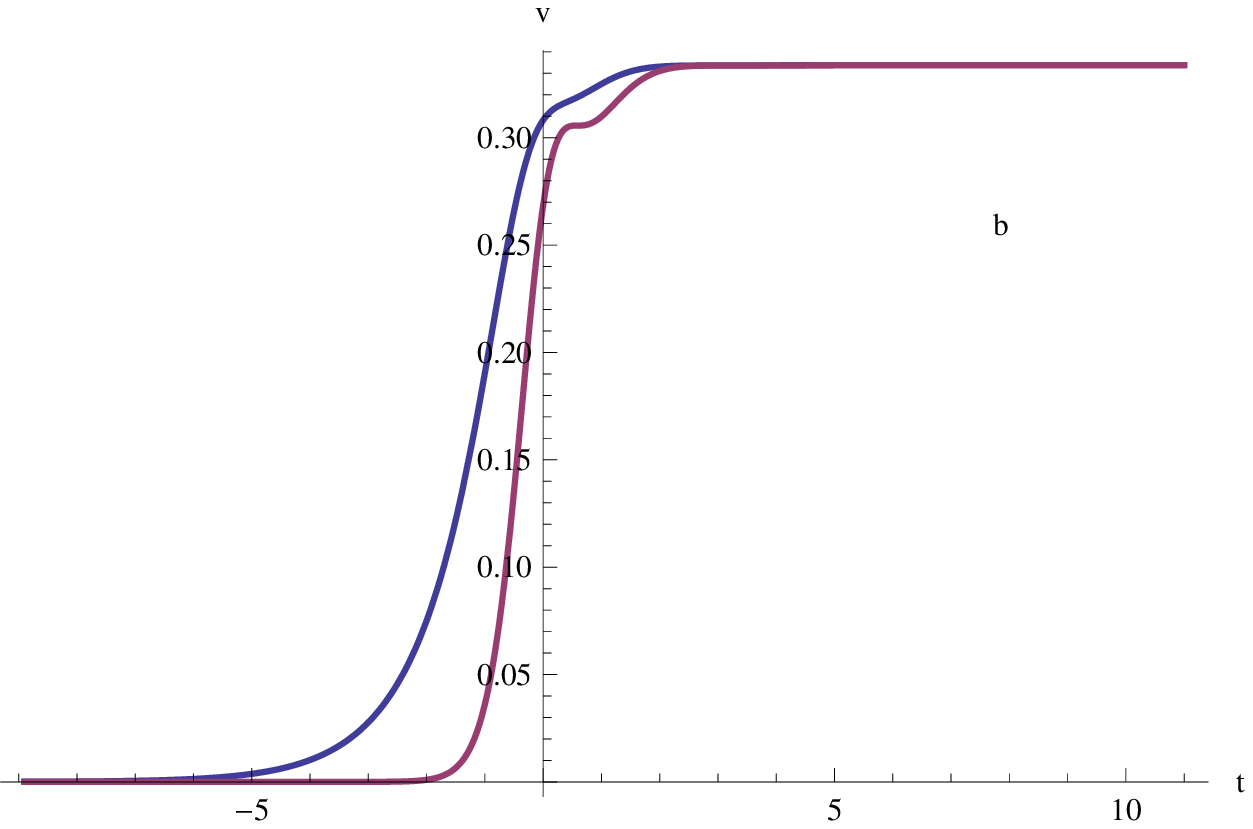}
      \includegraphics[width=3in]{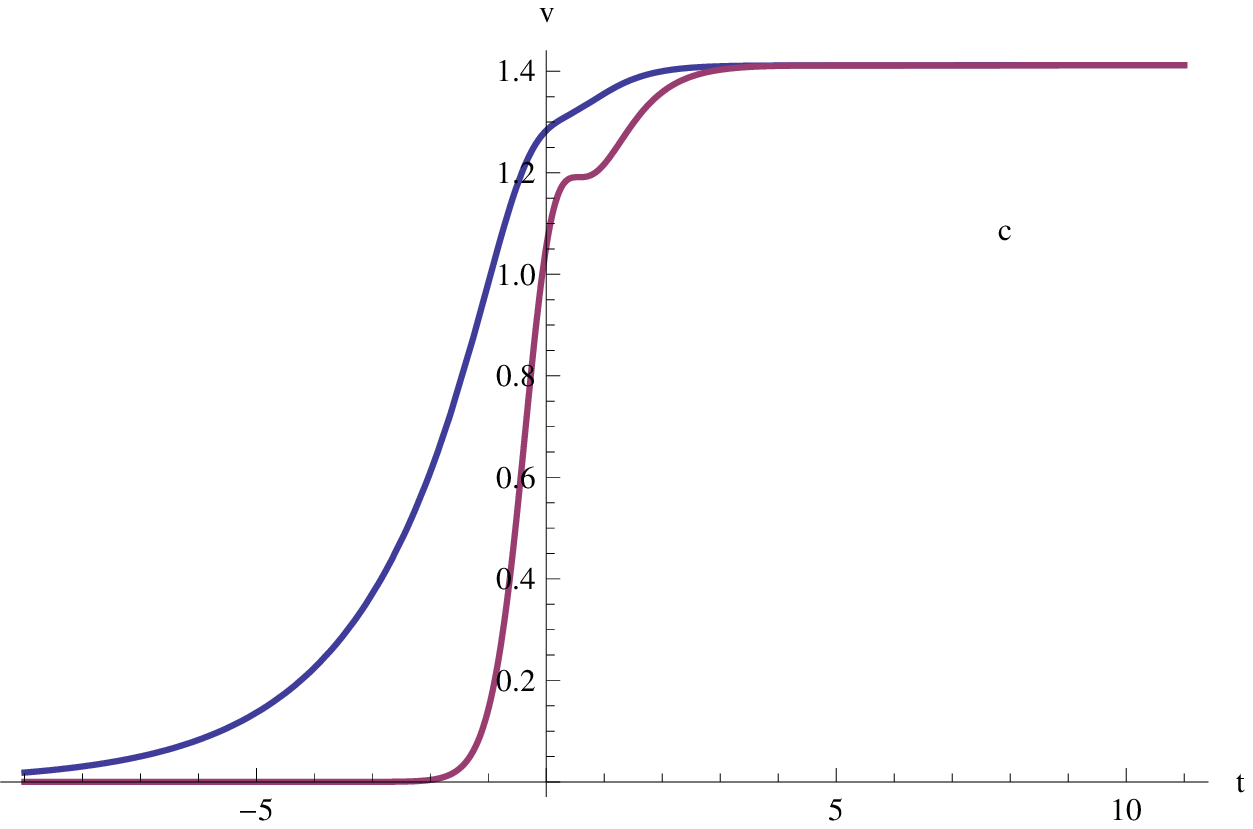}
        \includegraphics[width=3in]{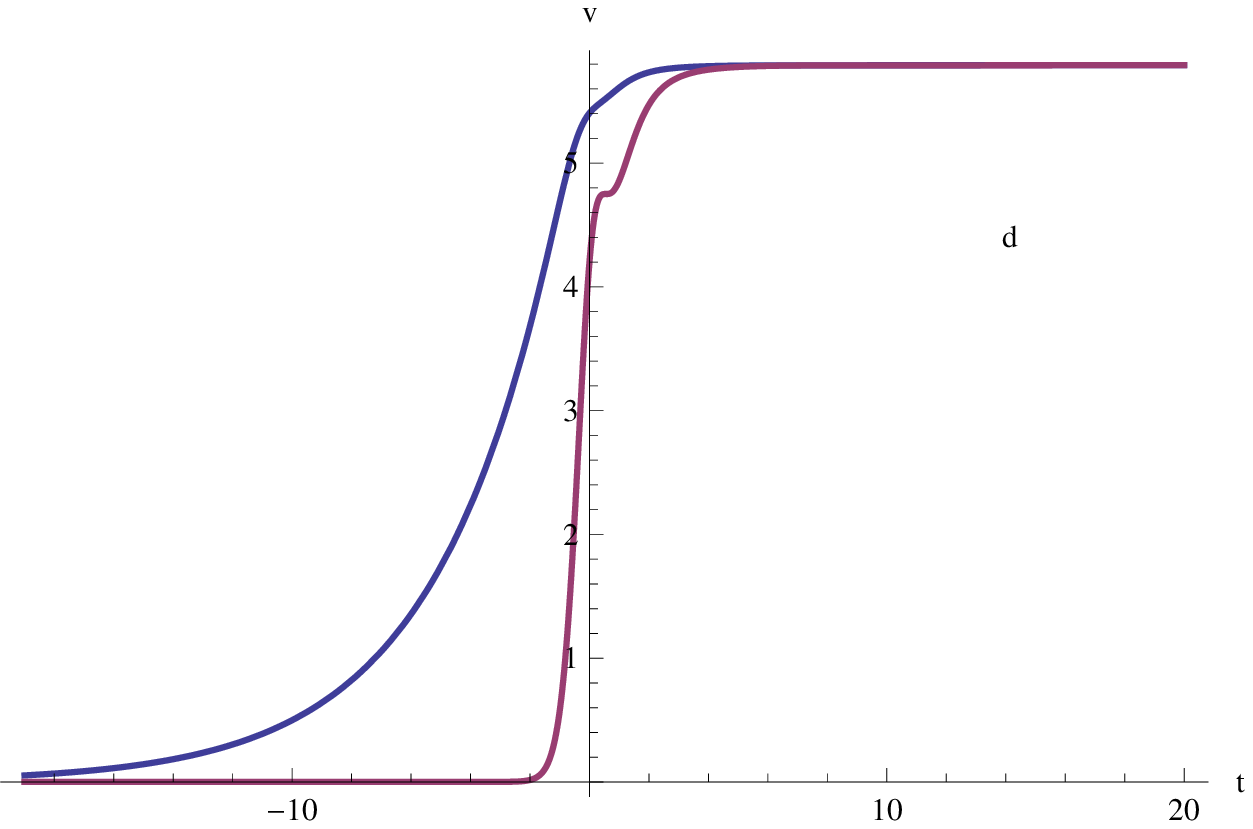}
\end{center}
  \caption{(Colour online) Plots of the evolution of the perturbation of the area of the event horizon (blue)
  and the apparent horizon (purple) for various quenching times $\a$.  The plots are (from left
  to right, top to bottom) for $\a=1,\frac{1}{2},\frac{1}{4}$ and $\frac{1}{8}$,
  respectively.  Note that the areas of both the apparent horizon and  the
  event horizon necessarily increase monotonically with time.}  \label{horfig}
\end{figure}

\begin{figure}
\begin{center}
\psfrag{x}{{$\scriptstyle{\t}$}}
\psfrag{y}{{$\scriptstyle{\dd V_{a}}$}}
  \includegraphics[width=4in]{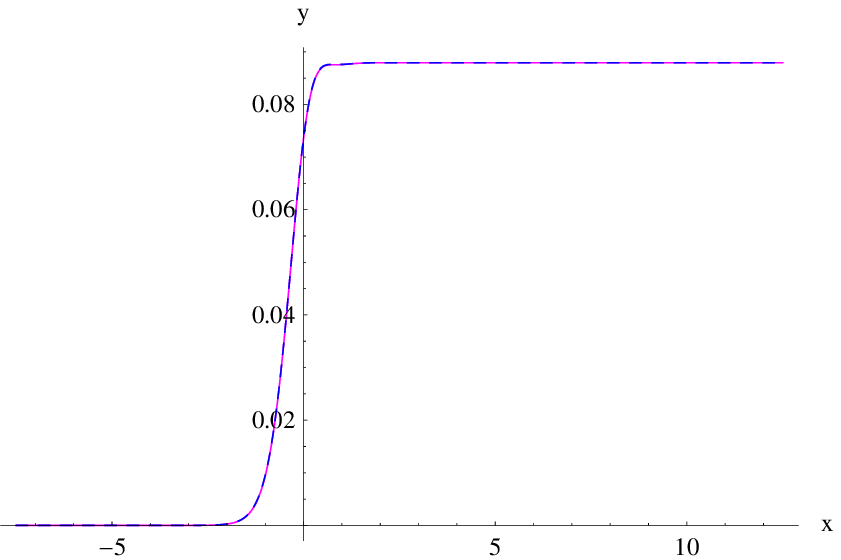}
\end{center}
  \caption{(Colour online)  Evolution of the perturbation of the area of the apparent horizon $\dd V_a$ 
for quenching time $\a=1$ computed using \eqref{aaold} (blue, dashed) and 
\eqref{dvdt} (magenta, solid). The agreement provides a highly nontrivial 
test of the constraint equation \eqref{neq5}.  }  \label{apphorfig}
\end{figure}

We are now ready to calculate and compare the evolution of the area density of the
apparent and event horizons.  See figure \ref{horfig} for the compared evolution of
the apparent and event horizons for various quenching times $\a=\{1, \ft 12, \ft 14, \ft 18\}$.
As we argue in appendix \ref{appA4}, $\a=\ft 18$ essentially corresponds to abrupt quenches.
Note that
the perturbation of the event horizon always has a larger area density than the
apparent horizon, as expected.  As $\a$ decreases, \ie the quenches become faster, we know
that more energy gets pumped into the geometry \cite{bmv} and the final area density of the
perturbed horizon also grows.  Both apparent and event horizons equilibrate
to the same area density (\ie the same radius) towards the end of the evolution. 
Finally, as we demonstrate in figure \ref{apphorfig} for a quenching time $\a=1$, 
there is an excellent agreement with the perturbation of the density of the area of the 
apparent horizon $\dd V_a$ computed with \eqref{aaold} or with the equivalent
expression \eqref{dvdt}. 
 This,
in addition to the convergence tests of the code (see appendix \ref{appA3}) gives us confidence
that our numerical evolution captures the correct evolution of the radial profile of the
metric perturbation's evolution.

\begin{figure}
\begin{center}
\psfrag{t}[Br][tl]{{$\scriptstyle{\t}$}}
\psfrag{v}[c]{{$\scriptstyle{f_{th}(\t)}$}}
\psfrag{a}[cc][][1.5]{{$\scriptstyle{\a=1}$}}
\psfrag{b}[cc][][1.5]{{$\scriptstyle{\a=\frac{1}{2}}$}}
\psfrag{c}[cc][][1.5]{{$\scriptstyle{\a=\frac{1}{4}}$}}
\psfrag{d}[cc][][1.5]{{$\scriptstyle{\a=\frac{1}{8}}$}}
  \includegraphics[width=3in]{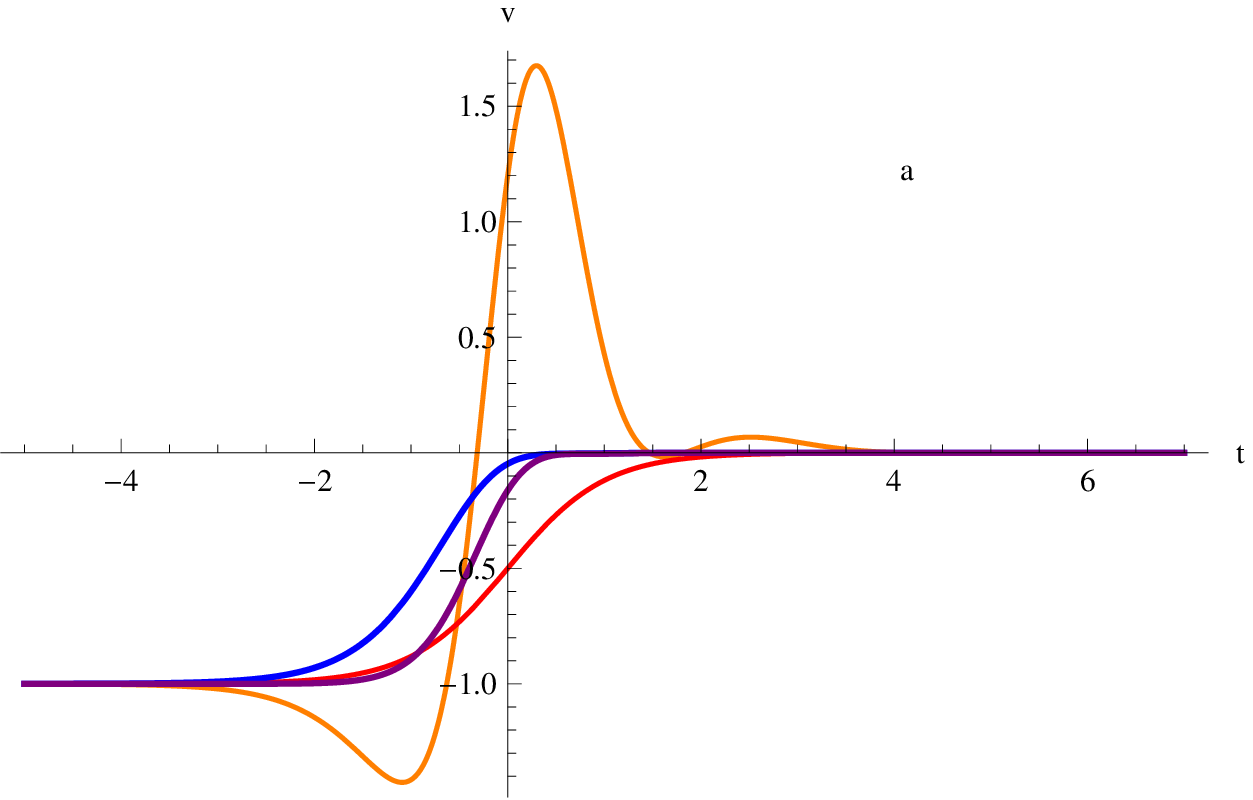}
    \includegraphics[width=3in]{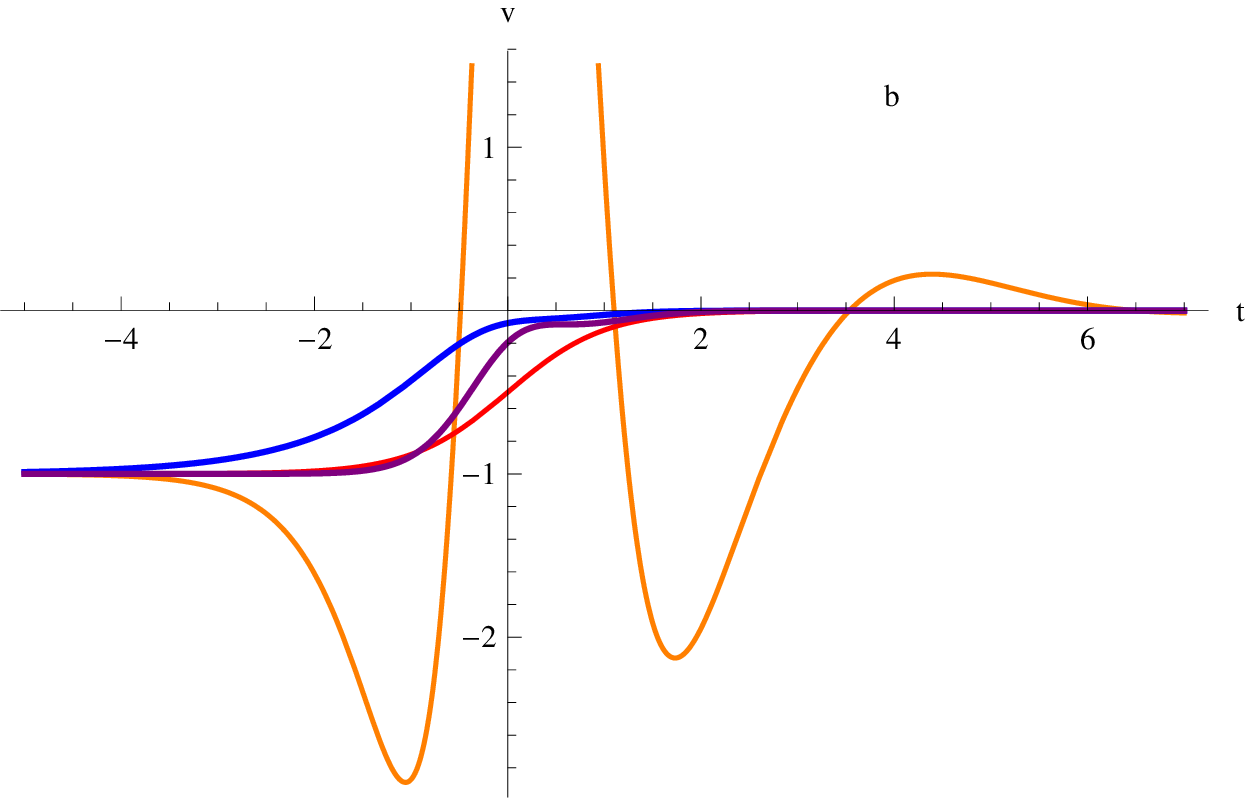}
      \includegraphics[width=3in]{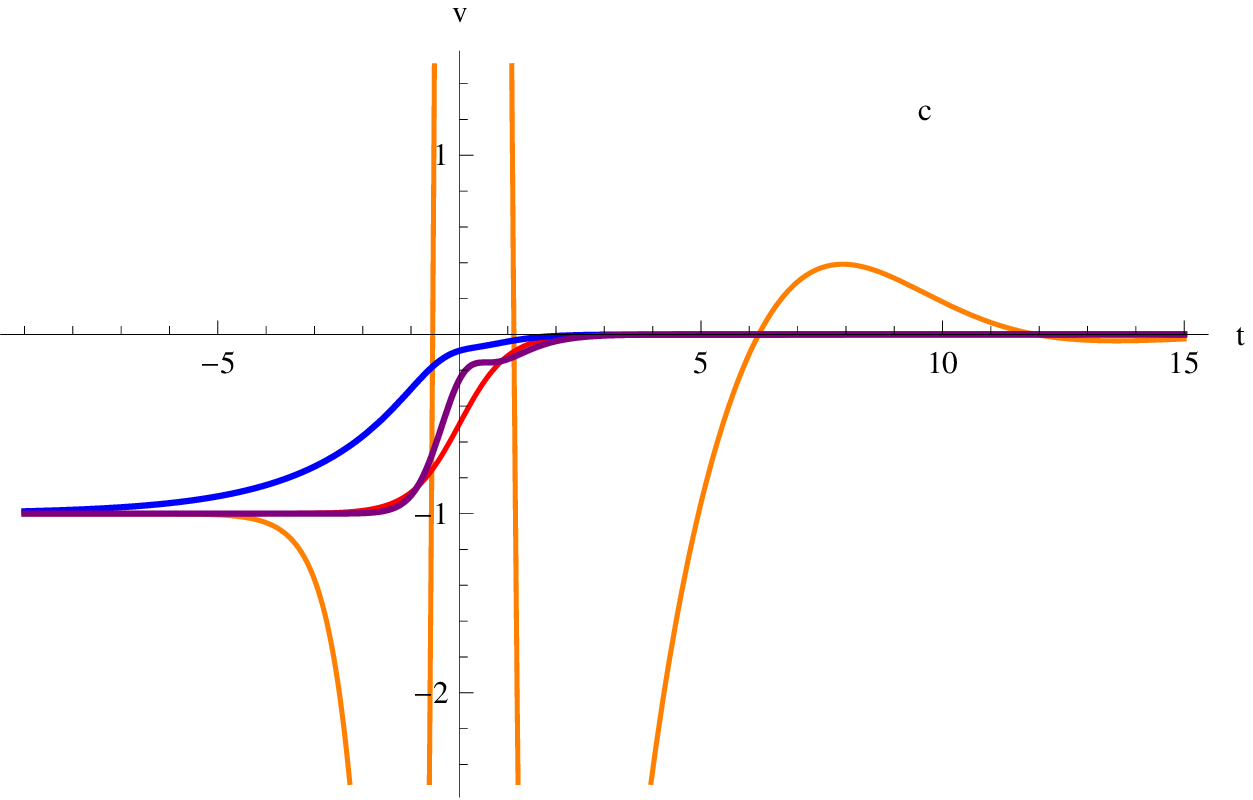}
        \includegraphics[width=3in]{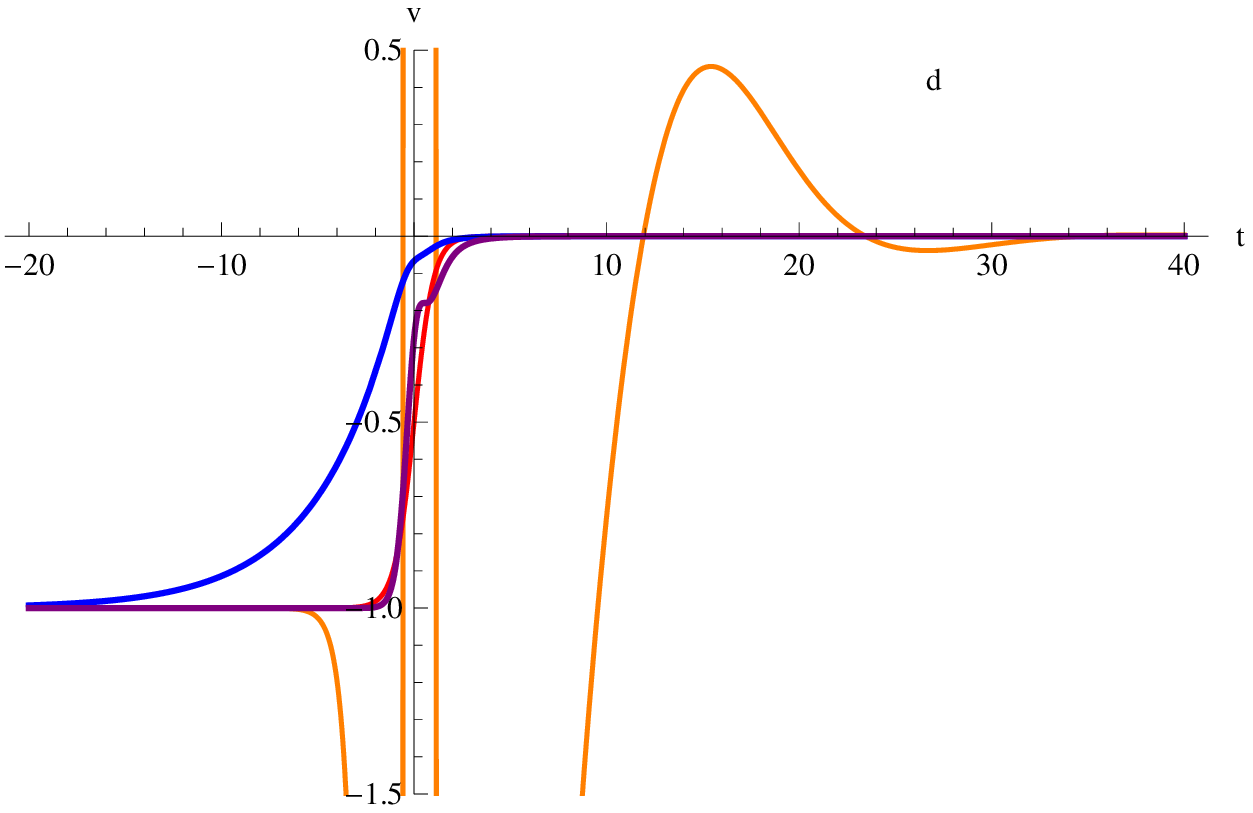}
\end{center}
  \caption{(Colour online) Here we compare the thermalization measure \eqref{fth} for the
apparent and event horizons ($\dd V_a$ in purple and $\dd V_e$ in blue) with
the thermalization measure for the normalizable mode $p_2$ of the scalar field
(orange), as well as its non-normalizable mode $p_0$ in red.  The plots are (from left
  to right, top to bottom) for $\a=1,\frac{1}{2},\frac{1}{4}$ and $\frac{1}{8}$,
  respectively.  In all cases, the horizon thermalizes before the one-point function, and
this becomes more noticeable for smaller $\a$.}  \label{horfigth}
\end{figure}

Intuitively, the evolution of the perturbed horizons of the black hole should provide us with a
measure of the time required for the system to return to thermal equilibrium after the quantum quench. However, to
produce quantitative results, we need to provide a precise measure with which we can extract the thermalization time. Hence we define our thermalization measure for a general dynamical quantity $f(\t)$ as,\footnote{
Note that for example by equations \eqref{vafull} and \eqref{vefull}, $\dd V_{e/a}$
represents only a perturbative correction to the leading area density.  Hence if we were to
use the naive measure $\frac{V(\t)-V(\infty)}{V(\infty)}$, the
result would only be $\mathcal{O}(\ell^2)$.  This reflects the fact that the system is actually only goes barely out of equilibrium, given our perturbative approach. Hence the equation \eqref{fth} gives a more reasonable
measure of the thermalization for our present study. Of course, similar comments also apply for the observables considered in the following. \label{footrob}}
\begin{equation} \eqlabel{fth}
f_{th}(\t)=\frac{f(\t)-f(\infty)}{f(\infty)-f(-\infty)}\,,
\end{equation}
which we will apply throughout the following, \ie both here in examining the horizon behaviour and also in considering various nonlocal probes in the following sections. From the above definition of the thermalization measure above, we see that $f_{th}(-\infty)=-1$, and $f_{th}(\infty)=0$. Throughout the following, our criterion
for saying that a quantity has thermalized will be that the corresponding measure comes within $2\%$ of its final value, \ie the thermalization time $\t_{th}$ will be defined with $|f_{th}(\t)|\le0.02$ for $\t\ge\t_{th}$.

Figure \ref{horfigth} shows the thermalization measure (\ref{fth}) for $\dd V_{e}$ and $\dd V_{a}$, \ie the entropy densities on the event and apparent horizons, as a function of time. 
For comparison, we also plot the thermalization measure for the expectation value of the fermionic mass operator $\langle\mathcal{O}_3\rangle$, \ie for the coefficient $p_2$ in the bulk scalar field. 
The corresponding thermalization or equilibration times determined with our $2\%$ criterion are given in table \ref{kappa}. In figure \ref{horfigth}, we also show the result of applying equation (\ref{fth}) 
to the source coefficient $p_0$. In the figure, we see that $p_{2(th)}$ makes excursions far beyond $(-1,0)$ while $\dd V_{e}$ and $\dd V_{a}$ remain within this range at all times. However, we note that although $p_2$ 
fluctuates much more than the horizon position, it reaches small values compared to its extrema before the horizons equilibrate. Nonetheless we see in table \ref{kappa} that the thermalization time is slower for the expectation 
value than for the equilibration of the horizons.  This relative difference becomes more pronounced as we make $\a$ smaller. Note that figure \ref{horfigth} is plotted in terms of the (dimensionless) rescaled time, as in eq.~\eqref{fastres1}, 
and so the thermalization times in table \ref{kappa} are measured in terms of the same rescaled time. The physical thermalization times\footnote{Note that the physical time and the quench parameter $\alpha$ are both dimensionful and are 
implicitly measured in units of $1/({\pi T})$, which is set to one in our conventions.} would carry an extra factor of $\alpha$, \ie $\t_{physical}=\a\,\t_{rescaled}$. In the table, we see that the equilibration times of the horizon become 
approximately constant for small $\a$ (although we see some variation for the event horizon), in terms of the rescaled time $\t$. Hence, in the physical time, the equilibration of the horizon perturbation therefore scales approximately as $\a$.  In contrast, as shown in the table, the equilibration 
time of $p_{2}$ becomes approximately constant when measured in the physical time, as previously noted in \cite{bmv,blmv}.
\begin{table}[ht]
\caption{The equilibration times of the area densities of the event and apparent horizons and the
thermalization time for the one-point correlator, which thermalizes as $p_2$, (as defined by the $2\%$ threshold of equation \eqref{fth}), for different values of the quenching parameter $\a$.  We also give $\a\,\t_{th}[\langle\mathcal{O}_3\rangle]$ which corresponds to the physical time, as discussed in the main text.\label{kappa}}
\centering
\begin{tabular}{c c c c c c c}
&&&\\
\hline
 $\a$ &\vline\vline& 1 & $\frac{1}{2}$ & $\frac{1}{4}$ & $\frac{1}{8}$ & $\frac{1}{16}$\\ [0.5ex]
\hline\hline
 $\t_{eq}[\dd V_{e}]$ &\vline\vline& .166 & 1.14 & 1.48 & 1.40 & 0.99\\ [0.5ex]
\hline
 $\t_{eq}[\dd V_{a}]$ &\vline\vline& .385 & 1.69 & 2.37 & 2.85 & 3.11\\ [0.5ex]
\hline\hline
 $\t_{th}[\langle\mathcal{O}_3\rangle]$ &\vline\vline& 3.41 & 6.16 & 15.25 & 30.46 & 61.14\\ [0.5ex]
 \hline
 $\a\,\t_{th}[\langle\mathcal{O}_3\rangle]$ &\vline\vline& 3.41 & 3.08 & 3.81 & 3.81 & 3.82\\ [0.5ex]
\hline
\end{tabular}
\end{table}

\subsection{Two-point correlators}

\subsubsection{Analytic expression for the correlator} \label{anexcor}

We now consider two-point correlators as probes of thermalization in the field theory.
More specifically, we mainly consider perturbations to the equal-time two-point correlator due to the quench.
This is because the mass coupling of the quenching operator in the field theory is small compared to
the thermal scale,
and therefore only perturbations of the correlator will be time-dependent
and contain information about thermalization (as noted in footnote \ref{footrob}).

For ease of computation on the AdS side, we will consider the correlator of an operator with
large conformal dimension (\ie not the quenching operator).  The correlator of such an operator can
be calculated in the geometric optics limit by the length of a boundary-to-boundary spacelike geodesic
\cite{geom1, geom2}.

Because it will turn out that the perturbations of the length of the geodesic remain finite, we
needn't concern ourselves with the regularization of the static geodesic length.  As a reminder to
the reader, the line element of the spacetime in dimensionless, rescaled coordinates is given by
\begin{equation}
ds^2=-a d\t^2 + s^2d\vec{y}^2 - \frac{2d\r d\t}{\r^2},
\end{equation}
where we have defined $\vec{y}=\mu\vec{x}/\a$ as the dimensionless boundary spatial directions.

To calculate the two-point correlator, we will calculate the length of a spacelike geodesic with endpoints at
$(\t=\t_\ast,y_1=-y_m,y_2=0=y_3)$ and $(\t=\t_\ast,y_1=y_m,y_2=0=y_3)$ (\ie with endpoints
at equal times, and symmetric in the $y_1$-axis.)  If we allow the geodesic to extend into the bulk,
it will have both $\r$ and $\t$ profiles that depend on $y_1$.  The length of the geodesic is given by
\begin{equation}
\mathcal{L} = \int^{y_m}_{-y_m}dy_1\sqrt{s^2(\t(y_1),\r(y_1))-
a(\t(y_1),\r(y_1))\t'(y_1)^2-\frac{2\r'(y_1)\t'(y_1)}{\r(y_1)^2}},
\eqlabel{length}
\end{equation}
where a prime denotes a derivative with respect to $y_1$.  The geodesic can then be viewed as the solution
of the Euler-Lagrange equation obtained from \eqref{length} when treating $\mathcal{L}$ as an action.

Expanding the metric coefficients in the perturbative parameter in
$\ell^2$ as given in equations \eqref{al2} and \eqref{bl2}, and the time and radial profiles of the geodesic as
\begin{eqnarray}
\t = \t_0 + \ell^2 \t_2, \nonumber\\
\r = \r_0 + \ell^2 \r_2, \label{coordpert}
\end{eqnarray}
the geodesic length can be written as $\mathcal{L}=\mathcal{L}_0 + \ell^2\mathcal{L}_2$.
Rescaling the coordinates as in section \ref{rescaling}, the length of the geodesic in the unperturbed geometry is expressed as\footnote{The various coordinates and fields are the rescaled
version of these fields, as explained in section \ref{rescaling}.
We will leave physical constants such as the geodesic half-width $y_m$ un-rescaled.
That is to say, the physical width of the surface in the rescaled coordinates is
$\Delta y_1 = 2\frac{y_m}{\a}$, and the physical height of the geodesic will be $\r_0=\frac{\r_m}{\a}$,
with the black hole horizon located at $\r=\frac{1}{\a}$.} \cite{periodic}
\begin{equation}
\mathcal{L}_0 = \int^{y_m/\a}_{-y_m/\a}dy_1\frac{\sqrt{D(\t_0,\r_0)}}{\r_0},
\eqlabel{length0}
\end{equation}
and a perturbation of that length given by
\begin{eqnarray}
&&\mathcal{L}_2 = \int^{y_m/\a}_{-y_m/\a}dy_1\frac{b-\a^2\r_0^2 \hat{a}(\t_0')^2/2}{\r_0\sqrt{D(\t_0,\r_0)}} \nonumber \\
&&+ \int^{y_m/\a}_{-y_m/\a}dy_1\left(-\frac{\sqrt{D(\t_0,\r_0)}}{\r_0^2}\r_2-\frac{\t_0'-2\a^3\r_0^3\t_0'^2}{\r_0\sqrt{D(\t_0,\r_0)}}
\r_2'-\frac{(1-\a^4\r_0^4)\t_0'-\r_0'}{\r_0\sqrt{D(\t_0,\r_0)}}\t_2'\right), \label{length2}
\end{eqnarray}
where
\begin{equation}
D(\t_0,\r_0) = 1-(1-\a^4\r^{4}_0)\t'^{2}_0-2\r'_0\t'_0.
\eqlabel{D}
\end{equation}
If we perform integration by parts, we can change the term involving $\r_2'$ to a term involving $\r_2$ plus a total derivative term.
In the case of a geodesic, this total derivative term vanishes when integrated.  We are then left with terms involving $\r_2$ and $\t_2'$.
It turns out that since we are perturbing around an extremal trajectory,
these two terms vanish by the equations of motion of $\r_0$ and
$\t_0$, and we therefore needn't consider perturbations of the radial and time profiles of the geodesic
in order to calculate the perturbations of its length \cite{periodic}.  Since the perturbations on the
shape of the geodesic $\t_2$ and $\r_2$ play no role in the calculation, we will for simplicity
refer to $\t_0$ and $\r_0$ as $\t$ and $\r$, respectively.  Because $\mathcal{L}_2$ depends
on the unperturbed profile of the geodesic, we must first solve for $\r$ and $\t$.  Since $y_1$ is an arbitrary transverse direction, we will simply refer to it as $y$.

As it turns out, it is useful to solve the problem by choosing $\r$ as our independent parameter,
and $\t$ and $y$ as our dependent parameters\footnote{Because of the fact that the
perturbations of the geodesic shape do not contribute to the two-point correlator at order $\ell^2$,
we only consider the static geodesic in our calculations.  As such, we can parameterize the geodesic
with either $y$ or $\r$, using the fixed endpoints $\pm\frac{y_m}{\a}$ and $\frac{\r_m}{\a}$,
respectively.  If $\t_2$ or $\r_2$ do contribute, as it does in the case of the
entanglement entropy, that integral must be evaluated in
$y$-coordinates, since $\r_m$ would change at order
$\ell^2$, while we would have to make the choice of keeping $y_m$ fixed.}.
We can find a closed form solution of $\t(\r)$.  The independence of the
integral in \eqref{length0} on constant shifts in $\t$ and the condition that
the geodesic be smooth at $y=0$ lead to the equation \cite{periodic}
\begin{equation} \eqlabel{lintime}
(1-\a^4\r^4)\t'+\r'= 0.
\end{equation}
Dividing equation \eqref{lintime} by $\r'$, and using the chain rule, the equation becomes
\begin{equation}
\t'(\r) = -\frac{1}{1-\a^4\r^4},
\end{equation}
with a general solution of
\begin{equation} \eqlabel{tausol}
\t(\r)=\t_\ast-\frac{\tan^{-1}(\a \rho)+\tanh^{-1}(\a \r)}{2 \alpha }.
\end{equation}
In the above solution, $\t_\ast$ is the arbitrary boundary time of the geodesic,
\ie the time of the equal-time correlator, which follows from the time translation
invariance of the equation for the geodesic.  While this solution for the time-profile
may look strange, it is in fact related to the change in  coordinates between Poincar\'{e}
coordinates and Eddington-Finkelstein coordinates.  The line-element of the Poincar\'{e}
patch is of the form (ignoring the transverse coordinates)
\begin{equation}
ds^2 = \frac{1}{z^2}\left(-f(z)dt^2+\frac{dz^2}{f(z)}\right).
\end{equation}
It turns out that the change of coordinates relates $\r=z$, and
\begin{equation} \eqlabel{tpoinc}
\t(t,z)=t-\frac{\tan^{-1}(\a z)+\tanh^{-1}(\a z)}{2 \alpha },
\end{equation}
where the Poincar\'{e} time will agree with the EF boundary time $\t_\ast$.  Replacing
$t$ with $\t_\ast$ and $z$ with $\r$, the above expression
is identical to the geodesic contour in time given in equation \eqref{tausol}.
The $\t(\r)$ of the geodesic therefore corresponds to a constant time slice in Poincar\'{e}
coordinates on the static background.  Inverting equation \eqref{tpoinc}, expressing $t$ as a function of
$\t$ and $z$, we see that constant $\t$ corresponds to an infalling null ray in the static Poincar\'{e}
geometry, and constant $\t$ in EF coordinates is actually the path of a light ray falling into the
black hole from the spacetime boundary.

Since the integrand in \eqref{length2} has no explicit dependence on $y$, we know that the
``Hamiltonian''
\begin{equation}
\r'\partial_{\r'}\mathcal{L}_0+\t'\partial_{\t'}\mathcal{L}_0-\mathcal{L}_0
\end{equation}
will be constant in $y$.  The equation simplifies to \cite{periodic}
\begin{equation} \eqlabel{dr2}
\a^2D(\t,\r)\r^2 = \r^2_m,
\end{equation}
$\r(y=0)=\frac {\r_m}{\a}$, \ie the maximum value of $\r$ on the geodesic.
Substituting in for $\t'(y)$ in $D(\t,\r)$ from \eqref{lintime}, \eqref{dr2} can be simplified to \cite{periodic}
\begin{equation} \eqlabel{rpy}
\r'(y)=-\frac{\sqrt{\left(\r^2_m-\a^2\r^2\right)\left(1-\a^4\r^4\right)}}{\a\,\r}.
\end{equation}

Changing the integration variable from $y$ to $\r$ in \eqref{length0} and \eqref{length2}, and substituting in from
equation \eqref{dr2}, the new expressions are
\begin{equation}
\begin{split}
\mathcal{L}_0=&2\int^{\r_m/\a}_{0}d\r\frac{\r_m}{\r\sqrt{\left(\r_m^2-\a^2\r^2\right)\left(1-\a^4\r^4\right)}},\\
\mathcal{L}_2=&-2\a\int^{\r_m/\a}_{0}d\r\frac{b-\a^2\r^2 (\t')^2\hat{a}/2}{\r'\r_m}.
\end{split}
\eqlabel{l2new}
\end{equation}
The expression for $\t'$ is given by \eqref{lintime}, \ie an expression depending on
$\r$ and $\r'$.  The $\r'$ terms in the integral can be substituted by an expression
depending on $\r$ and $\r_m$ from \eqref{rpy}.  The expression for the integral therefore
has no explicit $y$ dependence, and other than knowing the solution for $\t$ in terms of $\r$, we
need not know the profile of $y$ in terms of $\r$ at all!  The
additional factor of $2$ in \eqref{l2new} comes from the fact that the integration
limits correspond only to half the $y$-interval, $[0,y_m/\a]$.  Making the above substitutions
give us the perturbation of the two-point function as being
\begin{equation}\eqlabel{l2r}
\mathcal{L}_2=\a^2\int^{\r_m/\a}_{0}d\r\frac{\r\left(2\left(1-\a^4\r^4\right)b-
\a\left(\r_m^2-\a^2\r^2\right)\hat{a}\right)}
{\r_m\sqrt{\left(\r^2_m-\a^2\r^2\right)\left(1-\a^4\r^4\right)^{3}}}.
\end{equation}

\subsubsection{Numerical calculation of the perturbed two-point function} \label{numlin}

There is still some difficulty with integral \eqref{l2r}, namely that the integrand
diverges near $\r=\r_m$.  Although it is a one-over-square-root divergence, and the
integral itself will still be finite, this does pose a problem numerically.  In order
to avoid integrating a divergent quantity, we introduce a second change of variables:
\begin{equation} \eqlabel{qcor}
\r = \frac{\r_m}{\a}\left(1-q^2\right).
\end{equation}
Therefore the interval in $\r$, $[0,\r_m]$ corresponds to the reverse of the interval $[0,1]$ in $q$.
This transforms the integral into its final form
\begin{equation}
 \eqlabel{l2q}
\begin{split}
\mathcal{L}_2=&\int^{1}_{0}dq
\frac{2 \left(1-q^2\right)}
{ \sqrt{2-q^2} \left(1-\left(1-q^2\right)^4 \r_m^4\right)^{3/2}}\times \biggl( 2 \left(1- \r_m^4\left(1-q^2\right)^4 \right)b(\t(q),\r(q))\\
&-\r_m^2q^2 \left(2-q^2\right)\hat{a}(\t(q),\r(q))\biggr).
\end{split}
\end{equation}
This new integrand in terms of $q$ contains no divergences, and can easily be integrated
numerically.

With the numerical evolution of the scalar field-metric system, we found the profiles of
the metric components $\hat{a}$ and $b$, as described in appendix \ref{a1}.
This was done by solving at each timestep $\t_i$ for the coefficients
to the Chebyshev polynomials, and interpolating the values of the functions at the collocation points
(see appendix \ref{appA2}):
\begin{equation} \eqlabel{abcheby}
\begin{split}
\hat{a}(\t_i,\r) =&-\frac 16 (p_0'(\t_i))^2+a_{log}(\t_i,\r)+a_c(\t_i,\r),\qquad b(\t_i,\r)
=b_{log}(\t_i,\r)+b_c(\t_i,\r),\\
a_c(\t_i,\r)=&\left(\frac{L_\r}{2}\right)^2\  \sum_{j=1}^{N-2}
\calf_{g_c}^j(\t_i) \biggl[\sum_{s=1}^{j+2} \calc_{j,s}\  T_{s-1}
\left(\frac{2\r}{L_\r}-1\right)   \biggr],
\\
b_c(\t_i,\r)=& \sum_{j=1}^N \calf_{b_c}^j(\t_i)\ T_{j-1}\left(\frac{2\r}{L_\r}-1\right),
\end{split}
\end{equation}
where $L_\r$ is a numerical domain of the radial coordinate defined in \eqref{numdom},
and the fixed coefficients $\calc_{j,s}$ are given by \eqref{cjsdef}.
$N$ is the number of collocation points at which we choose to solve the functions,
and also the number of Chebyshev polynomials we choose to use to model the functions
at each timestep.  By recording the coefficients $\calf_{g_c}^j$ and $\calf_{b_c}^j$ of the polynomials, we can
calculate $\hat{a}$ and $b$ for any values of $\t$, by simply interpolating
between the values of the coefficients for intermediate times.  In equation \eqref{abcheby} we can
therefore  replace $\t_i\to \t$.

In order to calculate the value of \eqref{l2q}, we discretize the $q$-interval, and simply integrate
the interpolating function calculated by Mathematica.  Doing this for multiple values
of $\t_\ast$, we can see the full time evolution of the perturbed two-point function.  Doing this for multiple values
of $\r_m$ and the quenching parameter $\a$, we can see how the system thermalizes at different
length scales at different quenching rates.

\begin{figure}
\begin{center}
\psfrag{t}[Br][tl]{{$\scriptstyle{\t_\ast}$}}
\psfrag{p1}[c]{{$\scriptstyle{\mathcal{L}_{2(th)}(\t_\ast)}$}}
\psfrag{a}[cc][][1.5]{{$\scriptstyle{\r_m=0.1\r_h}$}}
\psfrag{b}[cc][][1.5]{{$\scriptstyle{\r_m=0.5\r_h}$}}
\psfrag{c}[cc][][1.5]{{$\scriptstyle{\r_m=0.9\r_h}$}}
\psfrag{d}[cc][][1.5]{{$\scriptstyle{\r_m=0.99\r_h}$}}
\psfrag{e}[cc][][1.5]{{$\scriptstyle{\r_m=0.999\r_h}$}}
  \includegraphics[width=3in]{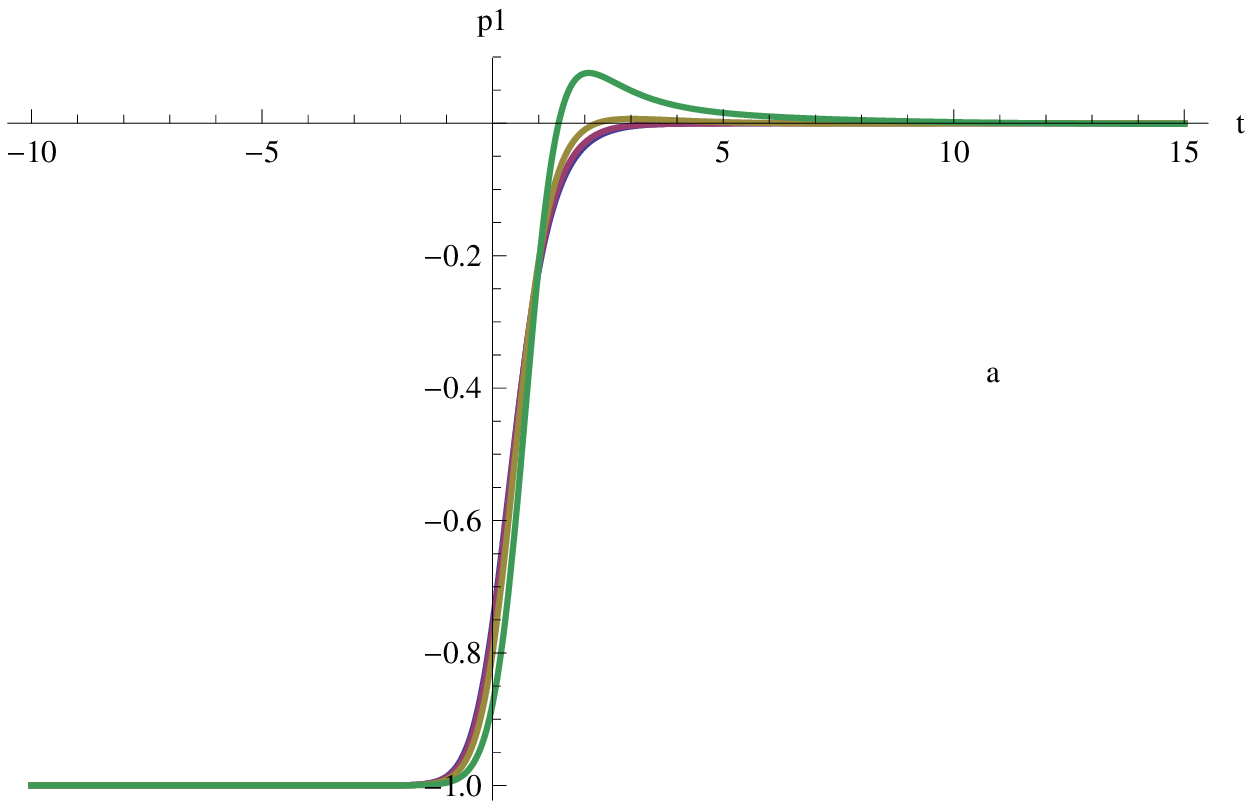}
  \includegraphics[width=3in]{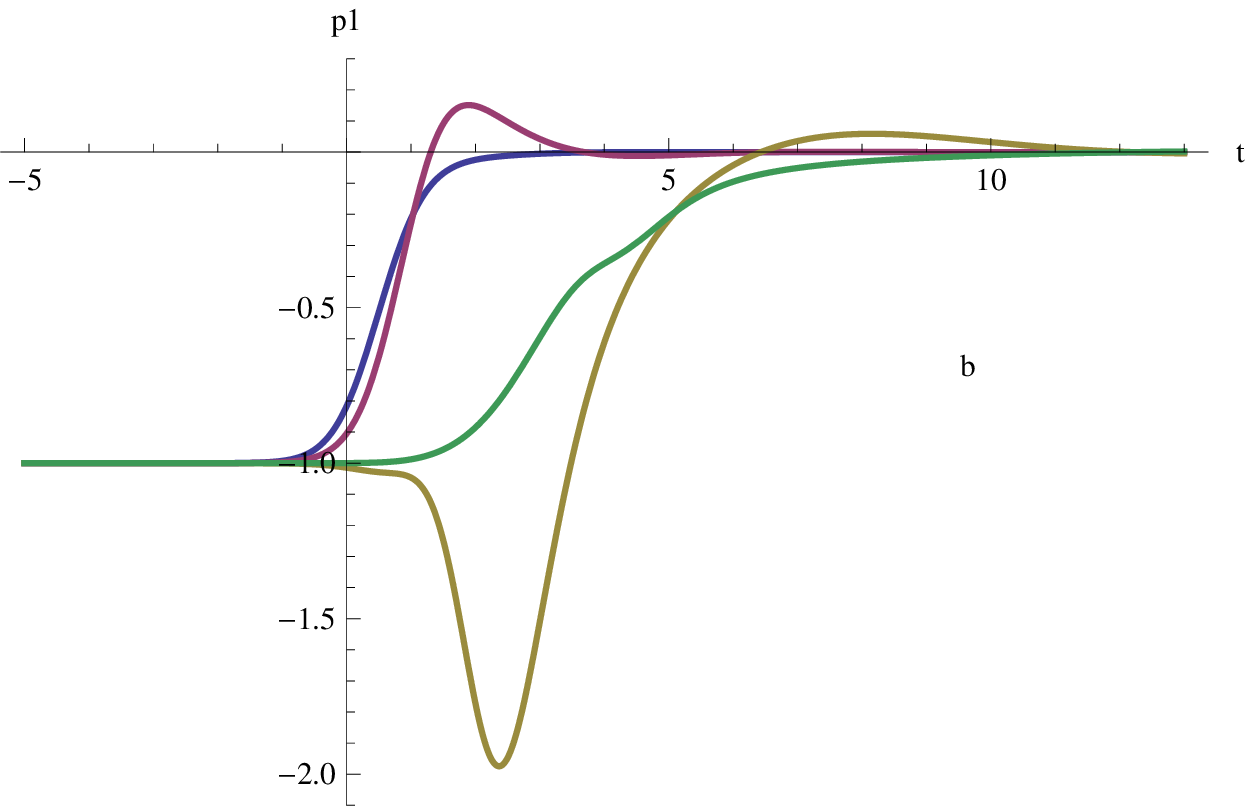}
  \includegraphics[width=3in]{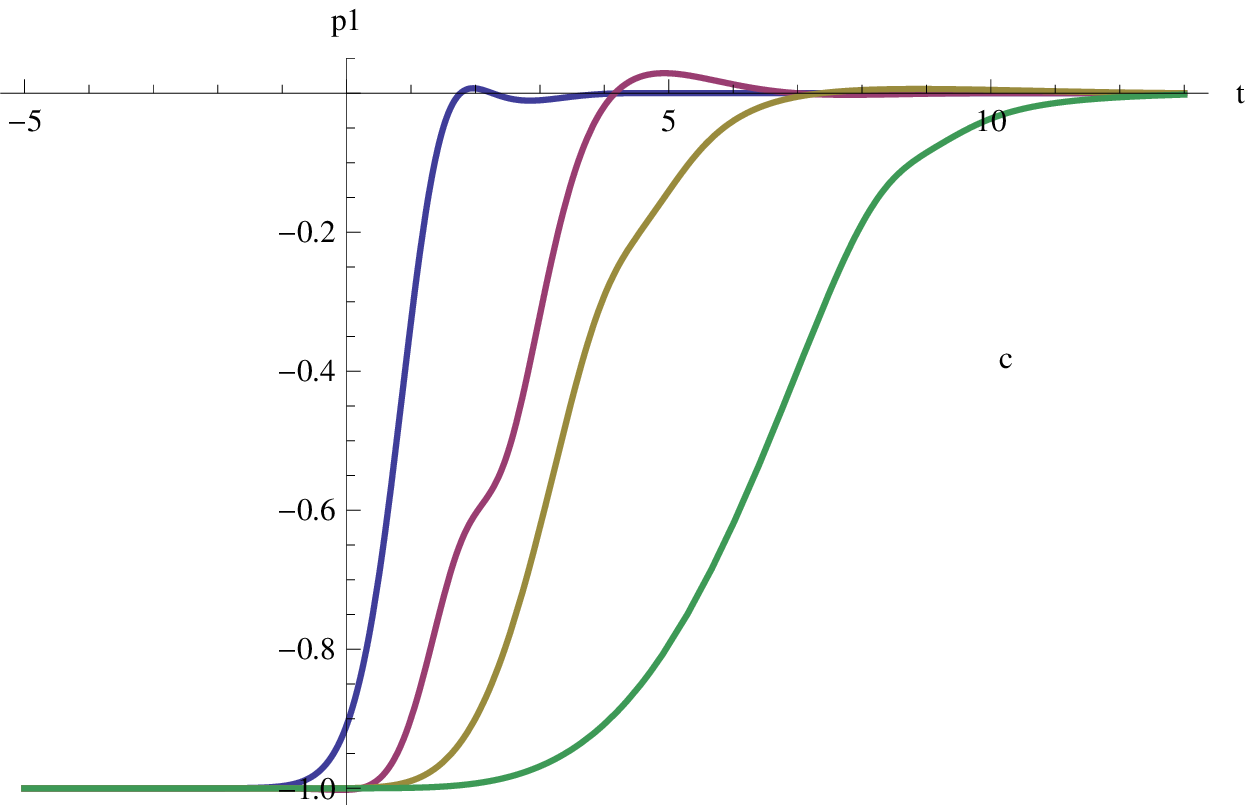}
  \includegraphics[width=3in]{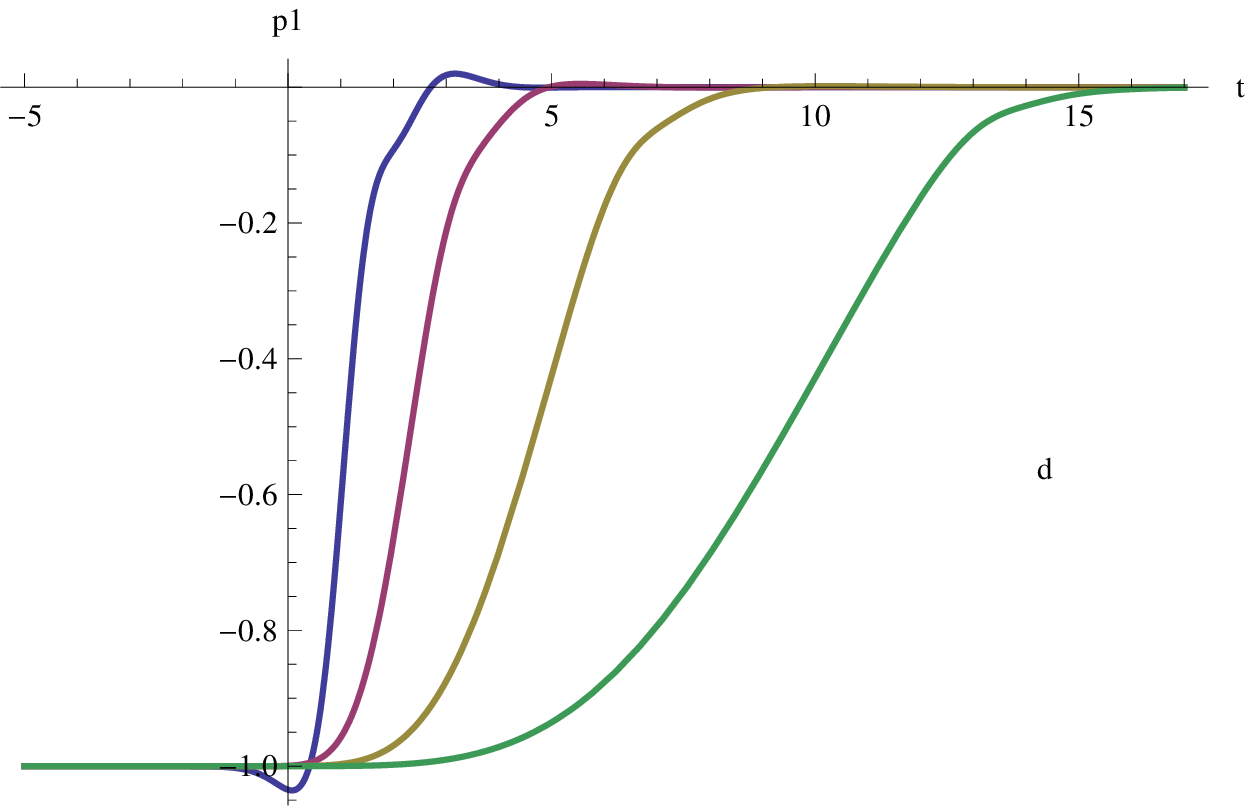}
  \includegraphics[width=3in]{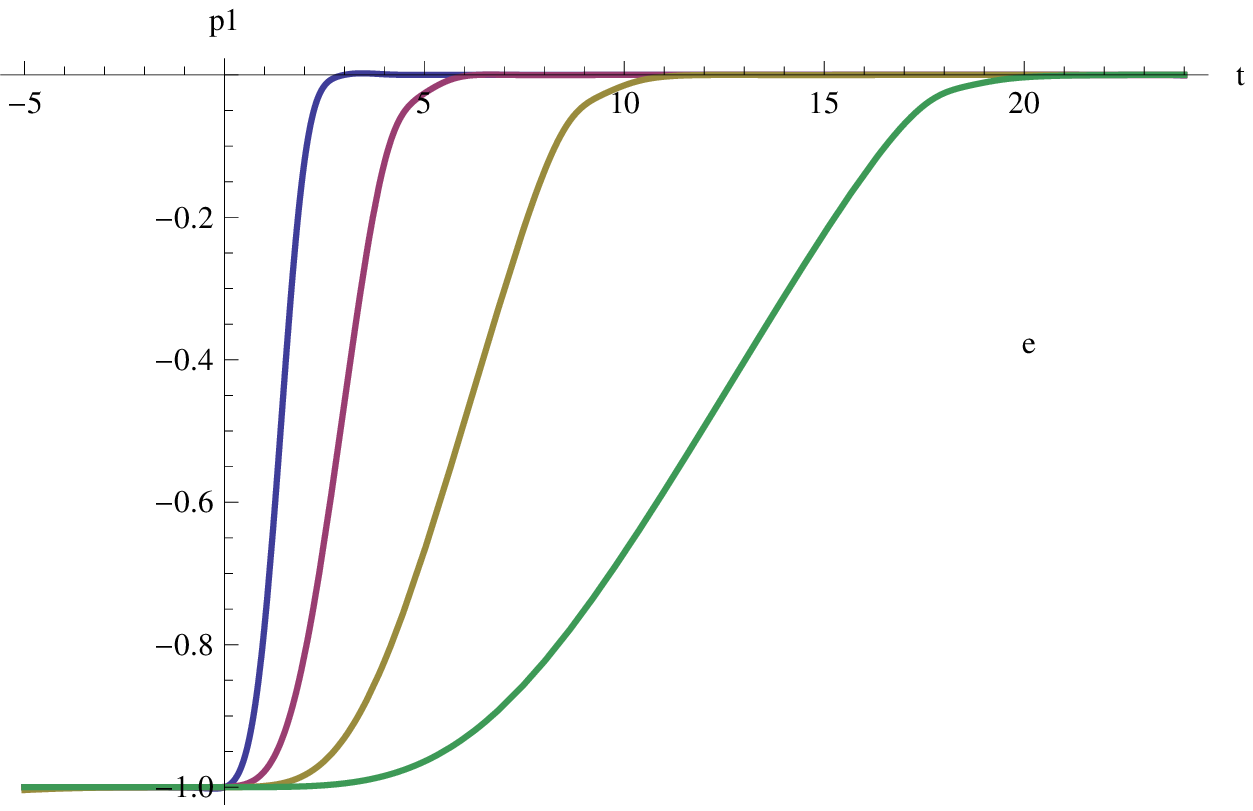}
\end{center}
  \caption{(Colour online) The thermalization measure as defined in \eqref{fth} of the perturbation of the two-point functions for different-sized
geodesics.  The evolution is a function of the rescaled boundary time $\t_\ast$.
The plots are, from left to right, top to bottom, for $\r_m=0.1\r_h$, $0.5\r_h$, $0.9\r_h$
$0.99\r_h$ and $0.999\r_h$.  In each plot the thermalization measure
is shown for quenching parameters $\a=1$ (blue), $\a=\frac 12$ (purple), $\a=\frac 14$ (brown) and $\a=\frac 18$ (green).
Note that the smaller $\a$ is, the longer equilibration takes, in this rescaled boundary time.}  \label{linpfig}
\end{figure}

\begin{figure}
\begin{center}
\psfrag{t}[Br][tl]{{$\scriptstyle{\a\t_\ast}$}}
\psfrag{p1}[c]{{$\scriptstyle{\mathcal{L}_{2(th)}(\t_\ast)}$}}
\psfrag{a}[cc][][1.2]{{$\scriptstyle{\r_m=0.1\r_h}$}}
\psfrag{b}[cc][][1.2]{{$\scriptstyle{\r_m=0.5\r_h}$}}
\psfrag{c}[cc][][1.2]{{$\scriptstyle{\r_m=0.9\r_h}$}}
\psfrag{d}[cc][][1.2]{{$\scriptstyle{\r_m=0.99\r_h}$}}
\psfrag{e}[cc][][1.2]{{$\scriptstyle{\r_m=0.999\r_h}$}}
  \includegraphics[width=2in]{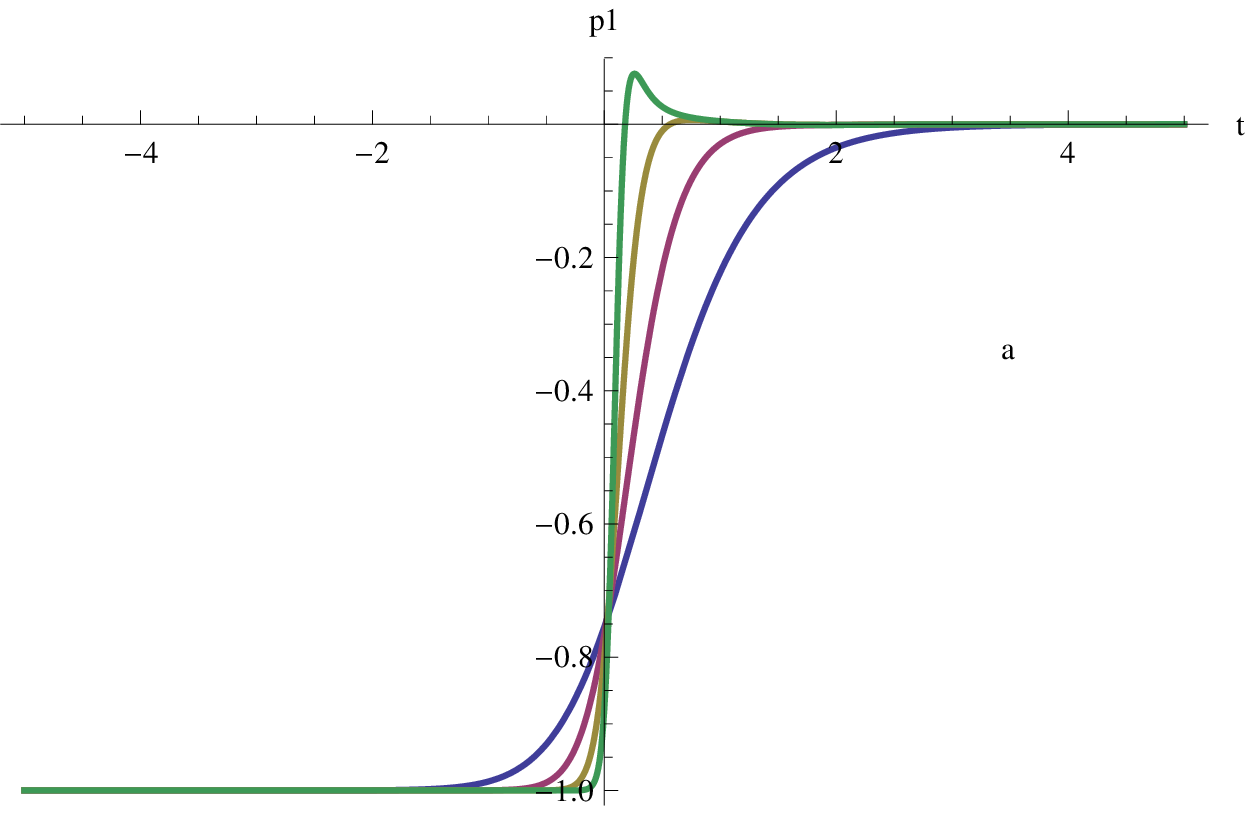}
  \includegraphics[width=2in]{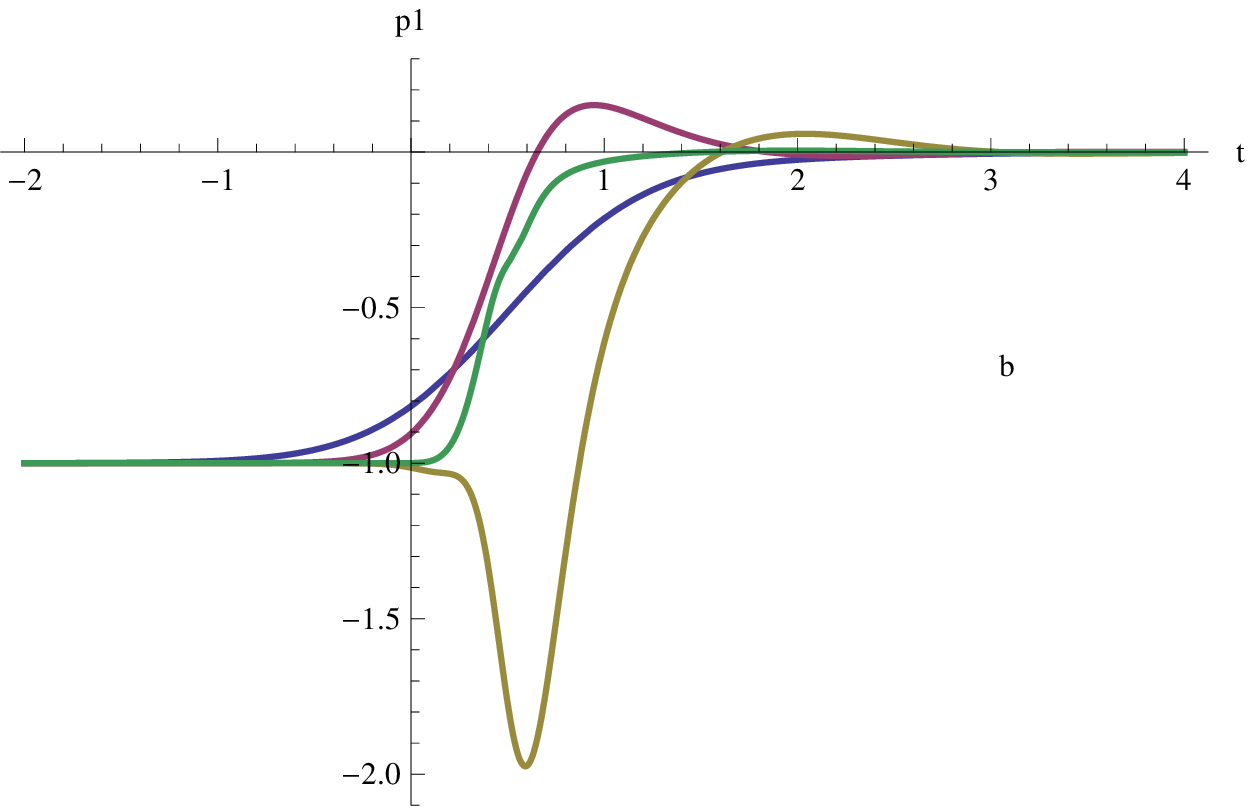}
  \includegraphics[width=2in]{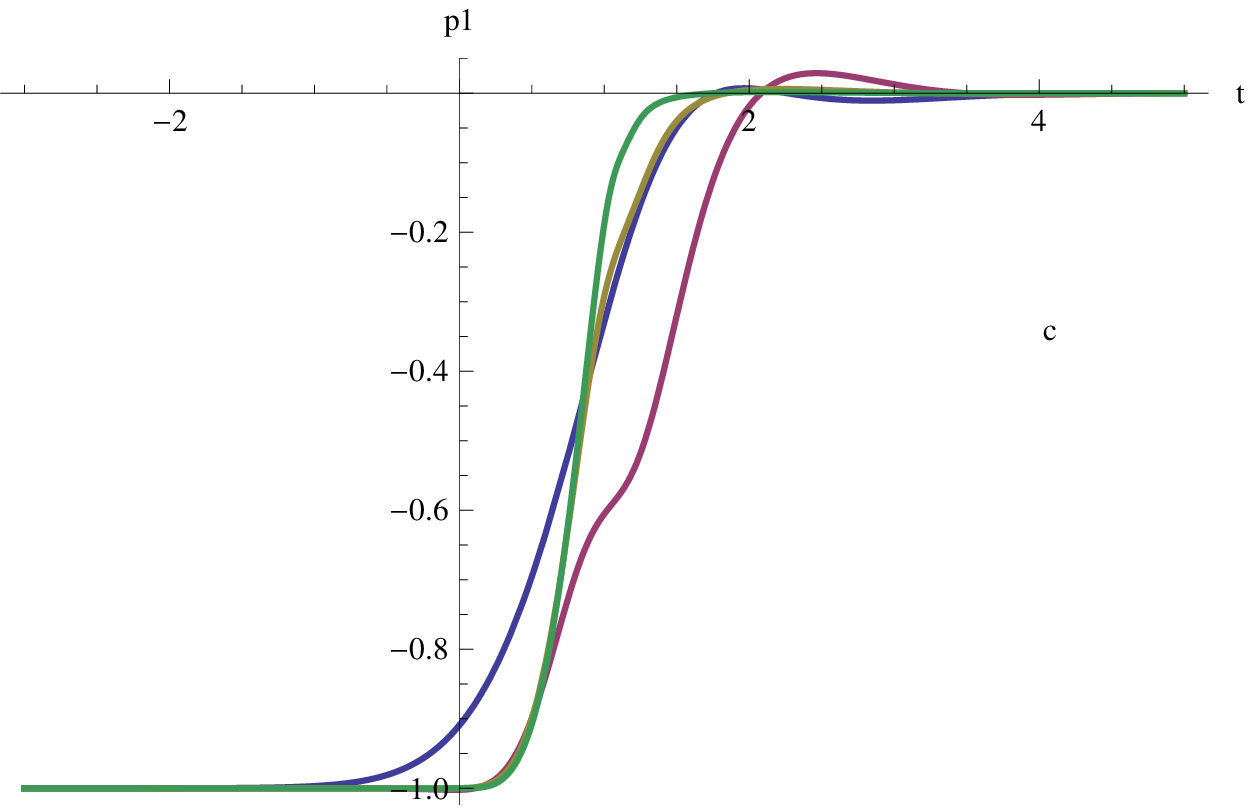}
  \includegraphics[width=2in]{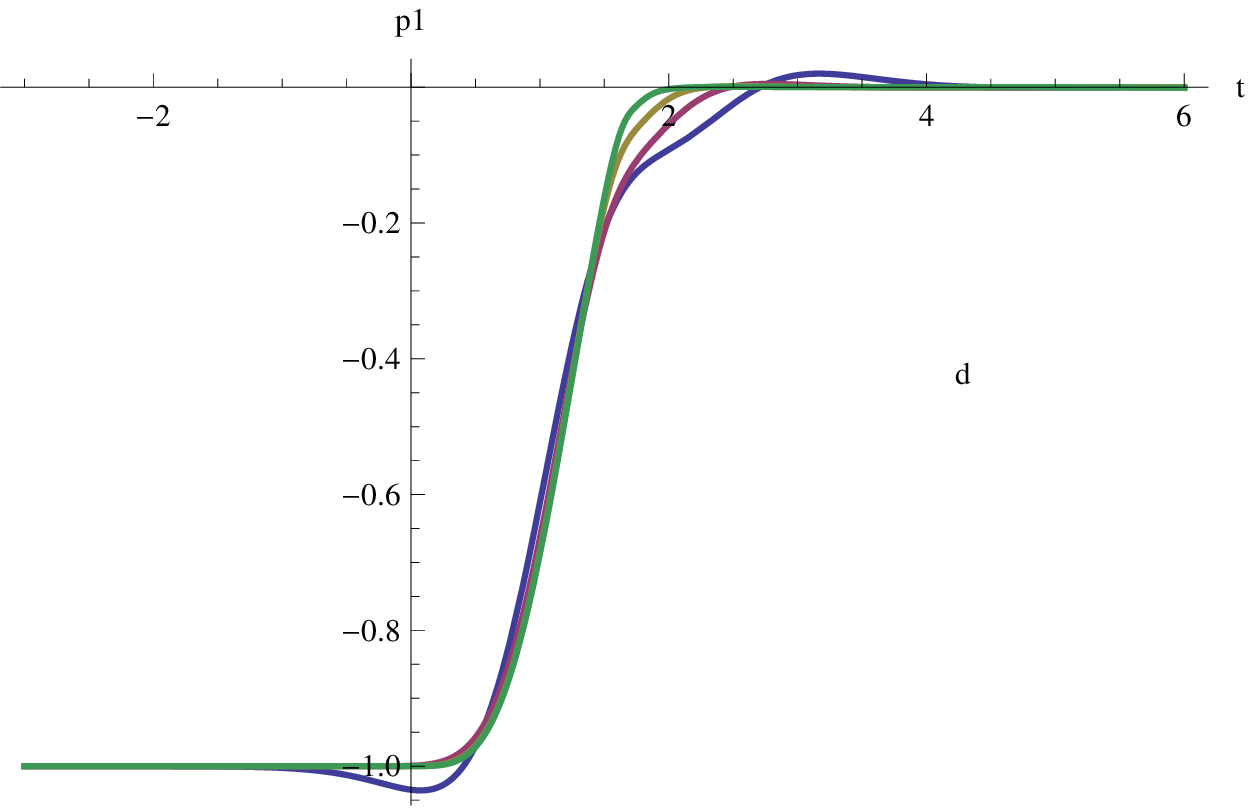}
  \includegraphics[width=2in]{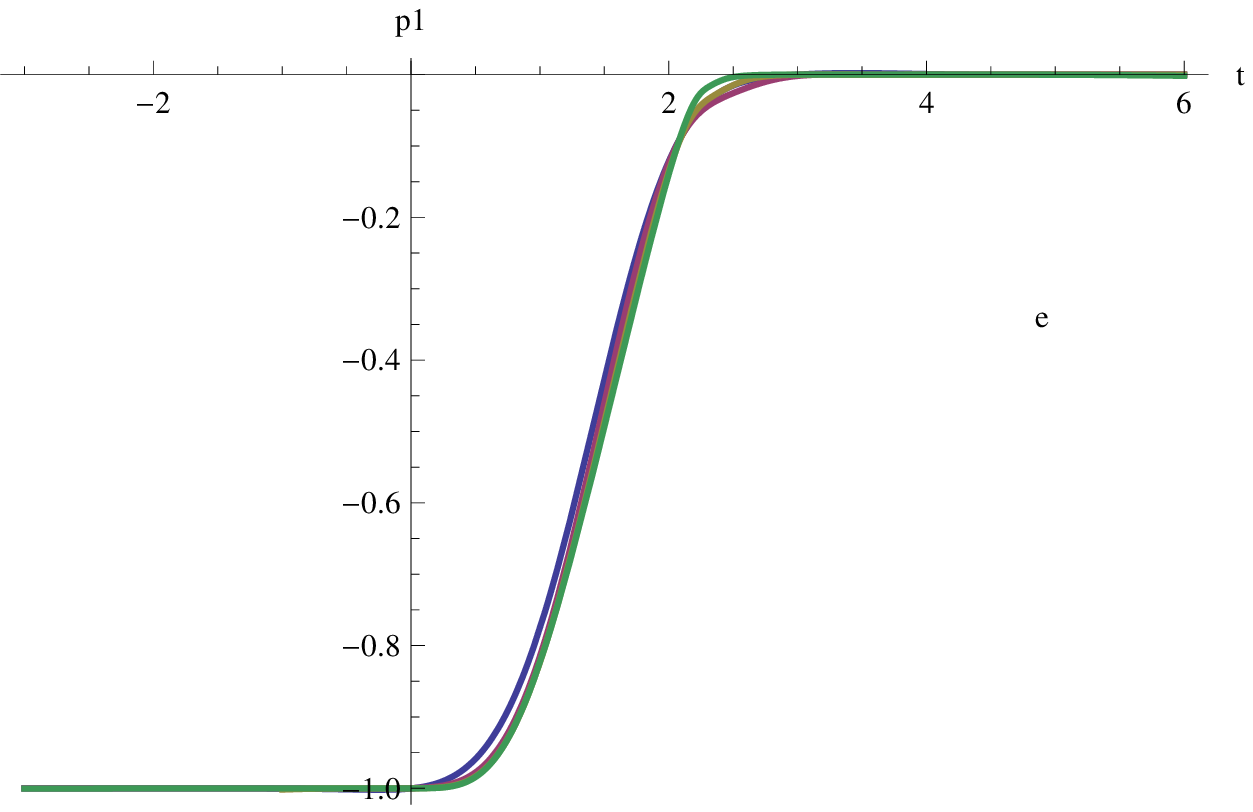}
\end{center}
  \caption{(Colour online) An alternative view of figure \ref{linpfig}.  The same plots are shown, but with the
thermalization measures being functions of the un-rescaled boundary time $\a\t_\ast$.  In this case one can see that
the smaller $\a$ is, the shorter equilibration tends to  take, from an absolute point of view.}  \label{linpafig}
\end{figure}

\begin{figure}
\begin{center}
\psfrag{aa}[Br][tl]{{$\scriptstyle{\frac{1}{\a}}$}}
\psfrag{t}[c]{{$\scriptstyle{\t_{(th)}}$}}
\psfrag{at}[c]{{$\scriptstyle{\a\t_{(th)}}$}}
    \includegraphics[width=3in]{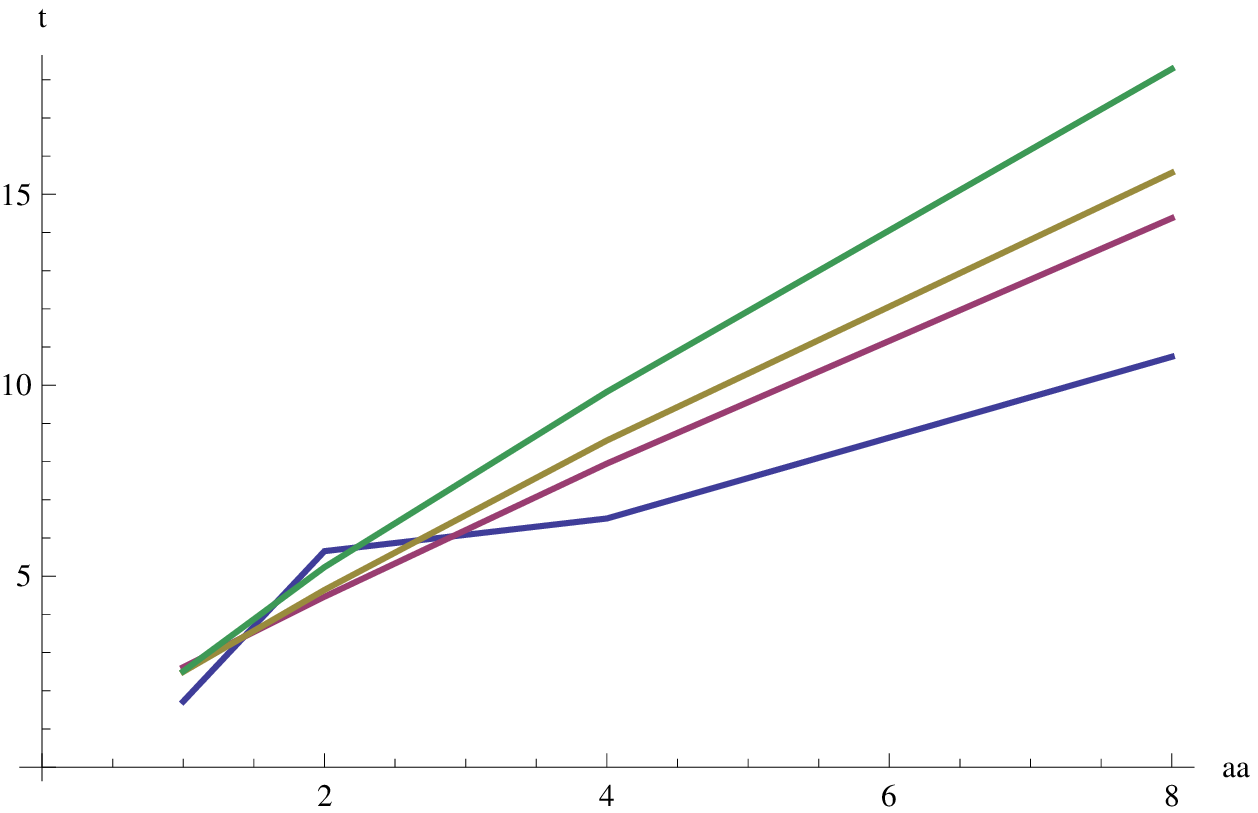}
        \includegraphics[width=3in]{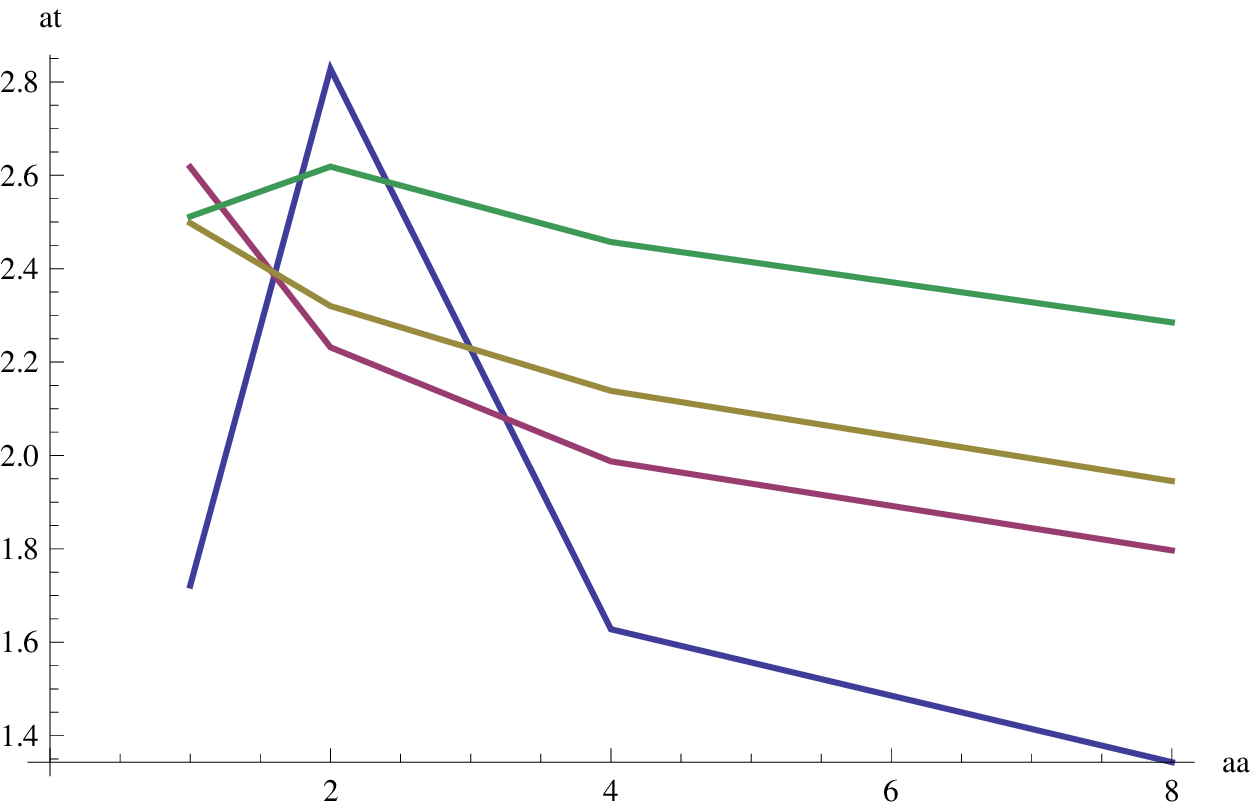}
\end{center}
  \caption{(Colour online) We show the equilibration times of $\mathcal{L}_{2}$ for various values of $\r_m$ as a function of the
  inverse of the quenching parameter $\a$, for $\a=1$, $\frac{1}{2}$, $\frac{1}{4}$ and $\frac{1}{8}$. On the left we show the rescaled equilibration time $\t_{(th)}$ as defined in \eqref{fth}, while on the right we show the same plot, but for the un-rescaled equilibration time $\a\t_{(th)}$.
  The blue, purple, yellow and green curves correspond to $\r_m=0.9\r_h$, $0.99\r_h$,
  $0.995\r_h$ and $0.999\r_h$ respectively.  Notice how the trends change sign from the left to the right plots.}  \label{linalphatherm}
\end{figure}

\begin{figure}
\begin{center}
\psfrag{t}[Br][tl]{{$\scriptstyle{\t_\ast}$}}
\psfrag{fth}[c]{{$\scriptstyle{f_{(th)}(\t_\ast)}$}}
  \includegraphics[width=3in]{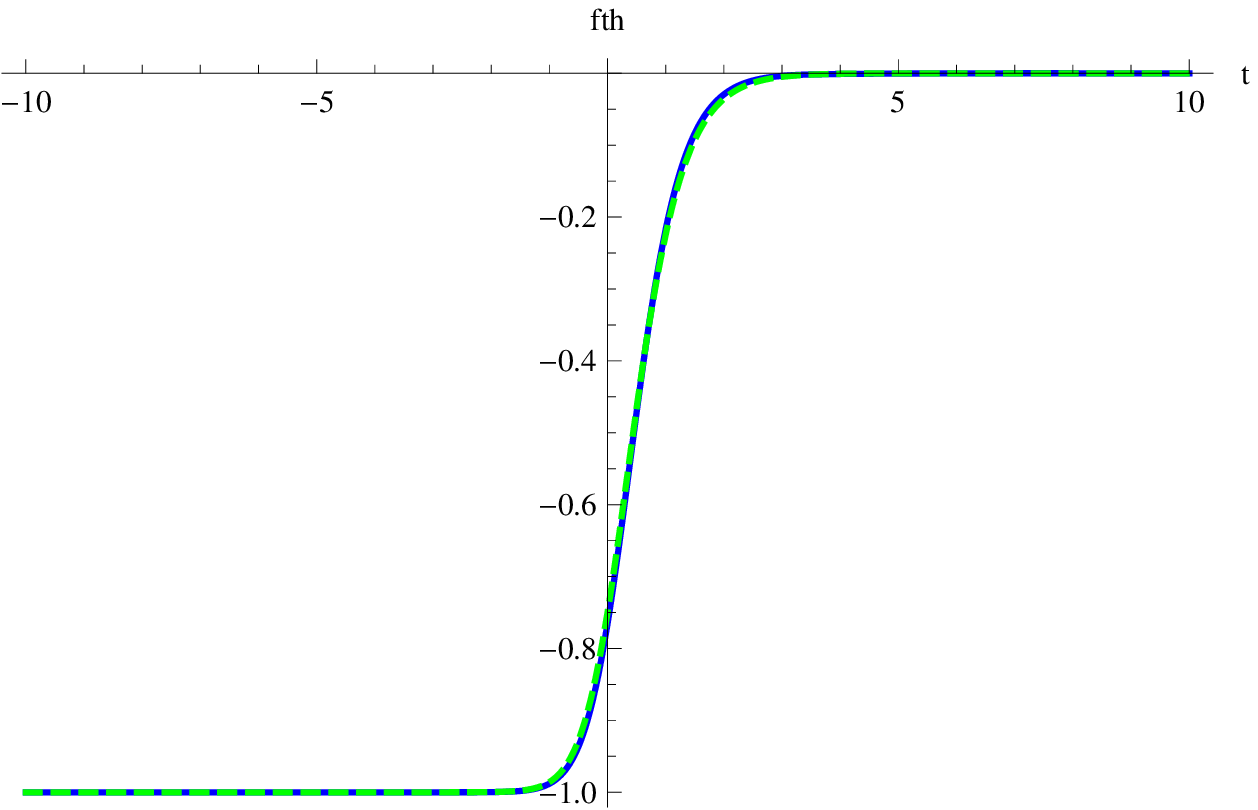}
\end{center}
  \caption{(Colour online) Here we show the thermalization measures
  as defined in \eqref{fth} of $\mathcal{L}_{2}$ in blue when $\r_m=0.1$ and $\a=\frac 12$,
  and of $p_0^2$ in the dashed green curve.  They closely coincide, since for such a small surface,
  the behaviour of the two-point function is dominated by the near-boundary metric.}
      \label{linvsp02}
\end{figure}

In figures \ref{linpfig} and \ref{linpafig}, we plotted the thermalization measure $\mathcal{L}_{2(th)}$ of the
two-point function (as defined in \eqref{fth}) for various values $\r_m=\r_h\times
\{0.1,0.5,0.9,0.99,0.999\}$ 
 ($\r_h=1/\a$ being the horizon position in the current coordinates) of the depth that the geodesic
extends into the geometry.
It is straightforward to convert $\r_m$ into the corresponding separation of the end-points in the two-point
function, \ie $2y_m$, by making the replacement $\r=\frac{z\r_m}{\a}$ in  equation \eqref{rpy} and then integrating:
\begin{equation}
\begin{split}
&{2y_m}
= 2\r_m\
\int_{0}^1  \frac{z dz}{(1-z^2)^{1/2} (1- z^4\r_m^4)^{1/2}}\,, \\
&\{(\r_m,2y_m)\}\approx
\{(0.1,\,0.2),\, (0.5,\,2.0),\, (0.9,\,2.35),\, (0.99,\,4.00),\,
(0.999,\,5.56)\}.
\end{split}
\eqlabel{rmtoenergy}
\end{equation}
Of course, as we vary the rate of the quenches, $\a$ provides a natural scale with which to compare
these separations. In particular, we examined $\a = \{1,\ft 12 =0.5,\ft 14 =0.25,\ft 18 =0.125\}$.
Alternatively, we can associate an energy with the two-point correlators using
$E_{2pt}=1/y_m$, which is roughly the minimum energy scale to which these nonlocal probes are sensitive.
For the different quenches, we might then compare $E_{2pt}$ with the quenching rate $1/\a$.
In each plot, the thermalization measure is plotted for a range of
quenching times $\a$.  Since the two-point functions we calculate are for points on the boundary of
the spacetime $\r=0$, we plot the thermalization measure
against the boundary time $\t_\ast$.  We see in figure
\ref{linpfig} that compared to the time-scale $\a$ set by the quench, the faster the quench is, the longer
the two-point function takes to equilibrate.  We notice in figure \ref{linpafig} that faster quenches
still equilibrate faster in the un-rescaled ``physical'' time $\a\t_\ast$.  That is to say, as
we increase the rapidity of the quench, $\a$ decreases faster than the
thermalization time $\t_{therm}$ for a correlator with fixed width $2\,\frac{y_m}{\a}$.  This behaviour
is more accurately reflected in figure \ref{linalphatherm}, where we plot the thermalization times (both
rescaled and un-rescaled) for different $\r_m$, as a function of $\a$.  We see that as $\a$ decreases
($\frac{1}{\a}$ increases) the slopes of the monotonic curves change sign.

In figure \ref{linvsp02} we plot the thermalization measures of the perturbation of the two-point function for $\a=\frac 12$ and $\r_m=0.1$,
as well as the thermalization measure of the non-normalizable mode-squared.  They very closely match each other,
as one might expect for small separations.  For such small separations, the geodesic does not dip
very far into the bulk geometry, and only ``sees'' the near-boundary metric, and its perturbations.
Since in the near boundary limit, the metric perturbations $\hat{a}$ and $b$ are proportional to
$p_0^2$, so will $\mathcal{L}_{2}$ be, as can be seen from equation \eqref{l2new}.  In the dual field theory
picture, one would say that for small separations the two-point function probes the UV.

\begin{figure}
\begin{center}
\psfrag{t}[Br][tl]{{$\scriptstyle{\t_\ast}$}}
\psfrag{v}[c]{{$\scriptstyle{\mathcal{L}_{2(th)}(\t_\ast)}$}}
\psfrag{a}[cc][][1.5]{{$\scriptstyle{\a=1}$}}
\psfrag{b}[cc][][1.5]{{$\scriptstyle{\a=\frac{1}{2}}$}}
\psfrag{c}[cc][][1.5]{{$\scriptstyle{\a=\frac{1}{4}}$}}
\psfrag{d}[cc][][1.5]{{$\scriptstyle{\a=\frac{1}{8}}$}}
  \includegraphics[width=3in]{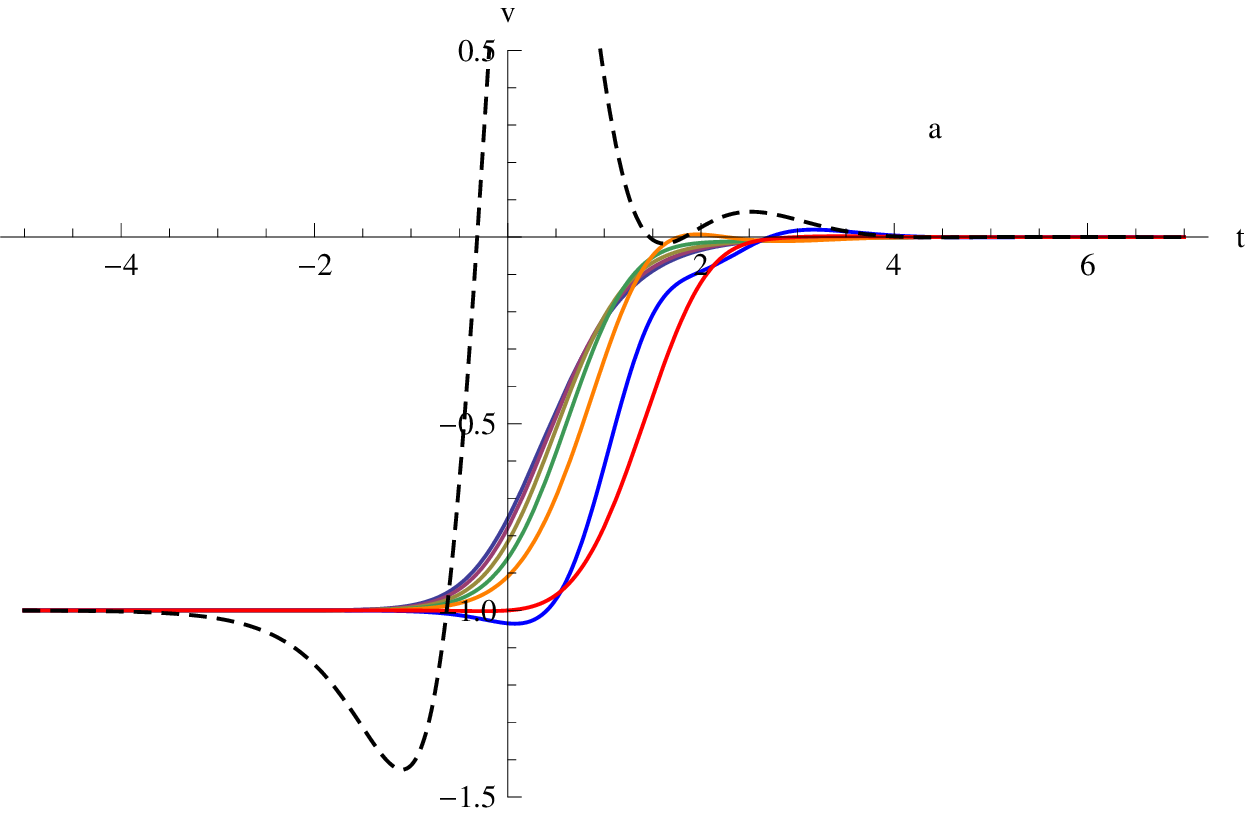}
  \includegraphics[width=3in]{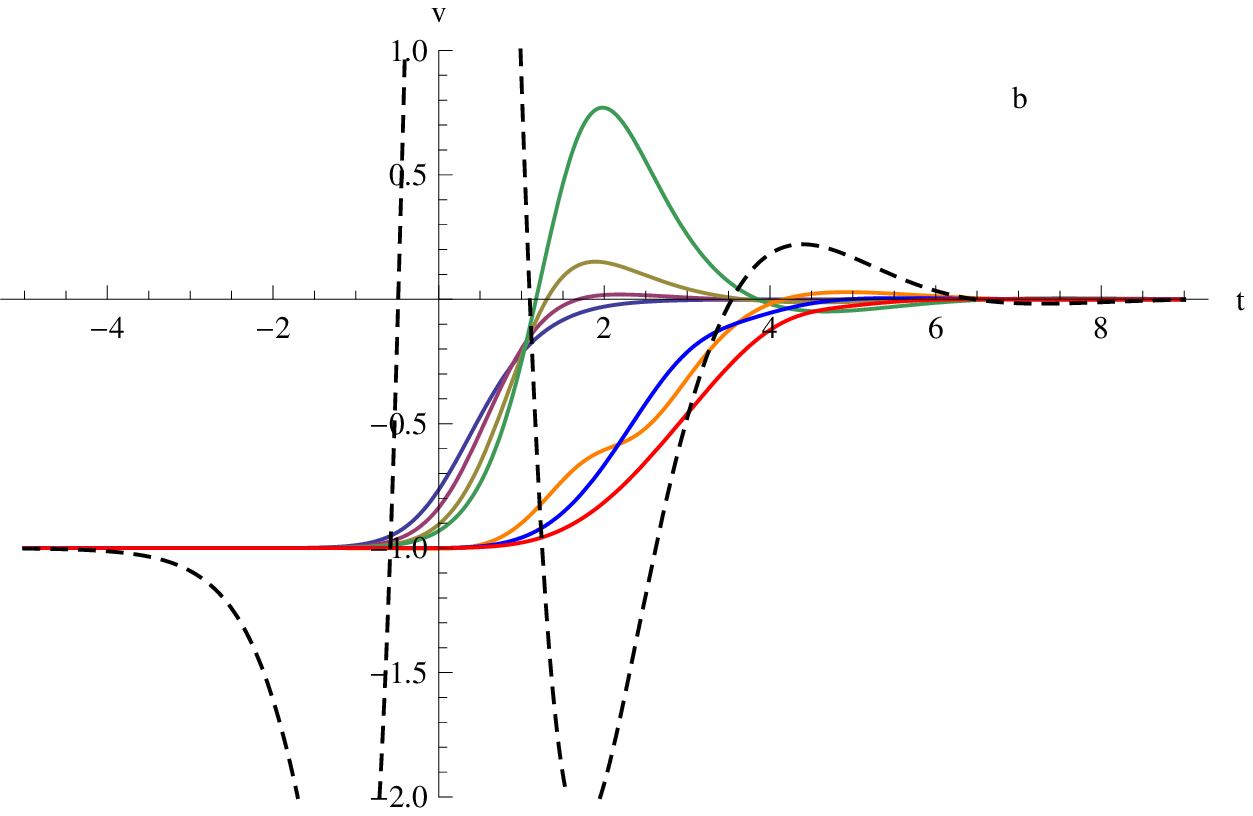}
    \includegraphics[width=3in]{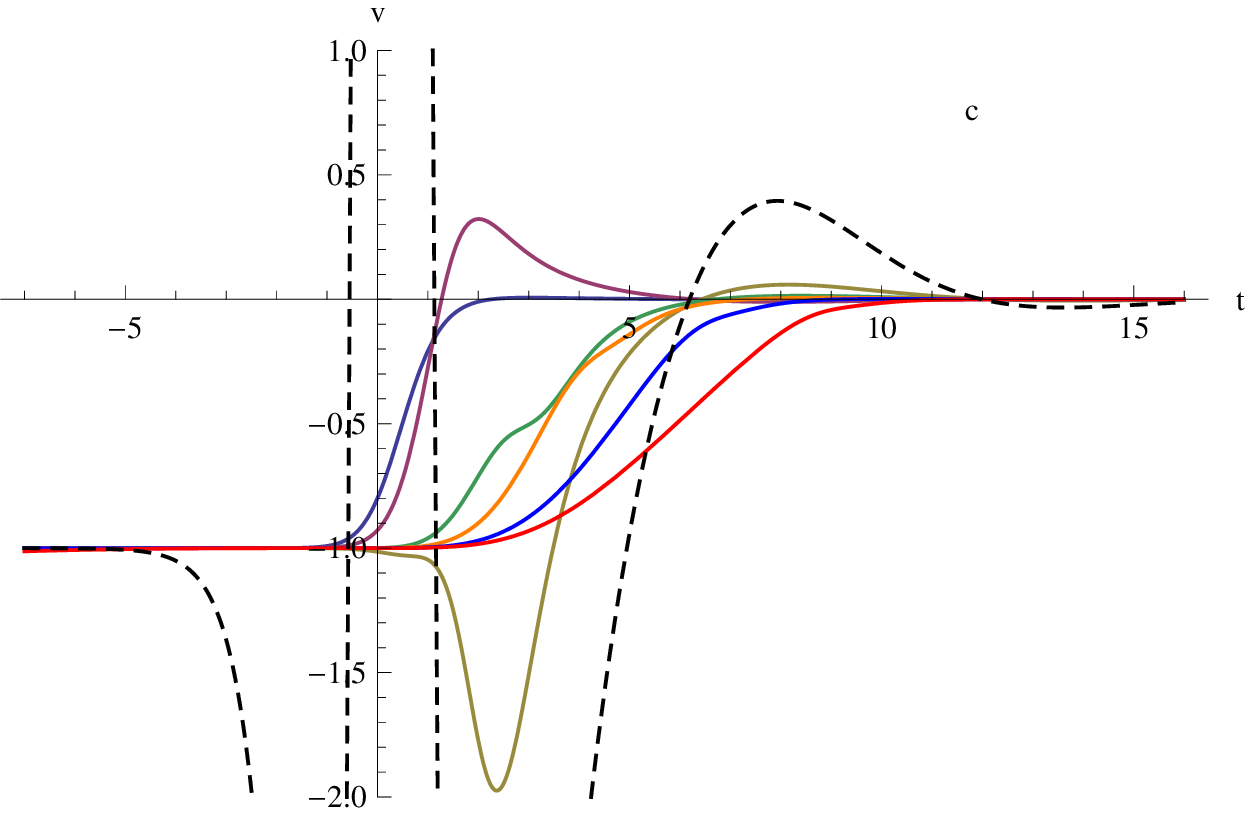}
        \includegraphics[width=3in]{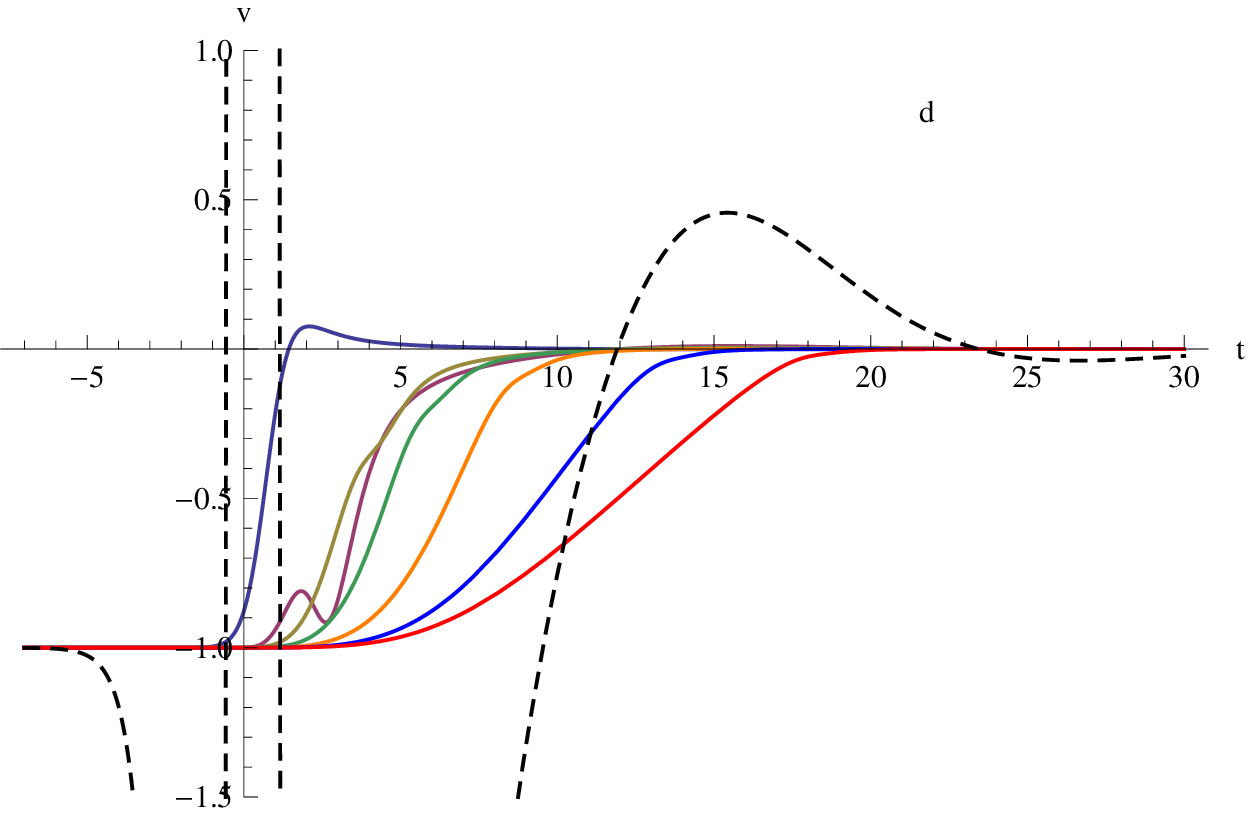}
\end{center}
  \caption{(Colour online) The evolution of $\mathcal{L}_{2(th)}$ as a function of the boundary
time of the two-point correlator. The plots are (from left
  to right, top to bottom) for $\a=1,\frac{1}{2},\frac{1}{4}$ and $\frac{1}{8}$,
  respectively.  Each figure contains the plot for an equal time two-point function
$\r_m=0.1\r_h$, $0.3\r_h$, $0.5\r_h$, $0.7\r_h$, $0.9\r_h$
$0.99\r_h$ and $0.999\r_h$, respectively.  The plots for $0.9\r_h$
$0.99\r_h$ and $0.999\r_h$ are orange, bright blue, and red, respectively.  We also plotted $p_{2(th)}$ in dashed
lines, to compare with the equilibration of the two-point functions.  We can see that the
larger the separation of the two points (\ie the depth $\r_m$), the longer the thermalization time is in
each case.}  \label{linafig}
\end{figure}

In figure \ref{linafig}, we compare the thermalization times for two-point functions with different
separations, but the same quenching parameter $\a$.  We notice that the larger the separation of the two points,
the longer the two-point function takes to equilibrate.

\subsection{Entanglement entropy}

\subsubsection{Analytic expression for the entanglement entropy}

Another useful scale-dependent probe of thermalization is entanglement entropy (EE).
An elegant method was proposed by Ryu and Takayanagi \cite{ent1, ent2} to calculate EE for holographic theories. In particular,
the Ryu-Takayanagi prescription involves evaluating the Bekenstein-Hawking formula \eqref{yoza}
on all bulk surfaces $\gamma$ which are homologous to the entangling region on the boundary of the bulk spacetime.
The holographic EE is then found by extremizing over all such bulk surfaces:
\begin{equation} \eqlabel{rt}
S_{EE}={\rm ext}\,\frac{A_{\gamma}}{4\,G}.
\end{equation}
Note that this prescription was originally proposed for static situations but has extended to consider
dynamical bulk geometries in \cite{hrt}. We will simplify our calculations by
evaluating the entanglement entropy for regions on constant time slices in the boundary.
Further in equation \eqref{rt}, $G$ would be the Newton's constant
in the bulk theory but for our purposes, we have set $4\,G=1$.
The EE depends on the size of the entangling region at the asymptotic AdS boundary.
Observing how fast the EE of the region stabilizes can therefore
serve as an indicator of thermalization at different length scales,
in analogy to the two-point function.

For simplicity, we consider a boundary entangling region $\Sigma$ with a strip-geometry.
That is, the region is three dimensional, and is infinite in the directions
$y_2$ and $y_3$ (regulated by $K$), but has a finite width
in the $y_1$-direction.  The metric for the bulk spacetime can be expressed as
\begin{equation}
ds^2=-a\, d\t^2 + s^2d\vec{y}^2 - \frac{2d\r d\t}{\r^2}.
\eqlabel{metric3}
\end{equation}
A surface $\gamma$  in the bulk spacetime connecting to the boundary of $\Sigma$,
has a surface area given by
\begin{eqnarray}
S_{\Sigma} &=& \int^{\infty}_{-\infty} dy_2\,dy_3 \int^{y_m}_{-y_m} dy_1\, s^{2} \sqrt{-a\left(\t'\right)^2+s^2-2\t'\r'/\r^2}
\nonumber \\
&=& 2\,K^2 \int^{y_m}_0 dy\, s^{2} \sqrt{-a\left(\t'\right)^2+s^2-2\t'\r'/\r^2}. \label{entropy}
\end{eqnarray}
In \eqref{entropy}
the factor of $2$ comes from us only integrating over half the
interval of $y_1$ (renamed to $y$) in the second line, since $\gamma$ is symmetric about $y=0$.
Following the Ryu-Takayanagi prescription \cite{ent1, ent2} described above, the appropriate surface $\gamma$ for calculating the EE is then the one that
minimizes $S_{\Sigma}$.  Once again, the quench
is treated as a perturbation on the spacetime, and as in the case of the correlator,
this splits the entropy into the static part plus a perturbation:\footnote{Strictly speaking, equation
\eqref{entfull} is valid only when considering the static extremal surface $\gamma$.
In fact, the perturbations on $\gamma$ due to the backreaction of the metric
causes a separate contribution at order $\ell^2$ from the first term which is different from the order
$\ell^2$ contribution coming from the metric perturbations in equation \eqref{eent2}.  We come back to
the contribution from the surface perturbations in section \ref{eepert}.}
\begin{equation} \eqlabel{entfull}
S_{\Sigma} = S_{\Sigma(0)} + \ell^2S_{\Sigma(2)}.
\end{equation}

In the rescaled coordinates, the entropy has a time-independent zeroth-order contribution
\begin{equation}
S_{\Sigma(0)} = 2K^2 \int^{y_m/\a}_0 dy \frac{\sqrt{D(\t_0,\r_0)}}{\r_0^{3}},
\eqlabel{eent0}
\end{equation}
(where $D$ was defined in \eqref{D}) while the time-dependent perturbation of the
EE, $\ell^2 S_{\Sigma(2)}$, is given by
\begin{equation}
S_{\Sigma(2)} =  2K^2\int^{y_m/\a}_0 dy \frac{1}{2\r_0^3\sqrt{D}}
\left(b\left(2+4\,D\right)-\a^2\r_0^2 \left(\t_0'\right)^2 \hat{a}\right).
\eqlabel{eent2}
\end{equation}
Once again we will drop the subscripts on the coordinates.  The perturbations on the
shape of the surface $\t_2$ and $\r_2$ {\it do} enter at order $\ell^2$ of the EE (although only
$\r_2$ actually contributes to the entropy), but we will be explicit when referring to them.

Notice that as in the case of the two-point function, the expressions for the zeroth and second-order
entanglement entropy have no explicit $y$-dependence.  This is due to the simple choice of geometry
of the entangling surface.  Treating \eqref{eent0} as an action, we can find the conserved charge
from time translation invariance, as well as the Hamiltonian like we did in the case of the two-point
function.

From time translation invariance, we find the conserved quantity \cite{periodic}
\begin{equation} \eqlabel{striptime}
(1-\a^4\r^4)\t'+\r'= C.
\end{equation}
The condition that the surface is closed and smooth at $y=0$, makes the choice $C=0$.  By dividing by
$\r'$,  the chain rule again leads to the solution for the time-profile of the minimal surface of
\begin{equation}
\t(\r)=\t_\ast-\frac{\tan^{-1}(\a \rho)+\tanh^{-1}(\a \r)}{2 \alpha }.
\end{equation}
The Hamiltonian to our action \eqref{eent0} will be constant because it has no $y$-dependence, and leads to
the identity
\begin{equation} \eqlabel{stripid}
D(\t,\r)\,\a^6\r^6 = \r_m^6.
\end{equation}
Using expression \eqref{D} for $D$, and substituting in for $\t'$ from \eqref{striptime},
we  find the equation for the radial profile
\begin{equation} \eqlabel{rprime}
\r' = - \frac{\sqrt{\left(1-\a^4\r^4\right)\left(\r_m^6-\a^6\r^6\right)}}{\a^3\r^3}.
\end{equation}
From the chain rule, we have that $\frac{d\r}{dy} = \left(\frac{dy}{d\r}\right)^{-1}$.  This gives us the additional
equation
\begin{equation} \eqlabel{yprime}
\frac{dy}{d\r} = - \frac{\a^3\r^3}{\sqrt{\left(1-\a^4\r^4\right)\left(\r_m^6-\a^6\r^6\right)}}.
\end{equation}

Using equations \eqref{striptime}, \eqref{stripid}, \eqref{rprime}, and \eqref{yprime}, we can express
\eqref{eent0} and \eqref{eent2} as integrals over $\r$ only.  This makes the calculation
much simpler since we do not need to solve for the $y$-profile of the surface $\gamma$ (\eg
we would need to solve for the $y$-profile in the case of a spherical entangling surface).
The expression for the EE at zeroth and second-order in perturbation theory then becomes
\begin{equation} \eqlabel{eent0b}
S_{\Sigma(0)} = 2K^2 \int^{\r_m/\a}_\epsilon d\r\frac{\r^3_m}{\r^3\sqrt{\left(1-\a^4\r^4\right)\left(\r_m^6-\a^6\r^6\right)}},
\end{equation}
and
\begin{equation} \eqlabel{eent2b}
S_{\Sigma(2)} =  2K^2\int^{\r_m/\a}_\epsilon d\r
\frac
{-\a^2\hat{a}\,\r^2\left(\r_m^6-\a^6\r^6\right)+2\,b\,\left(1-\a^4\r^4\right)\left(2\r_m^6+\a^6\r^6\right)}
{2\r^3\r_m^3\left(1-\a^4\r^4\right)\sqrt{\left(1-\a^4\r^4\right)\left(\r_m^6-\a^6\r^6\right)}},
\end{equation}
respectively.  In \eqref{eent0b} and \eqref{eent2b}, $\epsilon$ is the near-boundary cut-off, which
we introduced since both the leading-order and perturbative EE have UV divergences.

Both $S_{\Sigma(0)}$ and $S_{\Sigma(2)}$ have divergences close to the boundary of the
spacetime \cite{periodic}.  We can identify these divergences by using the perturbation series of
$\hat{a}$ and $b$ near the boundary of the spacetime (\ie in small $\r$):
\begin{eqnarray}
\hat{a}&=&-\frac{1}{6}p_0^2+\calo\left(\r^2\right),\label{hatapert}\\
b&=&-\frac{\a^2}{12}p_0^2\r^2+\calo\left(\r^3\right) \label{bpert}.
\end{eqnarray}
We then substitute these series into the expressions for the integrands in \eqref{eent0b} and
\eqref{eent2b}, and expand the integrand close to $\r=0$ to find its divergent parts.
Upon integrating, the divergence of the EE in terms of the cut-off is
\begin{equation} \eqlabel{sdiv}
\mathcal{S}_\text{div}=K^2\left(\frac{1}{\epsilon^2}+
\ell^2\frac{\a^2p_0^2\left(\t_\ast\right)}{6}\log\epsilon\right).
\end{equation}
It is worth noting that the logarithmic term is universal, as pointed out in \cite{hms,lms}.  This divergence
can be recast in the form
\begin{equation}
\chi\,A_{\Sigma}m^2_f\log(\epsilon),
\end{equation}
where $\chi$ is a universal numerical constant, $A_{\Sigma}$ is the area of the entangling surface $\Sigma$
on the boundary, and $m_f$ is the mass of the fermionic operator $\mathcal{O}_3$ (our quenching operator)
in the boundary CFT.  Comparing our divergence to the desired form,
we see that $2\a^2K^2$ is the surface area (density) of $\Sigma$
(the factor of $2$ coming form the fact that there is a surface of area density $K^2$ on
either side of the strip), $\ell\,p_0$ is the non-normalizable mode of our
scalar field, and by the holographic
duality (and our conventions) is equal to $m_f$.  This means that the remaining numerical constant in the divergence $\frac{1}{12}$
times some proportionality factor will be equal to the universal constant $\chi$.  In fact, this proportionality factor agrees with the above two references, and therefore the factor $\frac{1}{12}$ is our universal
coefficient and must agree for the logarithmic term for any entangling surface $\Sigma$.  As a further test that we have the correct constant, we calculate the logarithmic term for a spherical entangling
surface $\Sigma$.  Showing the final result, we obtain a logarithmic divergence in $\r$ of
\begin{eqnarray}
4\pi R^2\frac{\ell^2p_0^2}{12}\log(\epsilon),
\end{eqnarray}
$R$ being the radius of the sphere on the boundary.  We see that since $4\pi R^2$ is the surface
area of the spherical entangling surface, we indeed obtain the same universal coefficient of $\frac{1}{12}$.

We should note that in our calculations of \eg the thermalization time associated with the entanglement entropy,
we will simply discard the divergent contributions in equation \eqref{sdiv} and work only with the finite part of the entanglement entropy
--- see equation \eqref{eentreg} below. In fact, given the definition of the entanglement measure in equation \eqref{fth}, the area
law divergence in equation \eqref{sdiv} will drop out, since these agree for the EE at all times. However, the logarithmic divergence
shown there will not cancel, since it is proportional to $p_0^2(\t)$, and are therefore different at early and late times. Hence one might worry that
the precise results will be sensitive to cut-off redefinitions. However, we do not expect this issue will effect the qualitative features determined
 in the following. This matter could be avoided altogether by using a renormalized version of the
entanglement entropy, \eg $y_m\,\partial_{y_m}\!S_\Sigma$ \cite{hoog,foog}. We hope to return to this approach in future work.

\subsubsection{Contribution from surface perturbation} \label{eepert}

As mentioned, the perturbation of the static surface $\gamma$, namely $\r_2$ and $\t_2$ does not
contribute to the entropy at order $\ell^2$, with the exception of the boundary term of the EE.
Expanding $S_{\Sigma(0)}$ in \eqref{eent0} in terms of $\r_2$ and $\t_2$ as
\begin{equation}
S_{\Sigma(0)} = S_{\Sigma(0,0)} + \ell^2\dd S(\r_2,\t_2),
\end{equation}
$S_{\Sigma(0,0)}$ depending only on the unperturbed profile $\r_0$ and $\t_0$, we obtain the integral
\begin{equation}
\delta S = 2\,K^2\ell^2\int^{y_m/\a}_{0}dy\left(\frac{2\t'^2\r^4-3D}{\a^4\r^4\sqrt{D}}\r_2
- \frac{\t'}{\a^3\r^3\sqrt{D}}\r_2' -
\frac{\left(1-\a^4\r^4\right)\t'-\r'}{\a^3\r^3\sqrt{D}}\t'_2\right). \eqlabel{bdyterm1}
\end{equation}
Notice that the numerator of the factor multiplying $\t'_2$ is the left hand side of equation
\eqref{striptime}, and is therefore zero on-shell.  We therefore only care about the terms involving
$\r_2$ and its derivative.  Performing integration by parts on the second term, we obtain a term
multiplying $\r_2$ plus a total derivative term.  The term involving $\r_2$ combines with the first
term in \eqref{bdyterm1} to give the equation of motion for $\r$, and is therefore zero on-shell.  All
that remains is the total derivative term, which we can integrate to evaluate at the limits of
integration:
\begin{equation}
\delta S = 2\,K^2\ell^2\left(\frac{\t'}{\a^3\r^3\sqrt{D}}\r_2\right)\Big|^{y_m/\a}_{0}.
\end{equation}
Solving perturbatively for $\r(y)$ in \eqref{rprime} in small $y$, we see that factor of $\r_2$ in
the above equation vanishes when $y=0$.  We can solve \eqref{rprime} using the new coordinate
\begin{equation}
x=(y_m/\a-y).
\end{equation}
In that case we can evaluate the coefficient of $\r_2$ near $x=0$ ($y=y_m/\a$).  We see that
the coefficient diverges as
\begin{equation} \eqlabel{r2coef}
-2\,K^2\ell^2\frac{1}{2\sqrt{2}\r_m^{9/4}\a^{2}\delta^{3/4}},
\end{equation}
where $\delta$ is the cut-off in the $x$-direction.  Since this factor contains a divergence, it is
necessary to find the small-$x$ expansion for $\r_2$ in order to find potential divergent and finite
contributions to the EE from the boundary term.

In order to solve for the perturbation $\r_2$ of the surface $\gamma$,
we derive its Euler-Lagrange equation from
$S_\Sigma$ in equation \eqref{entropy} for both $\r_2$ and $\t_2$.  The equations
\begin{eqnarray}
0&=& \delta_{\r_2}S_\Sigma-\frac{d}{dx}\left(\delta_{\r_2'}S_\Sigma\right),\\
0&=& \delta_{\t_2}S_\Sigma-\frac{d}{dx}\left(\delta_{\t_2'}S_\Sigma\right)
\end{eqnarray}
yield the equations of motion for $\r_0$ and $\t_0$ at order $\ell^2$, and the coupled linear
equations of motion for $\t_2$ and $\r_2$ at order $\ell^4$.  These equations involve $\hat{a}$, $b$,
$\t_2$ and $\r_2$ and their derivatives up to second-order.  These full equations
also contain nonlinearities in $\t_0$ and $\r_0$ and are too formidable to be
explicitly included in this paper.  Nonetheless, we can find the
leading order expansions for $\r_2$ and $\t_2$.  We do this by substituting in
for the perturbation series of $\hat{a}$ and $b$ in small $\r_0$ and
the asymptotic series for $\t_0$ and $\r_0$ in small $x$.

We find the leading order {\it degenerate} solutions in terms of the boundary time $\t_\ast$ to be
\begin{eqnarray}
\t_2(x) &=& m(\t_\ast) x^{3/4} + \dots, \\
\r_2(x) &=& n(\t_\ast) x^{3/4} + \dots,
\end{eqnarray}
where
\begin{equation}
m(\t_\ast)+n(\t_\ast) = -\frac{\sqrt{2}}{9}\a^2\r_m^{9/4}p_0^2(\t_\ast).
\end{equation}
This means that to leading order $\r_2$ has the right behaviour  in $x$ to cancel
the divergence in \eqref{r2coef} and have the boundary term make a finite contribution
to the entanglement entropy. Although the equation above does not tell us the exact
value of $n$, we can reasonably expect it to be of the form
\begin{equation} \eqlabel{nprop}
n(\t_\ast) \propto \a^2\r_m^{9/4}p_0^2(\t_\ast),
\end{equation}
with a purely numerical factor missing.

In order to solve for the numerical factor in \eqref{nprop}, we solve for $\r_2$ and $\t_2$ again in
a static background.  That is, we solve the system at large times, after it has fully equilibrated, and
$p_0(\t_\ast)=1$.  In that case, we can find a simpler equation of motion for $\t_2$, since the
metric perturbations have no explicit time-dependence.  The Euler-Lagrange equation for $\t_2$ becomes
\begin{equation}
\frac{d}{dx}\left(\delta_{\t_2'}S_\Sigma\right)=0.
\end{equation}
Furthermore, this equation implies that
\begin{equation}
\delta_{\t_2'}S_\Sigma=\kappa,
\end{equation}
where $\kappa$ is a constant.
By similar arguments as for the unperturbed time-profile of $\gamma$ in the
previous subsection, we can set $K=0$.  The resulting equation is much simpler than we obtained in the
time-dependent case, and contains only terms either independent of $\t_2$, or terms that are linear in
$\t_2'$.  It is therefore possible to solve for $\t_2'$ as
\begin{equation}
\t_2'(x)=F(\r_2,\hat{a},b;\r_0,\t_0).
\end{equation}
In the above equation, $F$ is linear in $\r_2$, $\hat{a}$, $b$ and
their derivatives, but is nonlinear in the unperturbed profile $\r_0$ and $\t_0$ of the minimal surface.

Substituting in for $\t_2'$ (and $\t_2''$) in the static Euler-Lagrange equation for $\r_2$, we
have an equation where the only unknown function is $\r_2$.  Solving again for $\r_2$ perturbatively in $x$,
we can find an exact solution for its leading coefficient.  In the static case
\begin{equation}
\r_2(x) \xrightarrow[\t_\ast\to\infty]{} -\frac{5}{18 \sqrt{2}}\a^2\r_m^{9/4}x^{3/4} + \dots.
\end{equation}
Knowing it's time-dependence on the source, we can write the leading solution for $\r_2$ as
\begin{equation}
\r_2(x) = -\frac{5 }{18 \sqrt{2}}\a^2\r_m^{9/4}p_0^2(\t_\ast)x^{3/4} + \dots.
\end{equation}
Substituting back for the leading-order solution of $\r_2$ evaluated at the cut-off $\delta$,
we see that the boundary term has a finite contribution to the EE of
\begin{equation} \eqlabel{bdytermfinal}
\dd S = \a^2\ell^2K^2\left(\frac{5}{36}p_0^2(\t_\ast)\right).
\end{equation}

We note that it is necessary to work with $\r_2(y)$ rather than $y_2(\r)$ in the treatment above.
To leading order in $\ell$ the profile of $\gamma$ remains static, and we can parameterize it with
either $\r$ or $y$.  The expressions \eqref{eent0} and \eqref{eent0b}, as well as \eqref{eent2} and
\eqref{eent2b} are therefore related by a change in coordinates.
However, when dealing with the perturbations of the minimal
surface, depending on whether one lets it fluctuate in the $y$-direction or $\r$-direction,
either $y_m$ or $\r_m$ will be corrected for by the perturbations.  We must therefore make a choice
of which of these parameters to keep fixed.  Since $y_m$ is the field theory observable, \ie it determines
the width of the strip in the boundary, we
choose the perturbations of $\r$ and $\t$ to be functions of $y$, allowing for $\r_m$ to be adjusted
at order $\ell^2$, though not affecting the results calculated on the static surface $\gamma$.

We further note that this is a very unexpected result,
since in the case of unperturbed holographic EE, the boundary term
is typically ignored as it is assumed to vanish from the equations of motion --- see for example
\cite{bdyterm,periodic,hms,lms}, in which the boundary term is implicitly set to zero in the presence of
a relevant perturbation.
 We point out that in the case of
a spherical entangling surface on the AdS boundary, the boundary term above contributes both a
$\log$-divergence as well as a finite contribution.  While the divergence in that case is readily
obtainable by the same perturbative treatment as we did for the strip, the boundary term also has a
finite contribution depending on a normalizable mode, which would require knowing the full
profile of $\r_2$ in $y$ in order to be extracted.  That is outside the scope of this paper.

\subsubsection{Regularization of the entanglement entropy}

In equations \eqref{eent0b} and \eqref{eent2b}, the integrands have inverse square-root divergences near
$\r=\frac{\r_m}{\a}$.  This is not a problem mathematically, but since we have to numerically integrate these
expressions, our results would be more accurate if the integrands didn't diverge at all.

In order to separate the EE into a finite and divergent part, we use the perturbation series of the
metric perturbations
\begin{eqnarray}
\hat{a}&=&-\frac{1}{6}p_0^2+\r\,\hat{a}_2,\label{hatapert2}\\
b&=&-\frac{\a^2}{12}p_0^2\r^2+\r^3b_2 \label{bpert2},
\end{eqnarray}
since only the leading-order terms in these series contribute to the divergence of the EE.
We therefore have a finite part of the EE,
\begin{equation} \eqlabel{sfin}
S_{\Sigma(2)(fin)} =  K^2\int^{\r_m/\a}_0 d\r
\frac
{-\a^2\hat{a}_2\left(\r_m^6-\a^6\r^6\right)+2\,b_2\,\left(1-\a^4\r^4\right)\left(2\r_m^6+\a^6\r^6\right)}
{\r_m^3\left(1-\a^4\r^4\right)\sqrt{\left(1-\a^4\r^4\right)\left(\r_m^6-\a^6\r^6\right)}},
\end{equation}
as well as the divergent part
\begin{equation} \eqlabel{sdiv2}
S_{\Sigma(2)(div)} = K^2\int^{\r_m/\a}_\epsilon d\r \,p_0^2(\t(\r))
\frac{-\left(\r_m^6-\a^6\r^6\right)+\left(1-\a^4\r^4\right)\left(2\,\r_m^6+\a^6\r^6\right)}
{6\,\r\,\r_m^3\left(1-\a^4\r^4\right)\sqrt{\left(1-\a^4\r^4\right)\left(\r_m^6-\a^6\r^6\right)}}.
\end{equation}
In order to regularize $S_{\Sigma(2)(div)}$, we add the counterterm
\begin{equation} \eqlabel{counter}
S_{counter}=\a^2K^2 \int^{\r_m/\a}_\epsilon d\r \frac{p_0^2(\t_\ast)}{6\,\r}.
\end{equation}
There is some ambiguity in the counterterm \eqref{counter}, namely that it need only have the right
asymptotic behaviour to cancel the divergence of the integrand in \eqref{sdiv2}.  That means we could \eg
use $p_0^2(\t(\r))$ instead of $p_0^2(\t_\ast)$ in \eqref{counter}, and it would still cancel the
divergence of the EE, but yield a different finite result for the regularized EE.  In order to
circumvent this ambiguity, we need to add back the finite contribution that gets subtracted in the
counterterm, that is, the non-divergent limit of the integral.  Therefore, we need to add back the
finite contribution
\begin{equation}
S_{cor}=-\frac{1}{6}\a^2K^2\log\left(\frac{\r_m}{\a}\right)\,p_0^2(\t_\ast),
\end{equation}
to obtain a finite EE that is invariant under this particular regularization scheme.

In order for these equations to be accurately integrable numerically, we make the same change of
coordinates \eqref{qcor} as we did for the two-point function, namely
\begin{equation}
\r=\frac{\r_m}{\a}\left(1-q^2\right).
\end{equation}
This change of equations yields the new full expression for the regularized EE at order $\ell^2$ of
\begin{eqnarray}
&&S_{\Sigma(2)} = S_{\Sigma(2)(fin)} + S_{\Sigma(2)(div)}
+ S_{counter} + S_{cor} + \dd S  \nonumber\\
&=&K^2\int^{1}_0dq\,2\frac{\r_m q}{\a}
\frac{2\,b_2\cdot\left(2+(1-q^2)^6\right)\left(1-\r_m^4(1-q^2)^4\right)-\a^2\hat{a}_2\cdot\left(1-(1-q^2)^6\right)}
{\sqrt{1-(1-q^2)^6}\left(1-\r_m^4(1-q^2)^4\right)^{3/2}} \nonumber\\
&&-\a^2K^2\int^{1}_{0}dq
\frac{2qp_0^2(\t(\r))[(1-q^2)^6-1+\left(2+(1-q^2)^6\right)\left(1-(1-q^2)^4\r_m^4\right)]}
{6\sqrt{\left(1-(1-q^2)^6\right)\left(1-(1-q^2)^4\r_m^4\right)}\left(1-q^2-(1-q^2)^5\r_m^4\right)}
\nonumber\\
&&+\a^2K^2p_0^2(\t_\ast)\left(\left[\int^{1}_{0}\frac{2\,q}{6(1-q^2)}\right]
-\frac{1}{6}\log\left(\frac{\r_m}{\a}\right)+\frac{5}{36}\right),
\eqlabel{eentreg}
\end{eqnarray}
$\dd S$ being the boundary term defined in equation \eqref{bdyterm1}, and its final expression before
change of coordinates shown in \eqref{bdytermfinal}.

\subsubsection{Numerical calculation of the entanglement entropy}

Using equation \eqref{eentreg}, we can calculate the evolution of the entanglement entropy for different
quenching rates $\a$ and for entangling surfaces with different widths $y_m$ (corresponding to different
surface heights $\r_m$).  The procedure is very similar to the one described for the
two-point function, and we give only a brief overview here.

We calculate the EE by discretizing the integrand in the first term of
\eqref{eentreg} in $q$, and then integrating the interpolating function instead.  This shows a speedup
in the numerical calculation, without noticeable loss of precision.  The other terms in \eqref{eentreg}
do not require such a discretization procedure, since they do not involve the numerical metric components
$\hat{a}_2$ and $b_2$.

The metric components are calculated from their numerically calculated Chebyshev coefficients, as described
in the appendix.

The evolution of the perturbation of the EE is seen by calculating it in a range of boundary times
$\t_\ast$.

\begin{figure}
\begin{center}
\psfrag{t}[Br][tl]{{$\scriptstyle{\t_\ast}$}}
\psfrag{p1}[c]{{$\scriptstyle{S_{\Sigma(2)(th)}(\t_\ast)}$}}
\psfrag{a}[cc][][1.2]{{$\scriptstyle{\r_m=0.1\r_h}$}}
\psfrag{b}[cc][][1.2]{{$\scriptstyle{\r_m=0.5\r_h}$}}
\psfrag{c}[cc][][1.2]{{$\scriptstyle{\r_m=0.9\r_h}$}}
\psfrag{d}[cc][][1.2]{{$\scriptstyle{\r_m=0.99\r_h}$}}
\psfrag{e}[cc][][1.2]{{$\scriptstyle{\r_m=0.999\r_h}$}}
  \includegraphics[width=2in]{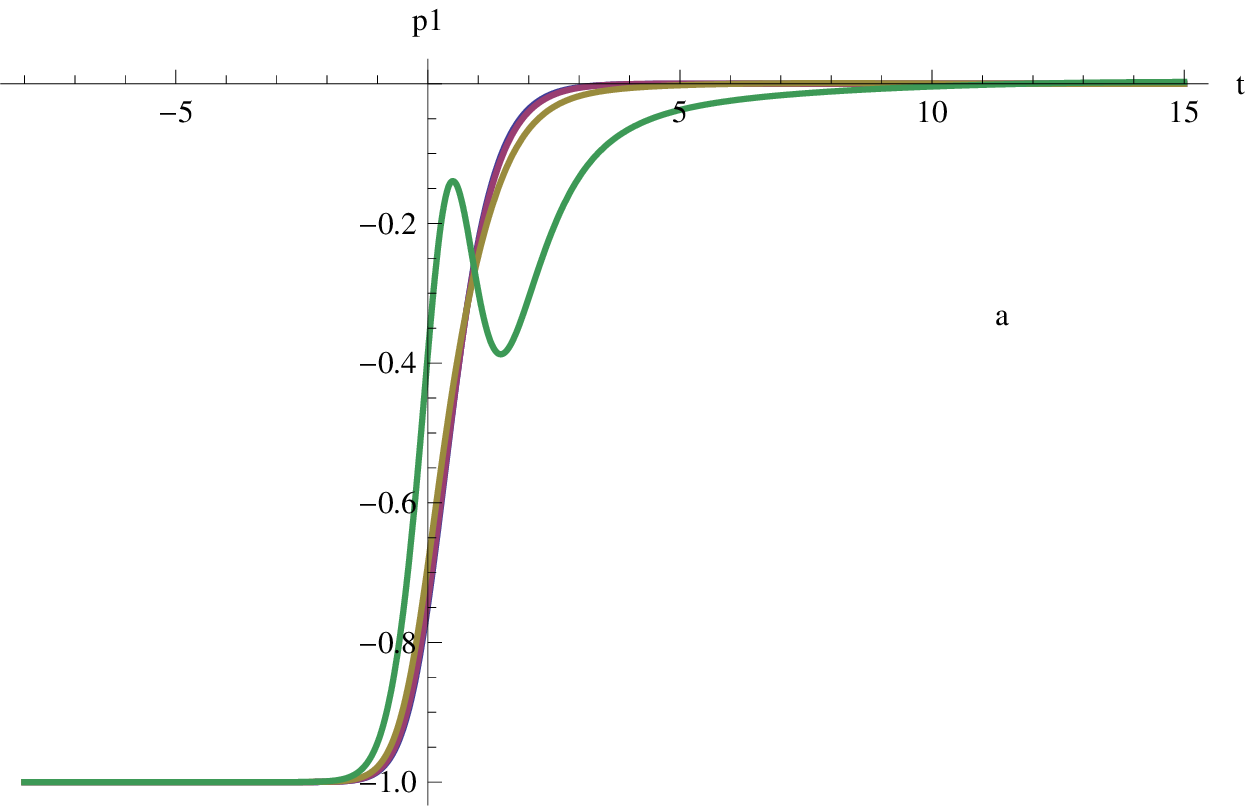}
  \includegraphics[width=2in]{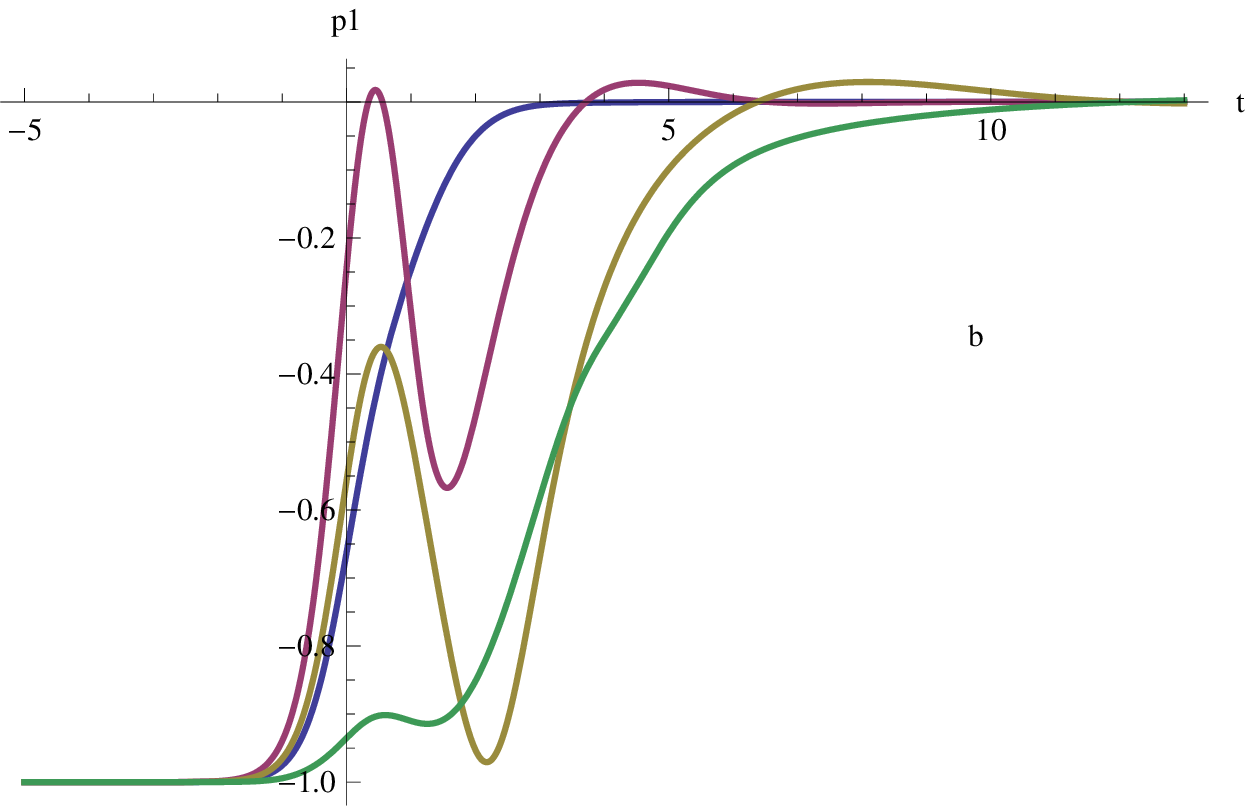}
  \includegraphics[width=2in]{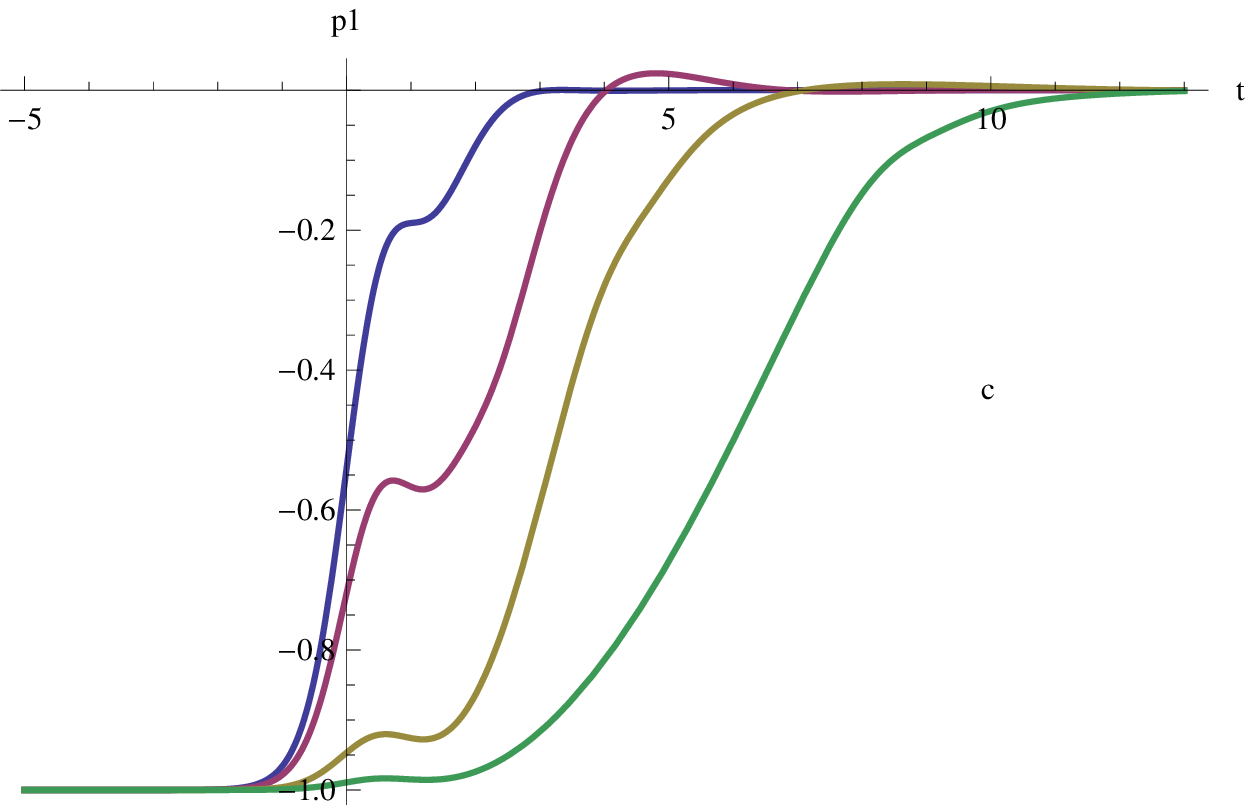}
  \includegraphics[width=2in]{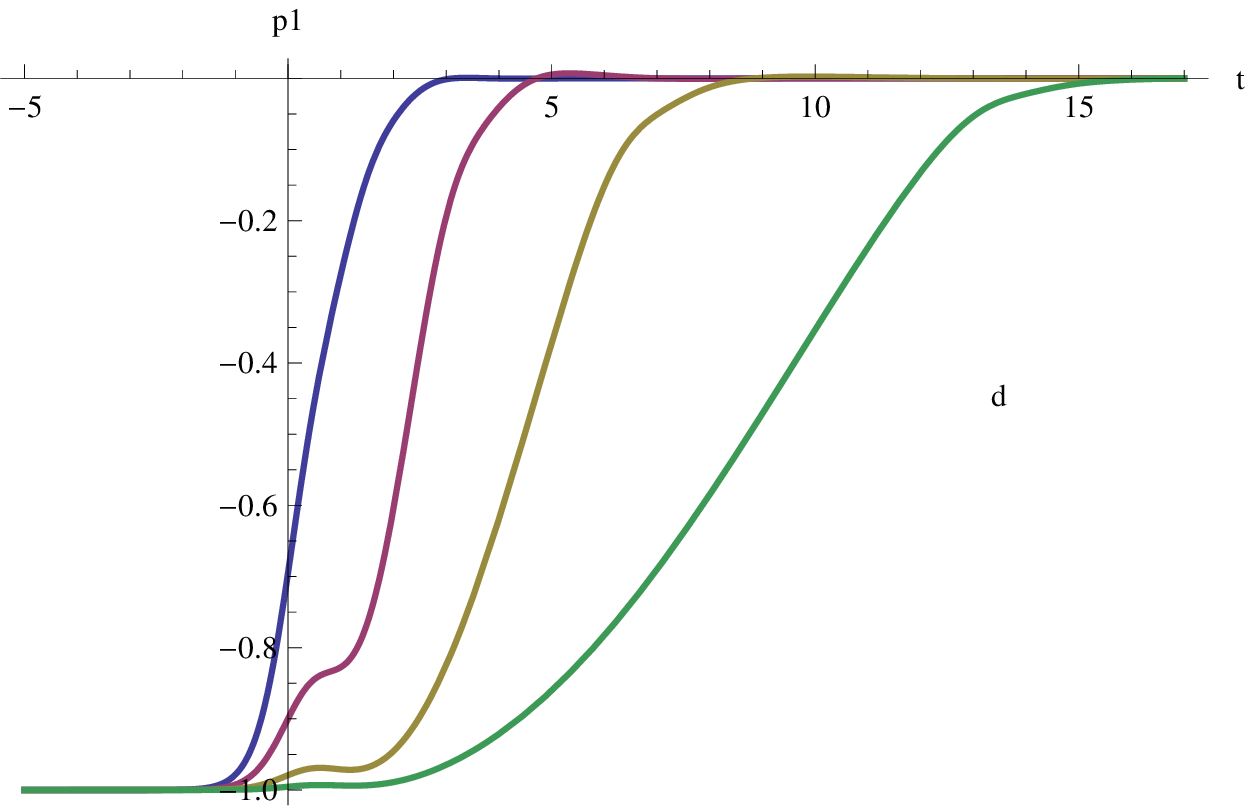}
  \includegraphics[width=2in]{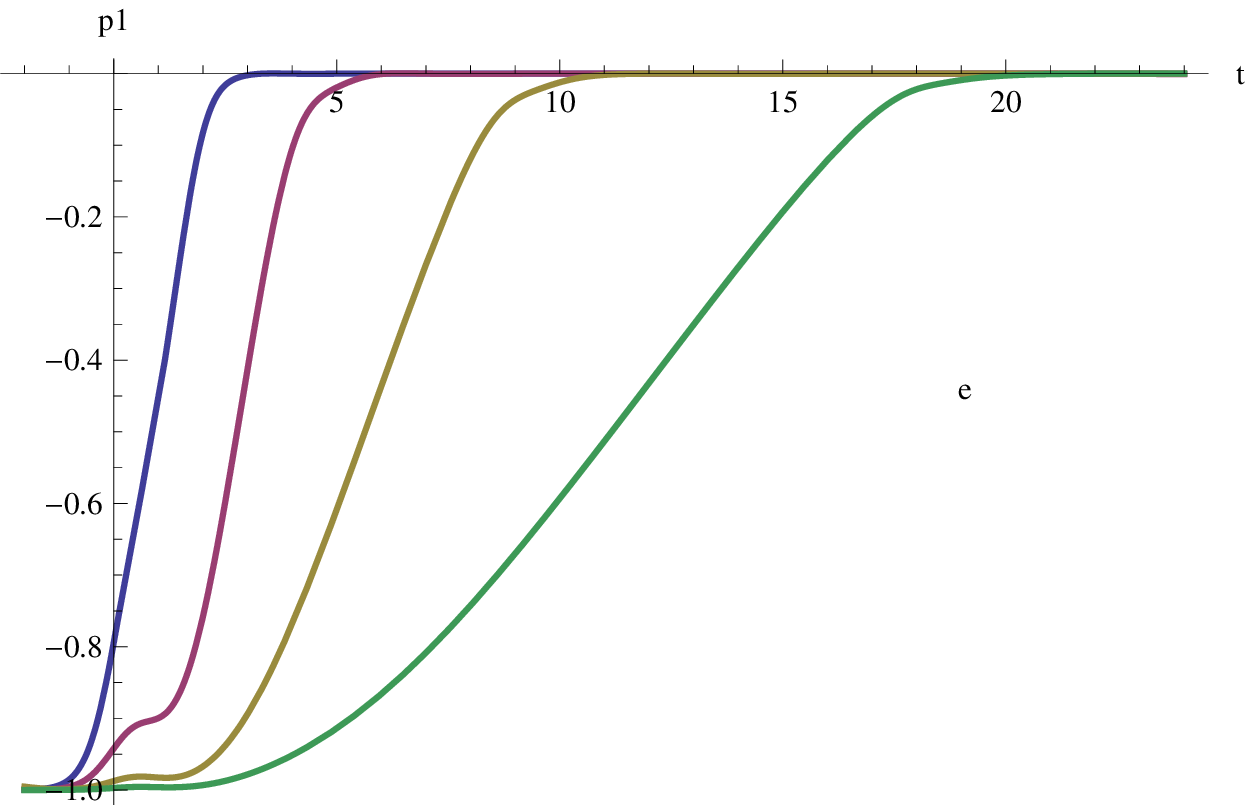}
\end{center}
  \caption{(Colour online) The thermalization measure of the perturbation of the entanglement entropy (as defined in \eqref{fth}) for different-sized
entangling regions.  The evolution is a function of the rescaled boundary time $\t_\ast$.
The plots are, from left to right, top to bottom, for $\r_m=0.1\r_h$, $0.5\r_h$, $0.9\r_h$,
$0.99\r_h$ and $0.999\r_h$.  In each plot the thermalization measure
is shown for quenching parameters $\a=1$ (blue), $\a=\frac 12$ (purple), $\a=\frac 14$ (brown) and $\a=\frac 18$
(green).
Note that the smaller $\a$ is, the longer thermalization takes, in this rescaled boundary time.}  \label{entpfig}
\end{figure}

\begin{figure}
\begin{center}
\psfrag{t}[Br][tl]{{$\scriptstyle{\a\t_\ast}$}}
\psfrag{p1}[c]{{$\scriptstyle{S_{\Sigma(2)(th)}(\t_\ast)}$}}
\psfrag{a}[cc][][1.2]{{$\scriptstyle{\r_m=0.1\r_h}$}}
\psfrag{b}[cc][][1.2]{{$\scriptstyle{\r_m=0.5\r_h}$}}
\psfrag{c}[cc][][1.2]{{$\scriptstyle{\r_m=0.9\r_h}$}}
\psfrag{d}[cc][][1.2]{{$\scriptstyle{\r_m=0.99\r_h}$}}
\psfrag{e}[cc][][1.2]{{$\scriptstyle{\r_m=0.999\r_h}$}}
  \includegraphics[width=2in]{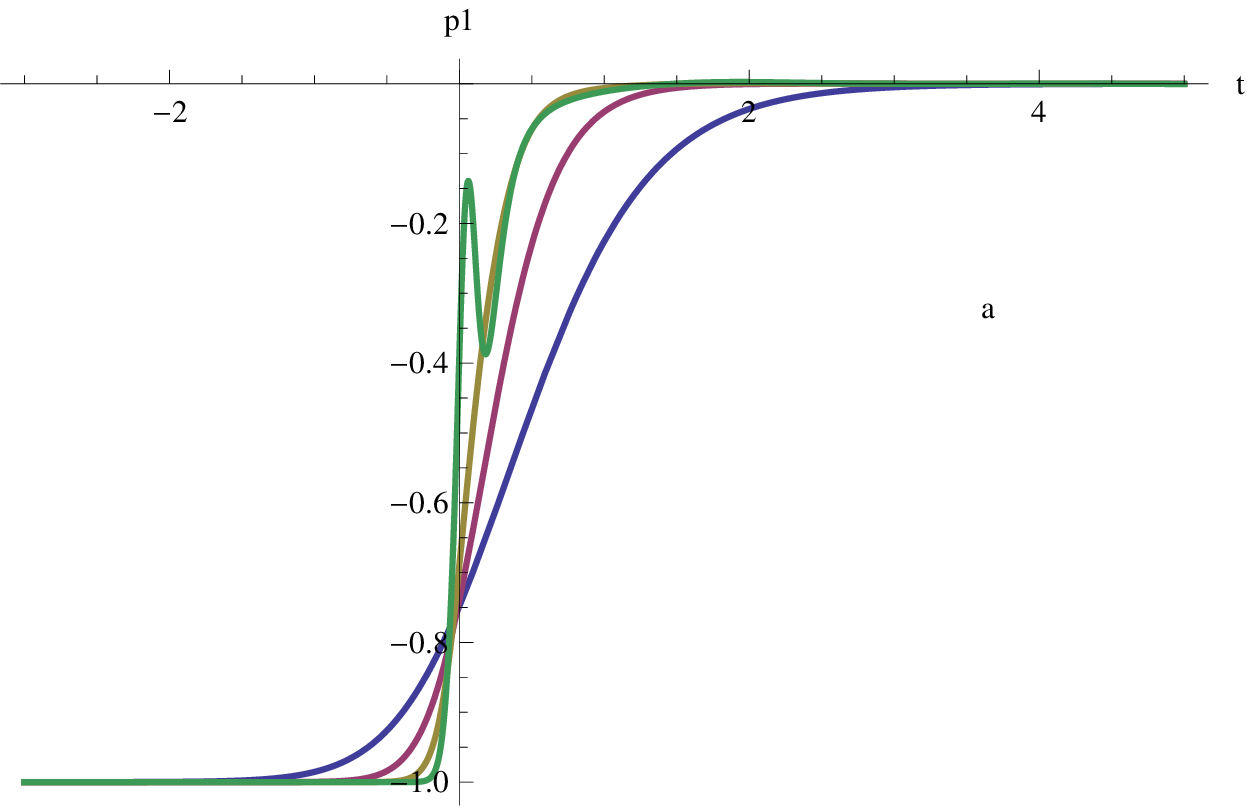}
  \includegraphics[width=2in]{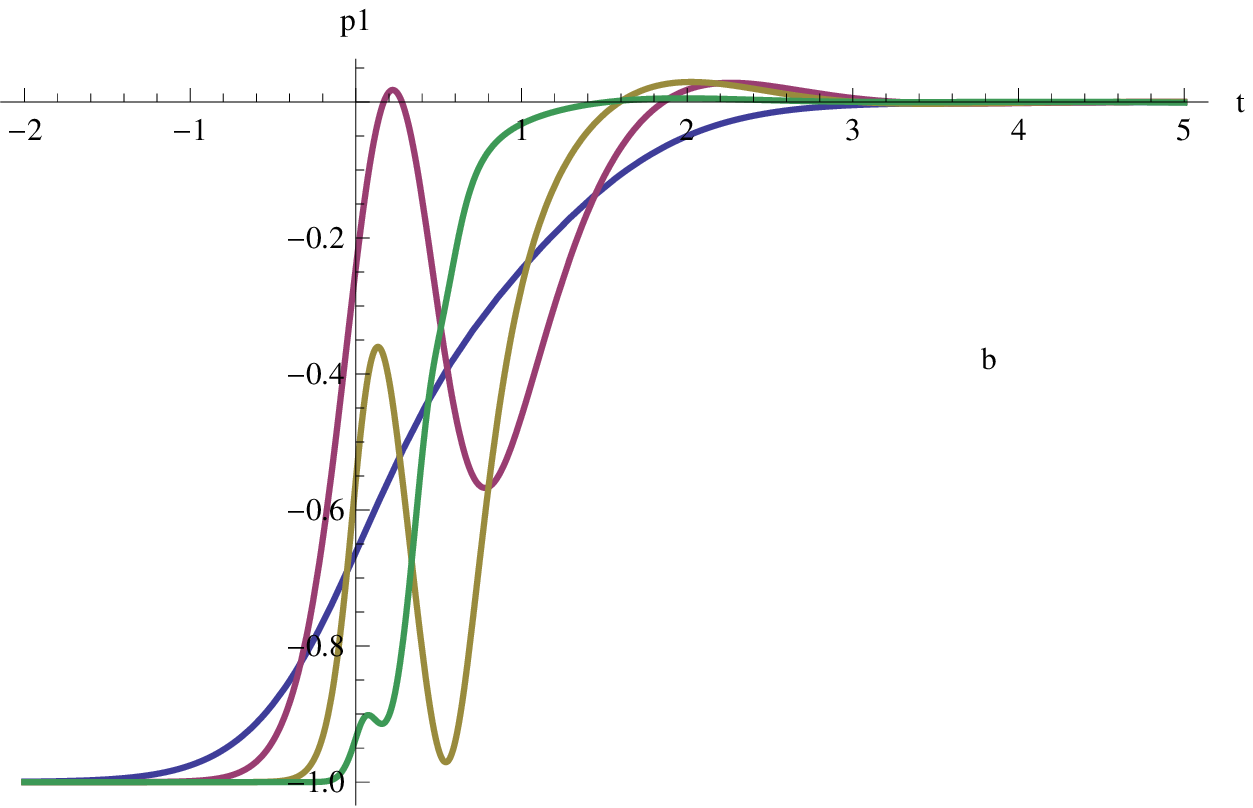}
  \includegraphics[width=2in]{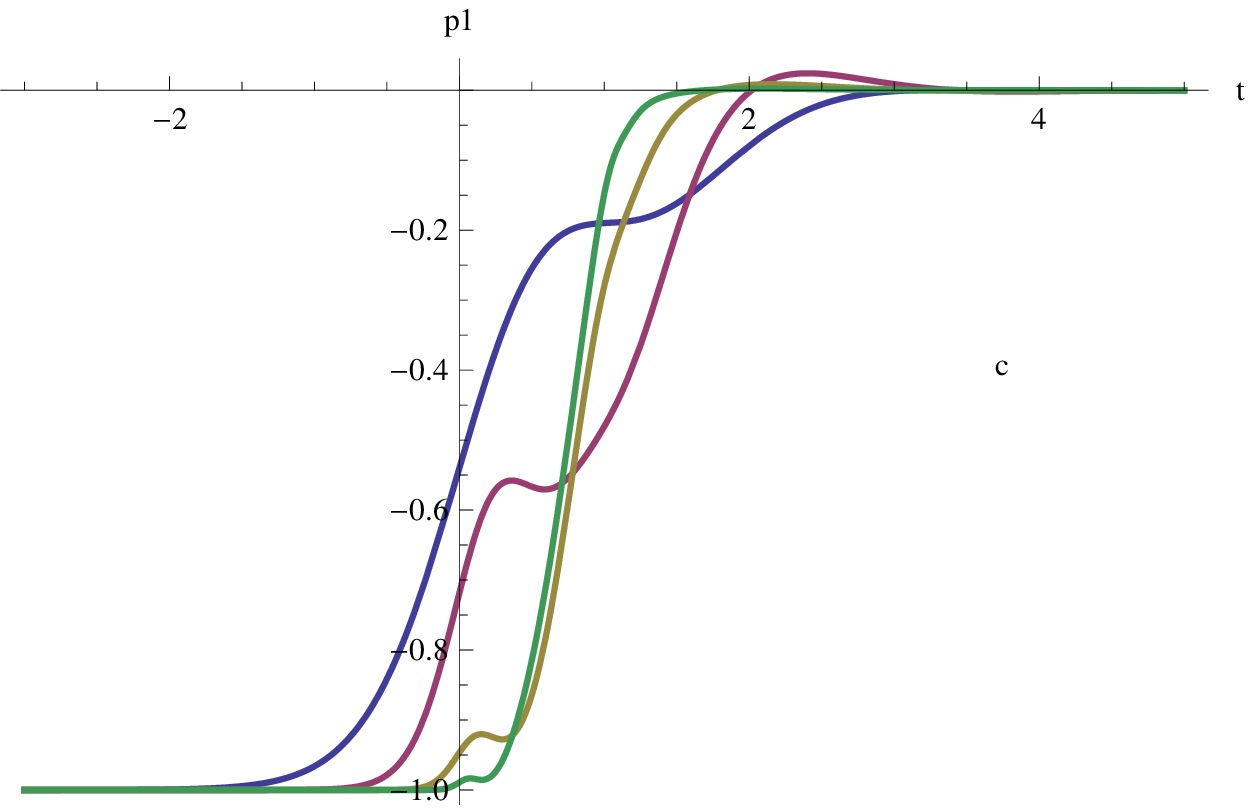}
  \includegraphics[width=2in]{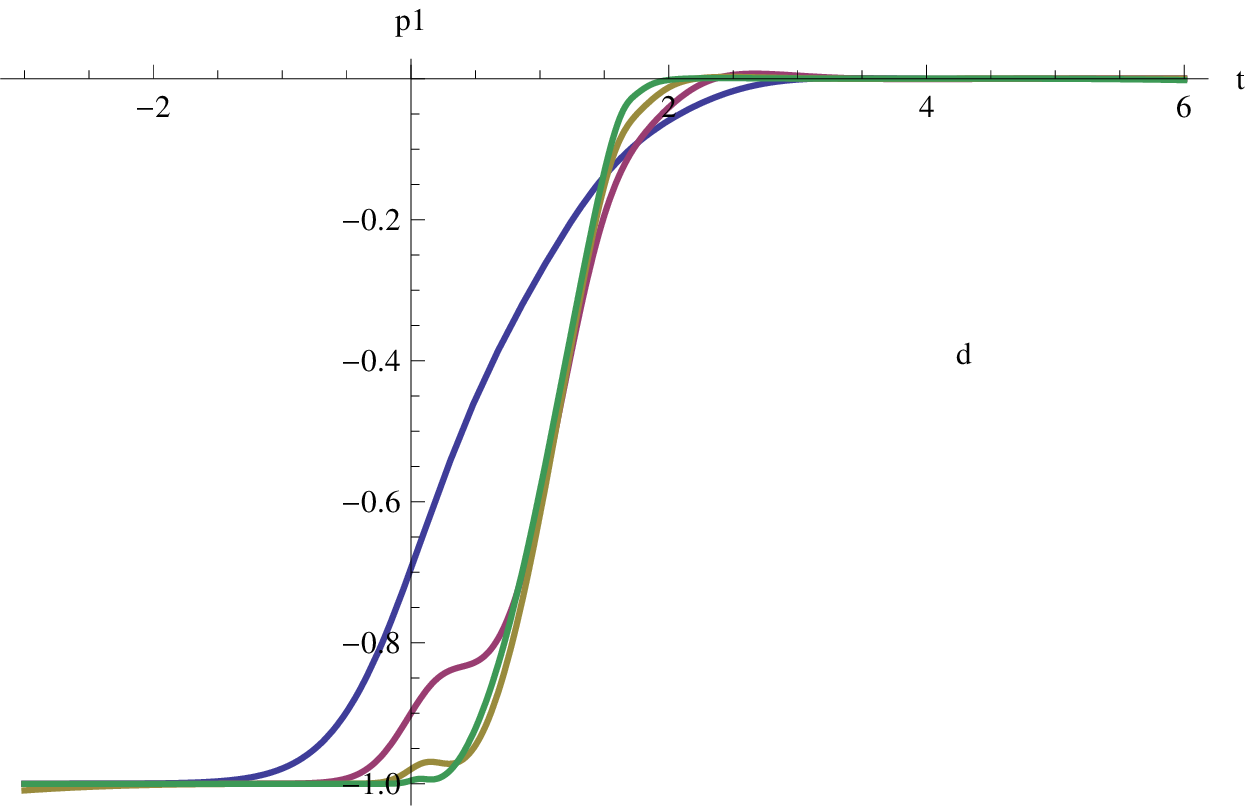}
  \includegraphics[width=2in]{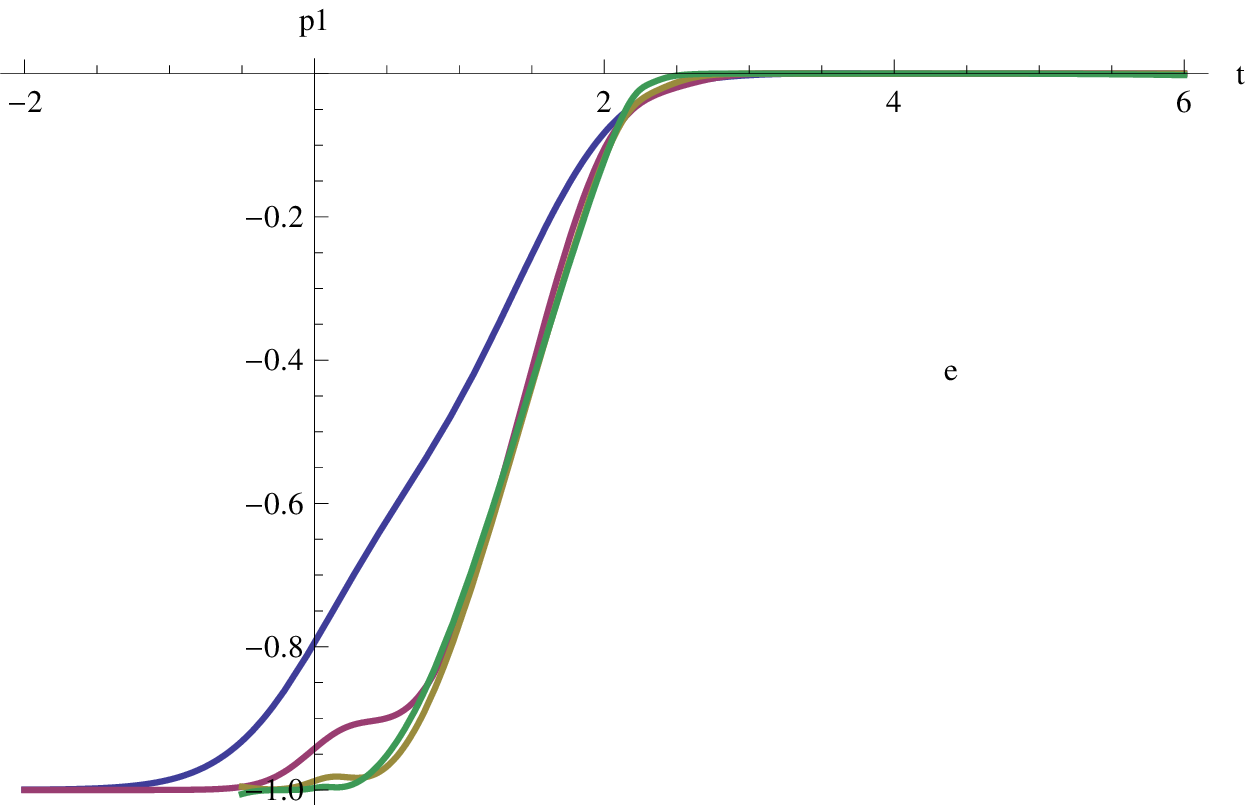}
\end{center}
  \caption{(Colour online) An alternative view of figure \ref{entpfig}.  The same plots are shown, but with the
thermalization measures being functions of the un-rescaled boundary time $\a\t_\ast$.  In this case one can see that
the smaller $\a$ is, the shorter thermalization of the entropy tends to take, from an absolute point of view.}  \label{entpafig}
\end{figure}

\begin{figure}
\begin{center}
\psfrag{aa}[Br][tl]{{$\scriptstyle{\frac{1}{\a}}$}}
\psfrag{t}[c]{{$\scriptstyle{\t_{(th)}}$}}
\psfrag{at}[c]{{$\scriptstyle{\a\t_{(th)}}$}}
    \includegraphics[width=3in]{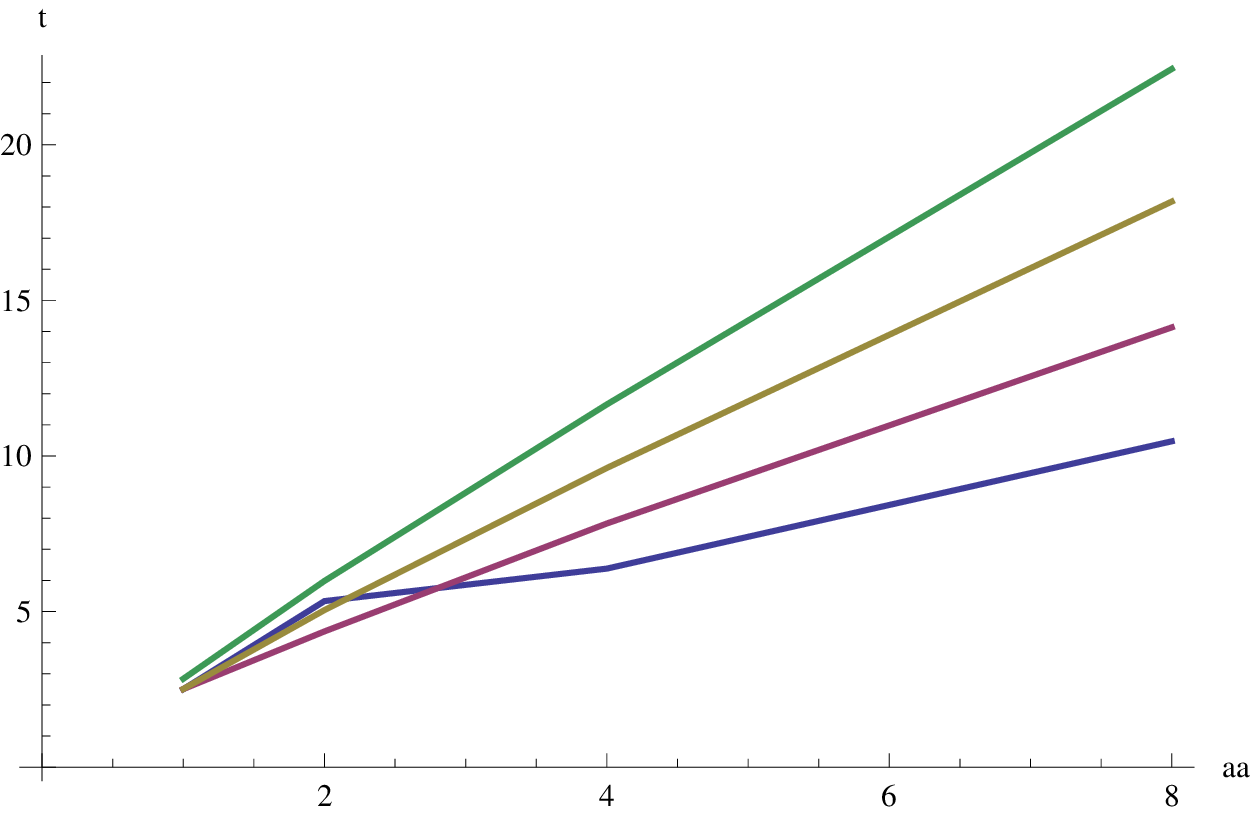}
        \includegraphics[width=3in]{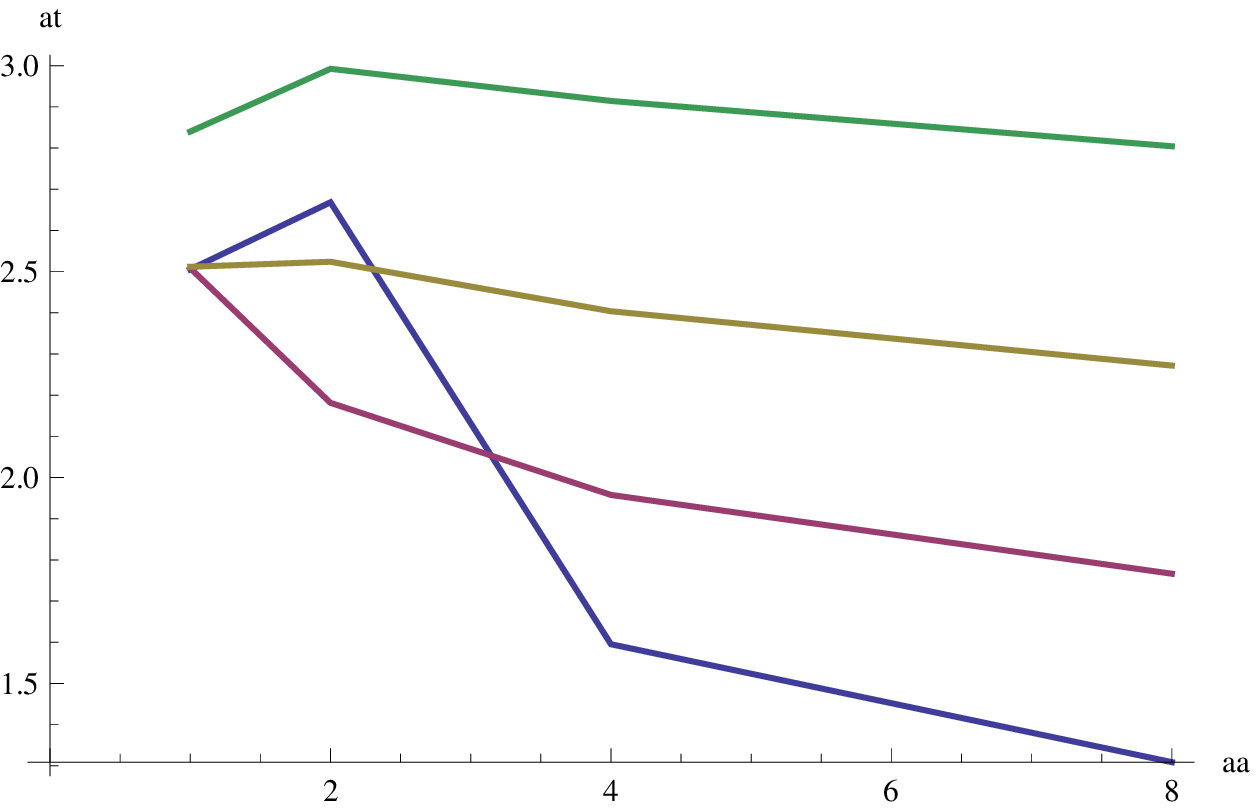}
\end{center}
  \caption{(Colour online) We show the thermalization times of $\mathcal{S}_{\Sigma(2)}$ for various values of $\r_m$ as a function of the
  inverse of the quenching parameter $\a$, for $\a=1$, $\frac{1}{2}$, $\frac{1}{4}$ and $\frac{1}{8}$. On the left we show the rescaled thermalization time $\t_{(th)}$, while on the right we show the same plot, but for the un-rescaled thermalization time $\a\t_{(th)}$.
  The blue, purple, yellow and green curves correspond to $\r_m=0.9\r_h$, $0.99\r_h$,
  $0.999\r_h$ and $0.9999\r_h$ respectively.  Notice how the trends change sign from the left to the right plots.}  \label{eealphatherm}
\end{figure}

\begin{figure}
\begin{center}
\psfrag{t}[Br][tl]{{$\scriptstyle{\t_\ast}$}}
\psfrag{v}[c]{{$\scriptstyle{S_{\Sigma(2)(th)}(\t_\ast)}$}}
\psfrag{a}[cc][][1.5]{{$\scriptstyle{\a=1}$}}
\psfrag{b}[cc][][1.5]{{$\scriptstyle{\a=\frac{1}{2}}$}}
\psfrag{c}[cc][][1.5]{{$\scriptstyle{\a=\frac{1}{4}}$}}
\psfrag{d}[cc][][1.5]{{$\scriptstyle{\a=\frac{1}{8}}$}}
  \includegraphics[width=3in]{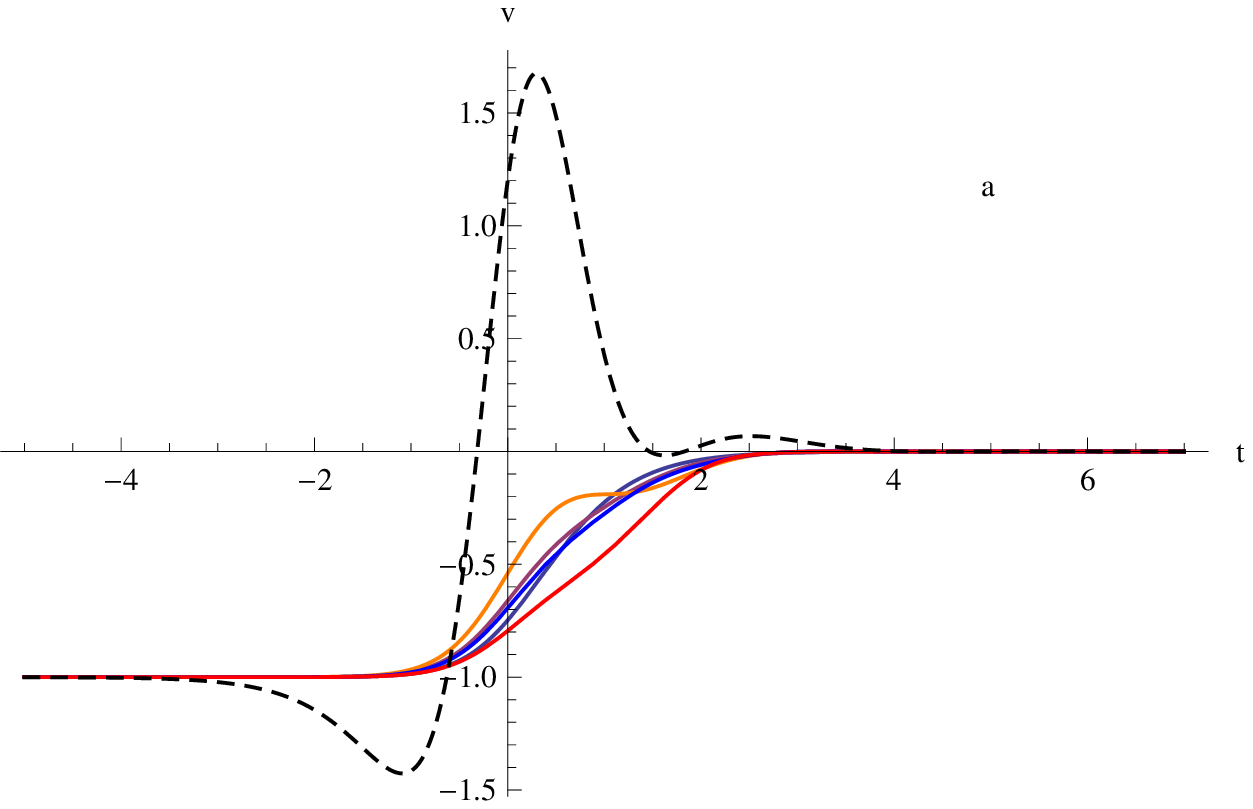}
  \includegraphics[width=3in]{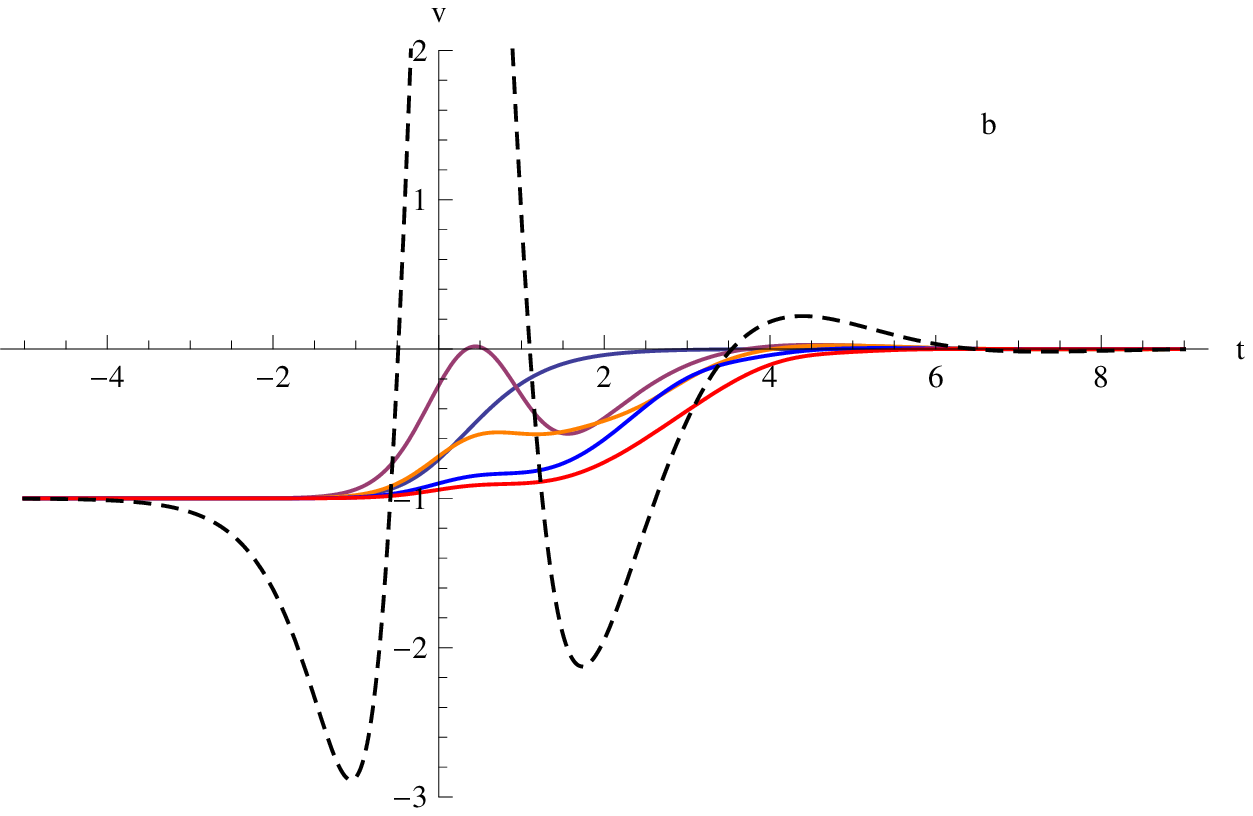}
    \includegraphics[width=3in]{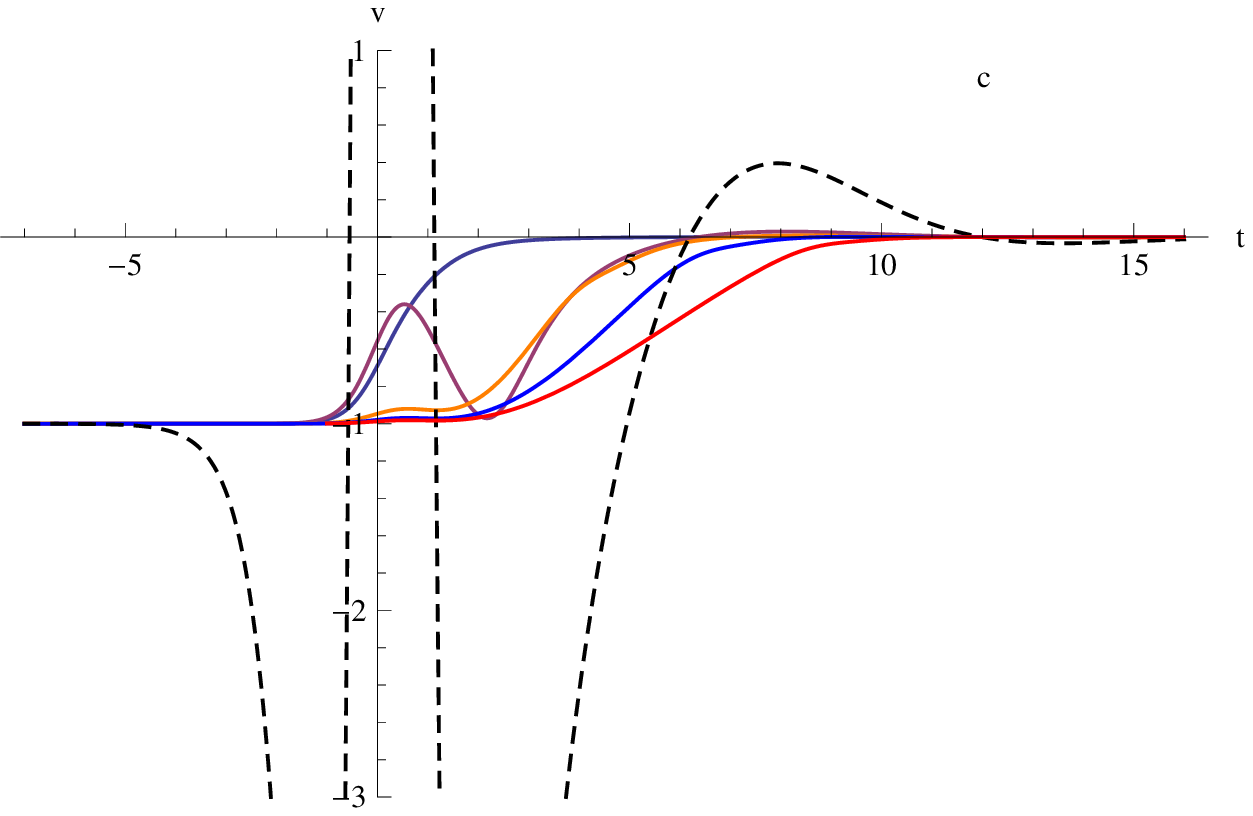}
        \includegraphics[width=3in]{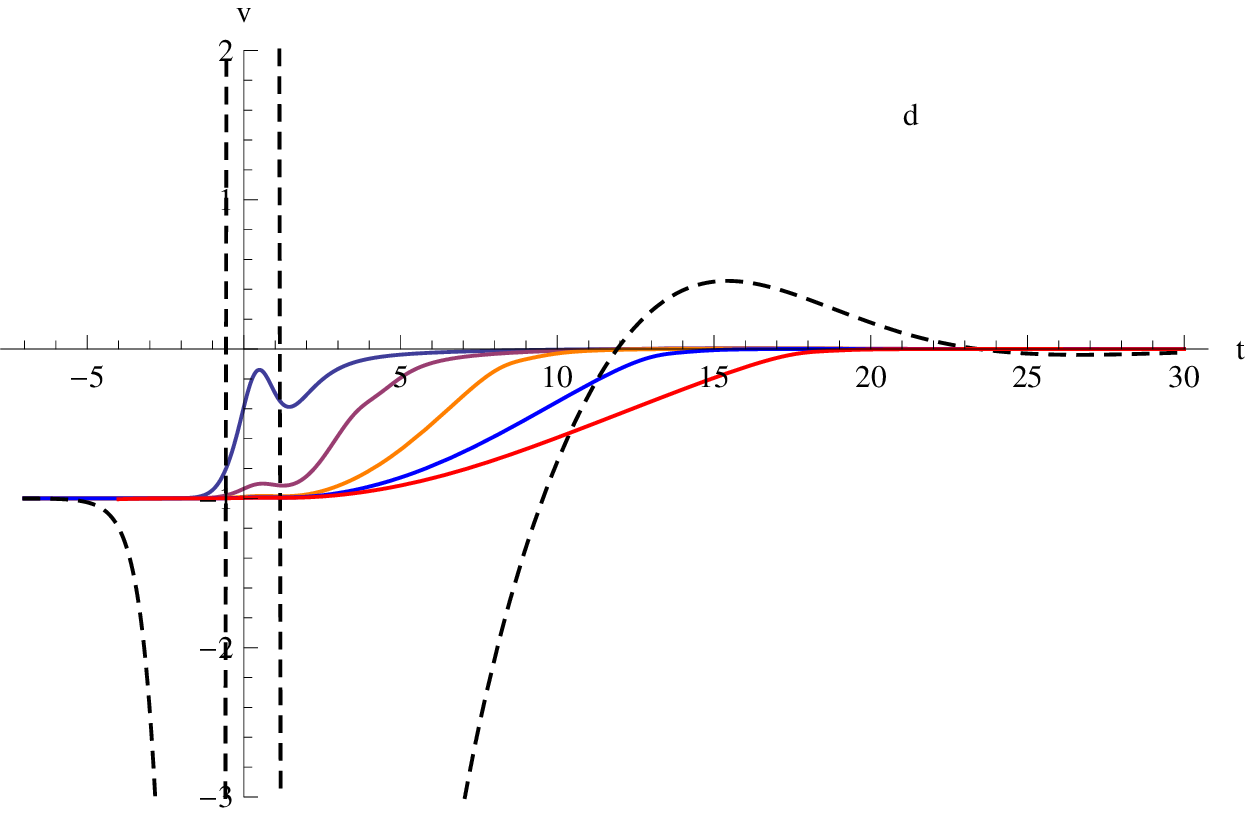}
\end{center}
  \caption{(Colour online) The evolution of $S_{\Sigma(2)(th)}$ as a function of the boundary
time. The plots are (from left
  to right, top to bottom) for $\a=1,\frac{1}{2},\frac{1}{4}$ and $\frac{1}{8}$,
  respectively.  Each figure contains the plot for a minimal surface of height $\r_m=0.1\r_h$, $0.5\r_h$, $0.9\r_h$,
$0.99\r_h$ and $0.999\r_h$, respectively.  The plots for $0.9\r_h$
$0.99\r_h$ and $0.999\r_h$ are orange, bright blue, and red, respectively.  We also plotted $p_{2(th)}$ in dashed
lines for comparison.  We can see that the
larger the entangling surface $\Sigma$ (\ie the depth $\r_m$), the longer the thermalization time of the entanglement entropy is in
each case.}  \label{entafig}
\end{figure}

We plotted the thermalization measure of the regularized EE at different quenching parameters $\a$ for the different sizes of
the entangling surface $\Sigma$ in figures \ref{entpfig} and \ref{entpafig},\footnote{Recall that the thermalization measure
is only calculated for the finite part of the entanglement entropy in \eqref{eentreg}.} as we did for the two-point
functions in section \ref{numlin}.  We see a similar behaviour as we saw for the two-point functions, namely
that the faster quenches have longer equilibration times as measured by the rescaled boundary time
$\t_\ast$ than the slower quenches for each surface size.  We also see that, as in the case of two-point
functions, the faster quenches have faster equilibration times when we measure the thermalization in
un-rescaled boundary time $\a\t_\ast$.  We also plot these opposite trends in figure \ref{eealphatherm}
as we did in the two-point function case.

We also plotted $S_{\Sigma(2)(th)}$ for each quenching parameter $\a$ separately in figure \ref{entafig},
but for the different sizes
of the entangling surface (measured by the depth that the minimal surface $\gamma$ extends into the bulk).
We again obtain similar results as for the two-point functions, namely that the EE of the larger entangling
regions equilibrates slower, but that the thermalization time for fixed $\r_m$ decreases at a slower
rate than $\a$.  Since the thermalization times
of the EE approaches that for the one-point function for
larger $\a$ (e.g. $\a=\frac{1}{2}$ comes close for the widest surface considered), we believe that it may be possible to obtain larger thermalization times for arbitrarily
small $\alpha$, if we let the entangling surface $\Sigma$ be large enough.

\subsection{Scaling of the thermalized correlator and entropy}

After reaching equilibrium, we expect our Yang-Mills plasma to satisfy
equilibrium thermodynamics.  At the level of the thermal entropy, we
know that \cite{blmv}
\begin{equation} \eqlabel{sf}
S_f \sim T_f^3 + T_f^3\left(\frac{\lambda}{T_f}\right)^2,
\end{equation}
meaning that up to constant prefactors, the above relation gives the
equilibrium behaviour for the system.  Here $T_f$ is the final
temperature of the system, and $\lambda$ is the field theory
coupling of the quenching operator.  It should also be noted that
$\lambda/T_f\propto\ell$ relates the small parameter in the AdS
picture to the coupling $\lambda$ in the field theory picture.

For wide entangling regions, as the ones we considered in the
previous subsection,the minimal surface $\gamma$ will become
wide, with the largest contribution coming from the part deep in
the bulk.  As the surface becomes wide in the transverse
$y$-direction, more of it will lie close to, and parallel with
the horizon of the AdS black brane. Most of its area will
come from a surface that almost coincides with a part of the
horizon of roughly the same width.  In the dual picture,
since the entanglement entropy is proportional to the area
of $\gamma$, an entangling region will have
the largest contribution of its EE be proportional to the
thermal entropy.  In the limit of infinitely wide surfaces,
the EE is simply equal to the thermal entropy \cite{nishioka}.

Equation \eqref{sf} now implies that both the zeroth-order and
and second-order  EE (expressed as $\calo(\ell^2)$) should be proportional to $T_f^3$.  In this
paper we have thus far kept the dependence on the temperature
hidden, by setting the black hole radius $\mu$ (in un-rescaled
$r$-coordinates) to $1$.  It happens that the temperature
is proportional to $\mu$, so we should reintroduce $\mu$, as
well as the AdS radius $L$, to see what behaviour to expect
from our EE.

It is easy to see that by reintroducing $L$ into our equations,
that $S_\Sigma\propto L^3$, which already has the correct units
for the area of $\gamma$.  Therefore we should introduce a
factor of $\mu$ for each other factor with units of length.
Since the profile of $\r$ in $y$ becomes proportional to $y_m$
for wide surfaces ($\gamma$ becomes a slab shape), we expect
$S_{\Sigma(0)}$ to scale as $L^2\,y_m$ (as we verified).  Therefore
we need a factor of $\mu^3$ to give the entropy the correct units.

\begin{figure}
\begin{center}
\psfrag{w}{{$y_m$}}
\psfrag{s0}{{$S_{\Sigma(0)}$}}
\psfrag{s}{{$S_{\Sigma(2)}$}}
  \includegraphics[width=3in]{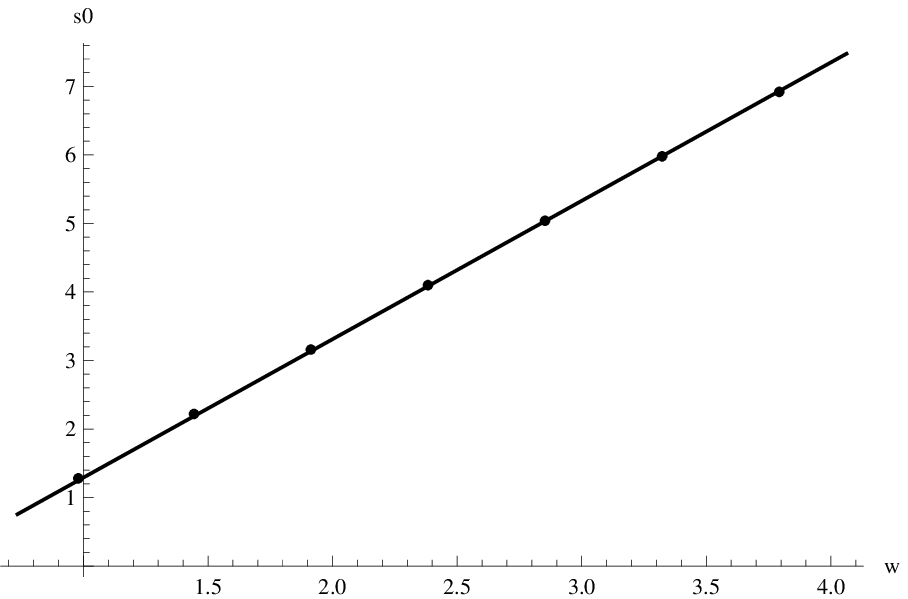}
    \includegraphics[width=3in]{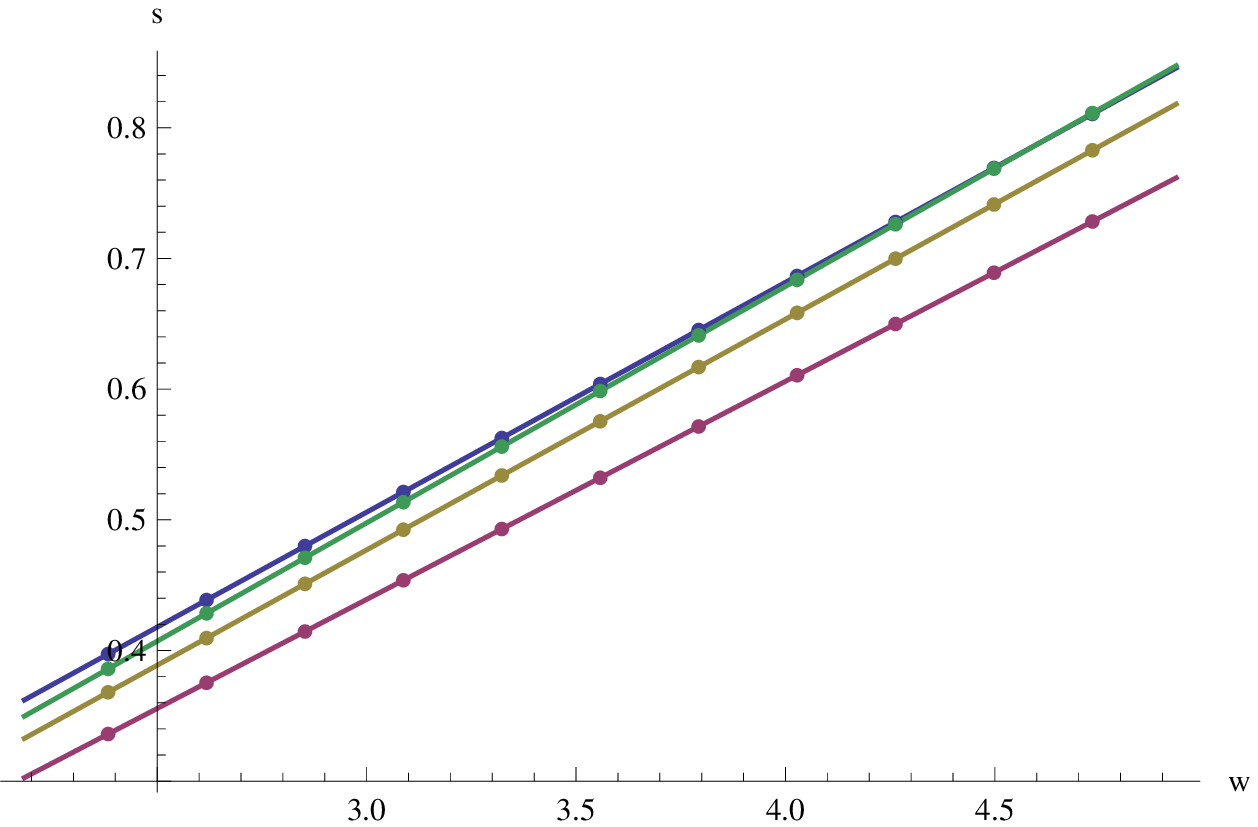}
\end{center}
  \caption{(Colour online)
Here we show the entanglement entropy as a function of the width of the entangling surface.
On the left, we show the unperturbed EE for various values of the width $y_m$.  We also show the
best-fit line $2.02 y_m - 0.73$ through the data.  On the right, we plot the perturbation of the
EE for different quenching rates, namely $\a=1$, $\frac 12$, $\frac 14$ and $\frac 18$ corresponding to the coloured
plots blue, purple, yellow and green, respectively.  We show the perturbations of the entropy for
different values of $y_m$, as well as the best-fit straight lines through the data.  The data in each
case is clearly well approximated by straight lines.} \label{eescale}
\end{figure}

The second-order EE, $S_{\Sigma(2)}$ also has a factor of $L^2$.
We should expect it to also scale linearly with $y_m$ in order to
balance the factor of $\mu^3$, as required by the arguments
above.  In figure \ref{eescale} we show that both $S_{\Sigma(0)}$,
and $S_{\Sigma(2)}$ scale linearly with $y_m$, as we would expect
it to.  Moreover, $S_{\Sigma(0)}$ has the correct slope of $2$
which we would expect because of $y_2$ being exactly half of the
width, and therefore being proportional to $\frac 12$ of the area of
the minimal surface.

We therefore see that the EE scales as $\mu^3\propto T_f^3$,
as predicted from equation \eqref{sf}.  Note that the additional
scaling of $T_f^{-2}$ in the EE at perturbative order is contained in
the perturbation parameter $\ell^2$.

\begin{figure}
\begin{center}
\psfrag{w}{{$y_m$}}
\psfrag{lin0}{{$\mathcal{L}_{\Sigma(0)}$}}
\psfrag{lin}{{$\mathcal{L}_{\Sigma(2)}$}}
  \includegraphics[width=3in]{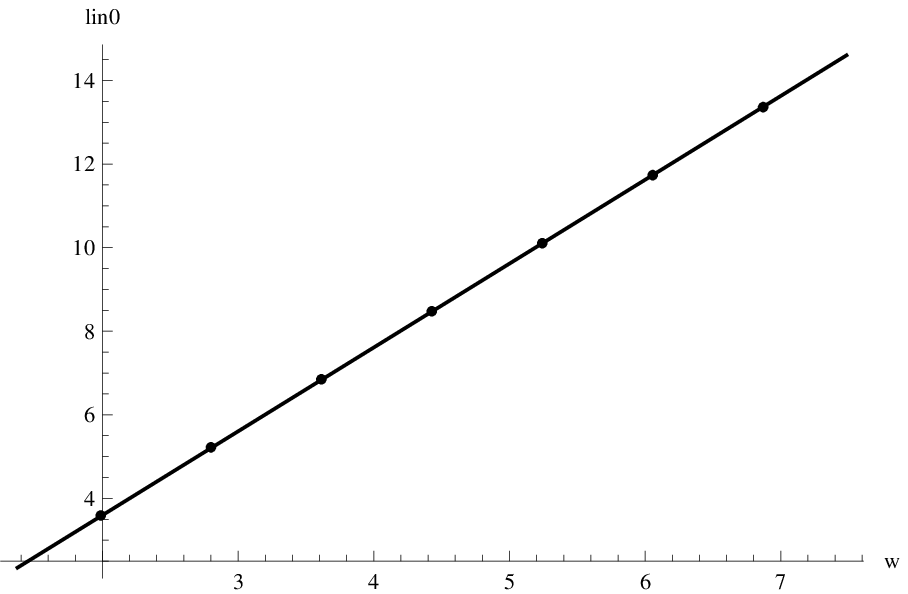}
    \includegraphics[width=3in]{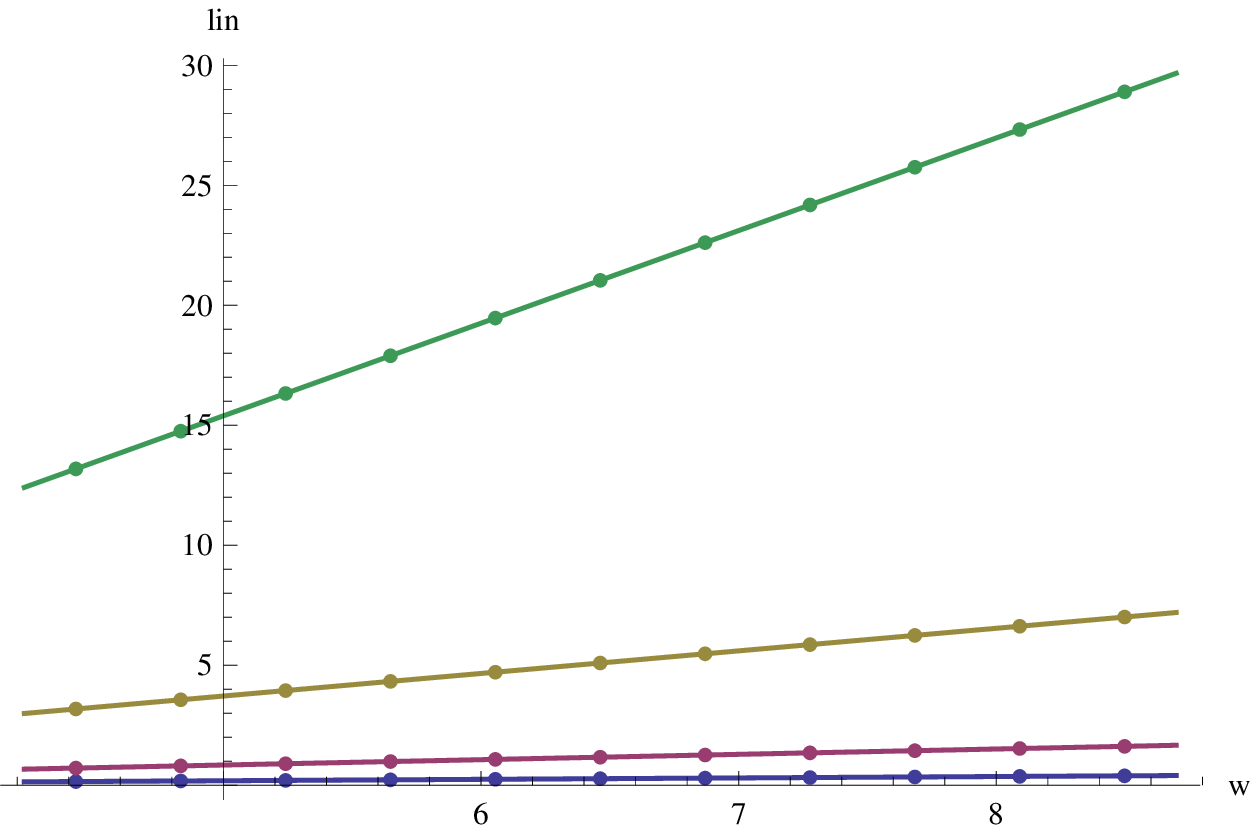}
\end{center}
  \caption{(Colour online)
Similar to figure \ref{eescale}, we show the two-point correlator as a function of the
separation of the points.  We see that the unperturbed correlator scales linearly with
the separation and with the correct slope of $\sim2$. The best fit line here is $2.01 y_m - 0.41$.
The perturbations of the correlator are also shown for various quenching rates $\a$ (same
colour scheme as figure~\ref{eescale}), and the data are clearly well approximated by straight lines.}
\label{linscale}
\end{figure}

By similar horizon arguments we can predict that the two-point
correlator should also scale linearly with $y_m$ for wide
separations.  In figure \ref{linscale} we see that both
$\mathcal{L}_0$ and $\mathcal{L}_2$ scale linearly with $y_m$
for wide separations, and moreover that $\mathcal{L}_0$ scales
with the correct slope of $2$.

\section{Thermalization}

We have so far discussed the different probes of the thermalization of the system.  In this section
we explore the mechanisms behind the thermalization behaviour seen in the two-point correlator and
entanglement entropy.

We first discuss the thermalization times for the different probes introduced, before
going on to examine how the different scales of the problem contribute to the observed thermalization.
The correlator and entropy are integrals
over the radius of the AdS spacetime, and different parts of the profile make different contributions.
We compare these contributions with the thermalization times of the integrands at fixed radii.
We then go on to see how the profile of the scalar field and different components of its
stress-energy tensor equilibrate.  We end this section by bringing all these observations together, and
speculate about the cause of thermalization at the different scales.

\subsection{Thermalization times of the entanglement entropy and two-point correlator} \label{thermtrend}

We can ask how long the two-point function and the entanglement entropy take to thermalize for different
separations of the points, or widths of the strip, respectively.  Here we show the plots of the
thermalization times of the EE and correlator as a function of the width of the surface and separation
of the points, respectively.  The thermalization time is determined by applying equation \eqref{fth} to
the EE and correlators, and choosing a thermalization threshold of $2\%$ of its final equilibrium value.

\begin{figure}
\begin{center}
\psfrag{width}[Br][tl]{{$\scriptstyle{y_m}$}}
\psfrag{t}[c]{{$\scriptstyle{\t_{(th)}}$}}
  \includegraphics[width=3in]{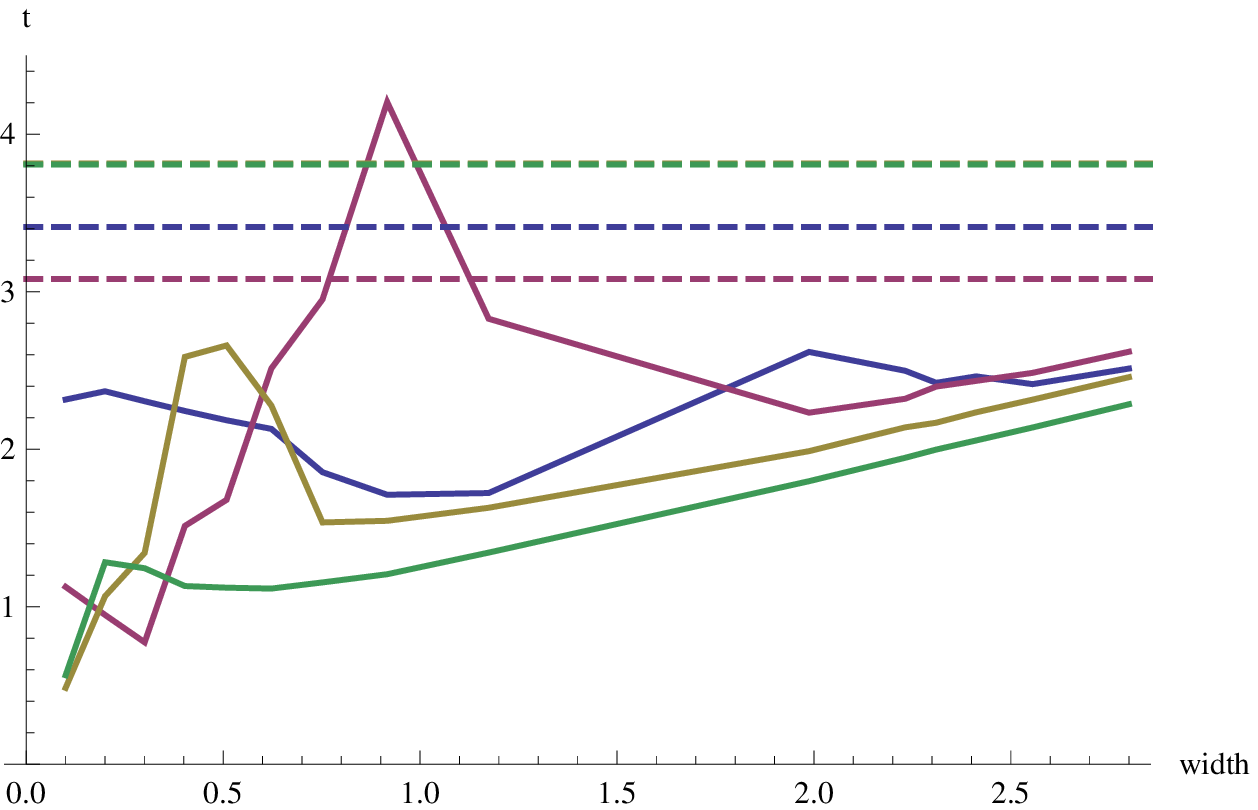}
  \includegraphics[width=3in]{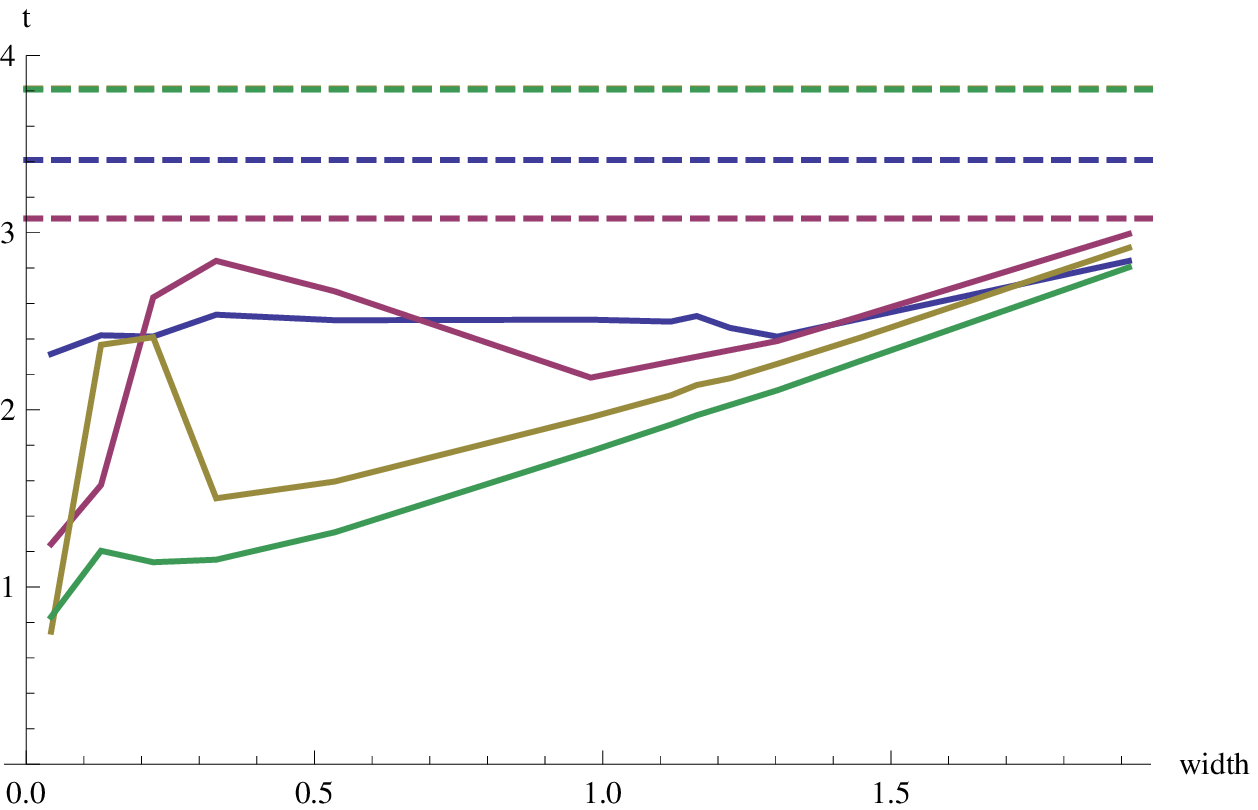}
\end{center}
\caption{(Colour online) The thermalization times for the two-point correlator on the left
and entanglement entropy on the right, respectively, as a function of the half-width
of the correlator and entanglement regions, respectively.
The blue, purple, yellow and green curves are for $\a=1$, $\frac 12$, $\frac 14$,
and $\frac 18$, respectively.  The thermalization times in un-rescaled boundary
time $\t_\ast$ are functions of the un-rescaled separations and widths, respectively.  We also show the
thermalization times of the one-point correlator $\langle\mathcal{O}_{3}\rangle$ for the various
values of $\a$ as the horizontal dashed lines with the same colour scheme.  Note that the thermalization
time for the one-point function is nearly the same for $\a=\frac{1}{4}$ and $\a=\frac{1}{8}$.
}
\label{lineeth}
\end{figure}

We plotted the thermalization times of both the correlator and EE for various values of $\a$
in figure \ref{lineeth}.  As one can see for narrow surfaces, the increase in the
thermalization time is not monotonic.  This occurs due to the fluctuations that occur in
the quasinormal modes, which are large compared to the size of the EE and correlator for
small widths. For wider surfaces we see a linear growth of thermalization time with the width of the
surface.  Although these thermalization times observed here
are smaller than that for the normalizable mode \ie the one-point function which are also shown in
figure \ref{lineeth} (at least for faster quenches),
its monotonic nature, and its linear nature, indicates that for wide
enough separations and widths, the two-point
correlator and EE should have longer thermalization times than  the normalizable mode.

\subsection{Equilibration of the correlator and entropy profiles} \label{thermprofile}

We would like to know how the two-point correlator and the entanglement entropy thermalize.  The
thermalization time of the previous subsection is informative, insofar as it tells us that wider surfaces
have longer equilibration times than narrow surfaces or separations, as well as the limiting behaviour
for wide surfaces.  The regions with wider separations have minimal surfaces or geodesics that
probe deeper into the bulk geometry.  This provides us with a clue as to what may be causing the
observed difference in thermalization time, namely that the part of the surface deeper in the geometry
equilibrates later than parts near the boundary.

In this subsection we will show how the thermalization of the EE and two-point correlators depend on
different parts of the dual minimal surfaces or geodesics at different depths in the AdS-geometry.
First, it turns out that the parts of the integrands of the correlator or
EE integrals corresponding to larger $\r$ in the regularized (\ie finite) version of integral
\eqref{eent2b} make larger contributions to the full integral.  This makes
sense for wide surfaces, since most of the area is near $\r_m$, close to the black brane horizon.
Secondly we show that it is at larger $\r$ that the integrand thermalizes last.

\begin{figure}[p]
\begin{center}
\psfrag{r}[Br][tl]{{$\scriptstyle{\tilde{\r}/\r_m}$}}
\psfrag{frac}[c]{{$\scriptstyle{S_{frac}}$}}
  \includegraphics[width=3in]{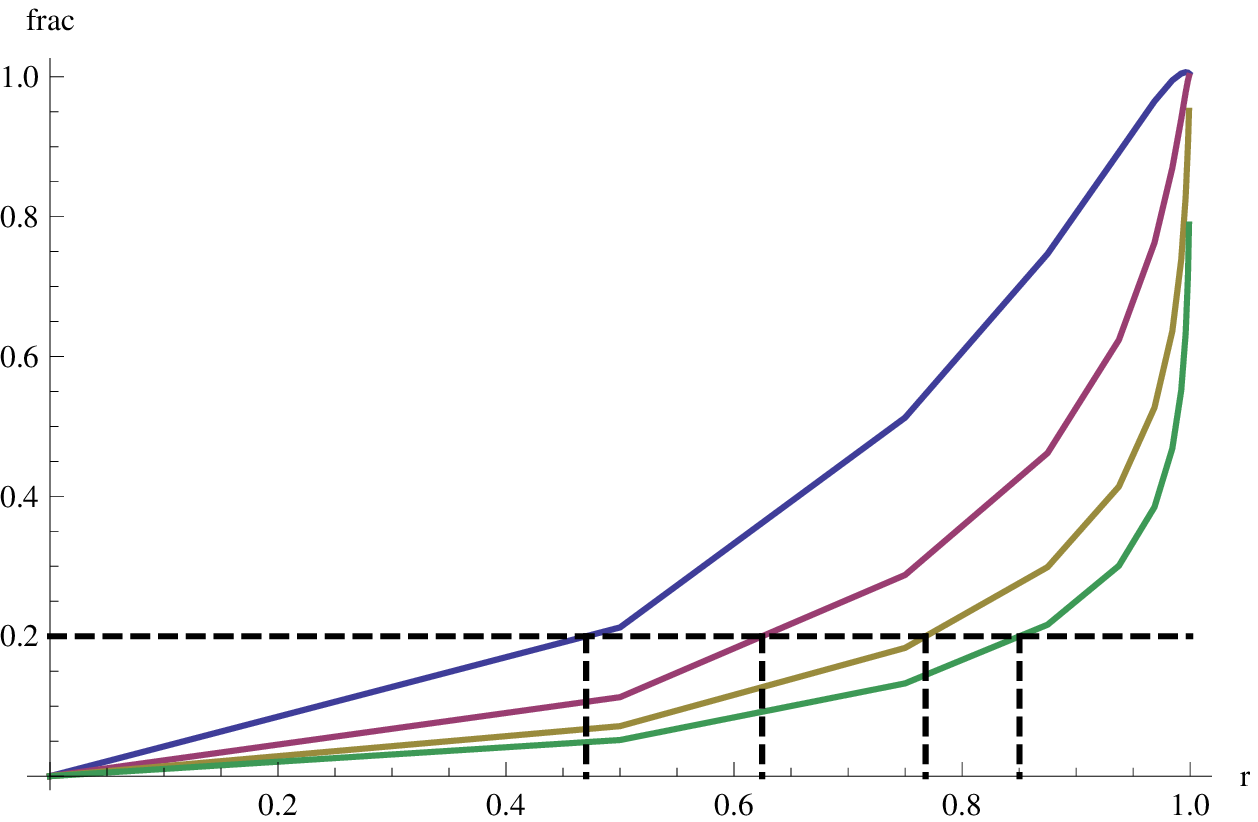}
\end{center}
\caption{(Colour online) The fractional contribution to the total renormalized
entanglement entropy (after thermalization)
that the integral in
\eqref{eent2b} has, when only integrated up to a particular fraction of the full integration region.
These particular
curves are for $\a=\frac 18$, and $\r_m=0.9\r_h$, $0.99\r_h$, $0.999\r_h$ and $0.9999\r_h$, when the curve is
blue, purple, yellow and green, respectively.  The analogous curves for the correlator are very similar, and therefore omitted here.
For wider surfaces, the deepest part of the integrand in the geometry contributes
significantly more to the full value of the EE, than the near-boundary part.  For comparison we show
the horizontal dashed line $S_{frac}=0.2$, and the vertical dashed lines where this line (roughly)
intersects each curve.  Note that when $\r_m=0.9999\r_h$,
integrating up to $\frac{\tilde{\r}}{\r_m}=0.85$ only contributes $20\%$ of the full regularized
entropy. \label{fracplot}}
\begin{center}
\psfrag{r}[Br][tl]{{$\scriptstyle{\r}$}}
\psfrag{t}[c]{{$\scriptstyle{\t_\ast}$}}
  \includegraphics[width=3in]{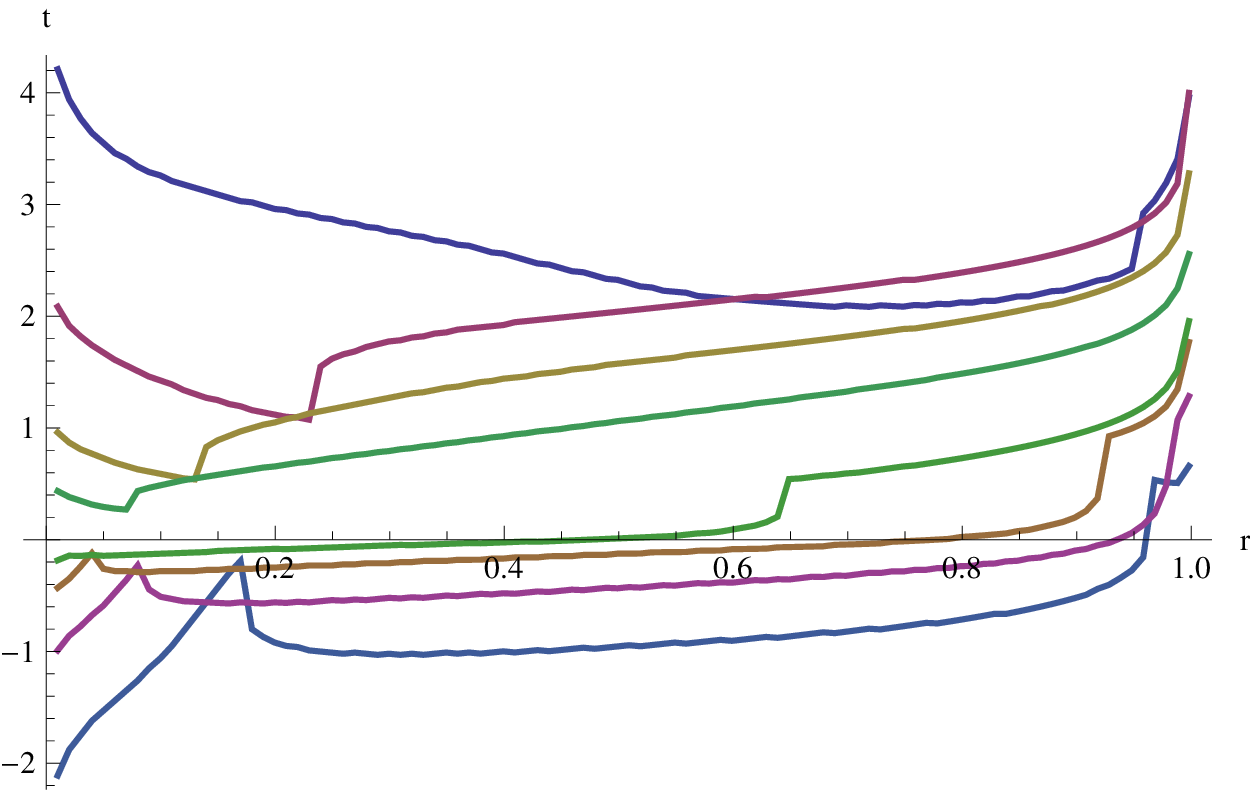}
\end{center}
\caption{(Colour online) Here we show the excitation and equilibration times of the integrand in \eqref{eent2b}
as a function of radius (both in un-rescaled coordinates).  The blue, purple, yellow and green
curves show the excitation (bottom) and equilibration boundary (top) times for $\a=1$, $\frac 12$, $\frac 14$ and $\frac 18$, respectively. \label{rtherm}}
\end{figure}

In figure \ref{fracplot}, we show the fractional contribution to the regularized entanglement entropy
\eqref{eent2b}
\begin{equation}
S_{frac} = \frac{\int^{\tilde{\r}}_{0}\left[\textrm{regularized integrand}\right]d\r}{S_{\Sigma(2)(finite)}},
\end{equation}
when integrating
up to a particular fraction of the full range of the integral.  Notice how especially for the wider
surfaces, most of the contribution comes from the deepest part of the integration interval.  As an
example, in the figure we show the line at which point the integral reaches $20\%$ of its final value.
As $\r_m$ increases, so does $\frac{\tilde{\r}}{\r_m}$ at which the $S_{frac}=20\%$ fraction is achieved. The roughly interpolated values of $\frac{\tilde{\r}}{\r_m}$ when this occurs are:
\begin{equation}
\{(\frac{{\r_m}}{\r_h},\frac{\tilde{\r}}{\r_m})\}\approx
\{(0.9,\,0.47),\, (0.99,\,0.62),\, (0.999,\,0.77),\, (0.9999,\,0.85)\}.
\end{equation}
That means that for $\r_m=0.9999\r_h$, approximately the last $15\%$ of the integration interval
contributes $80\%$ of the total regularized entropy.

Next, in figure \ref{rtherm}, we plot the excitation (\ie when the equilibration measure \eqref{fth}
of the integrand at a particular radius is more than $2\%$ of its final value away from its initial equilibrium value) and equilibration boundary time $\t_\ast$ of the
EE integrand (for various $\a$ and for a wide surface) as a function of its radial position, for the
minimal surface where $\r_m=0.999\r_h$.  Note that the we say the profile ``equilibrates'', rather
than thermalizes, since the integrand of the correlator or EE at a particular radius is not
a physical quantity in the boundary theory that can thermalize.  Rather, it comes to rest in
some equilibrium, after which it is equilibrated.

We can conclude from these plots that the parts of the surface that lie deeper in the geometry
are also generally the ones that thermalize the latest.\footnote{Note that in fig.~\ref{rtherm}, the
equilibration curve for $\a=1$ is an exception, equilibrating earlier at most points in the bulk than at the boundary (even the deepest part).   The effect of the integrand equilibrating later in the bulk than the boundary 
is visible for the smaller values of $\a$, which corresponds more to  the universal behaviour related to instantaneous quenches (see appendix \ref{appA4}).}  It is precisely this part of the
surface that contributes the most to the EE.  Although not shown, we see a similar behaviour in the case
of the correlator.

In the next subsection, we show why it may be that these deeper parts of the geodesics and minimal
surfaces thermalize later than the near-boundary part.

\subsection{Equilibration profile of the scalar field and its stress-energy} \label{thermscalar}

The scalar field encodes both the source and response of the field theory to the quench.  For this reason,
we will consider the scalar field and its stress-energy as an indicator of how the energy of the quench
enters the interior of the AdS bulk.  It should be remembered that $\frac 1\r$ is proportional to
the energy scale of the
field theory.  Therefore the propagation of energy into the bulk is in a sense dual to the energy of the
quench being distributed through the different
energy scales of the field theory -- from the UV down to the thermal scale.

\begin{figure}
\begin{center}
\psfrag{r}[Br][tl]{{$\scriptstyle{\r}$}}
\psfrag{t}[c]{{$\scriptstyle{\t}$}}
  \includegraphics[width=3in]{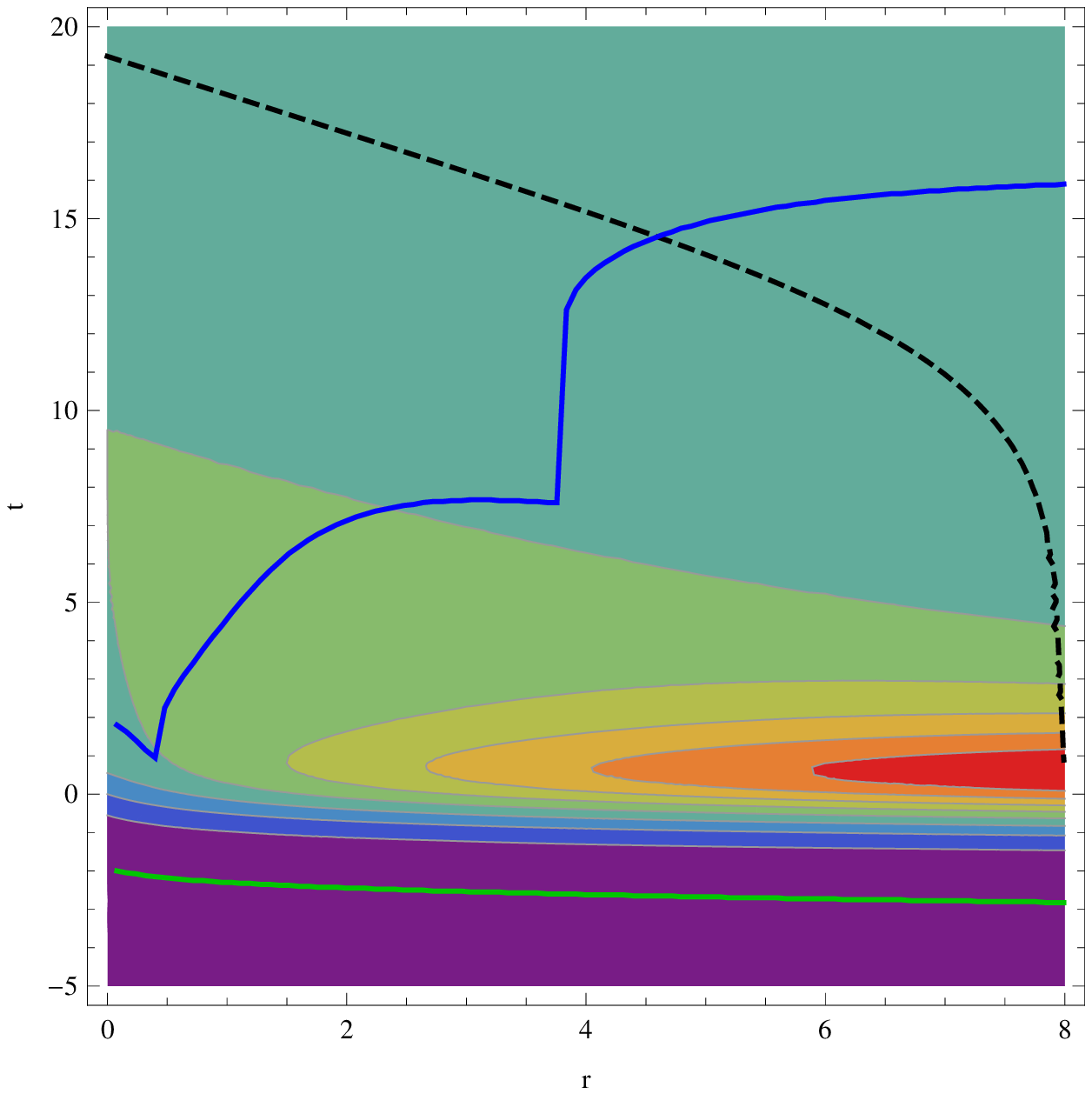}
    \raisebox{0.3\height}{\includegraphics[width=.3in]{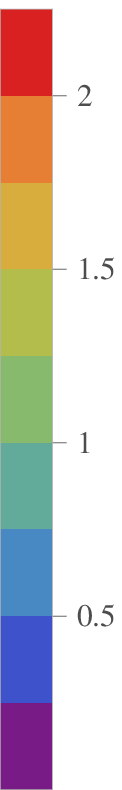}}
\end{center}
\caption{(Colour online) A contour plot of $\frac{\phi}{\r}$ with $\a=\frac 18$ as a function of $\r$ and
$\t_\ast$ (the boundary located at $\r=0$, the horizon at $\r=8$).  We add in the
excitation and equilibration curves in blue for the scalar field.  The bottom green curve represents the
time $\t$ at a particular radius $\r_p$ where $\left(\frac{\phi(\t,\r_p)}{\r_p}\right)_{(therm)}$ is outside
of the $2\%$ threshold for excitation.  The top blue curve represents the scalar field likewise being within
the $2\%$ threshold for equilibration at that radius.  The dashed curve shows the time contour for the
minimal surface with height $\r_m=0.999\,\r_h$, at which it time the EE thermalizes.
 \label{phitherm}}
\end{figure}

\begin{figure} [p]
\begin{center}
\psfrag{r}[Br][tl]{{$\scriptstyle{\r}$}}
\psfrag{t}[c]{{$\scriptstyle{\t}$}}
  \includegraphics[width=3in]{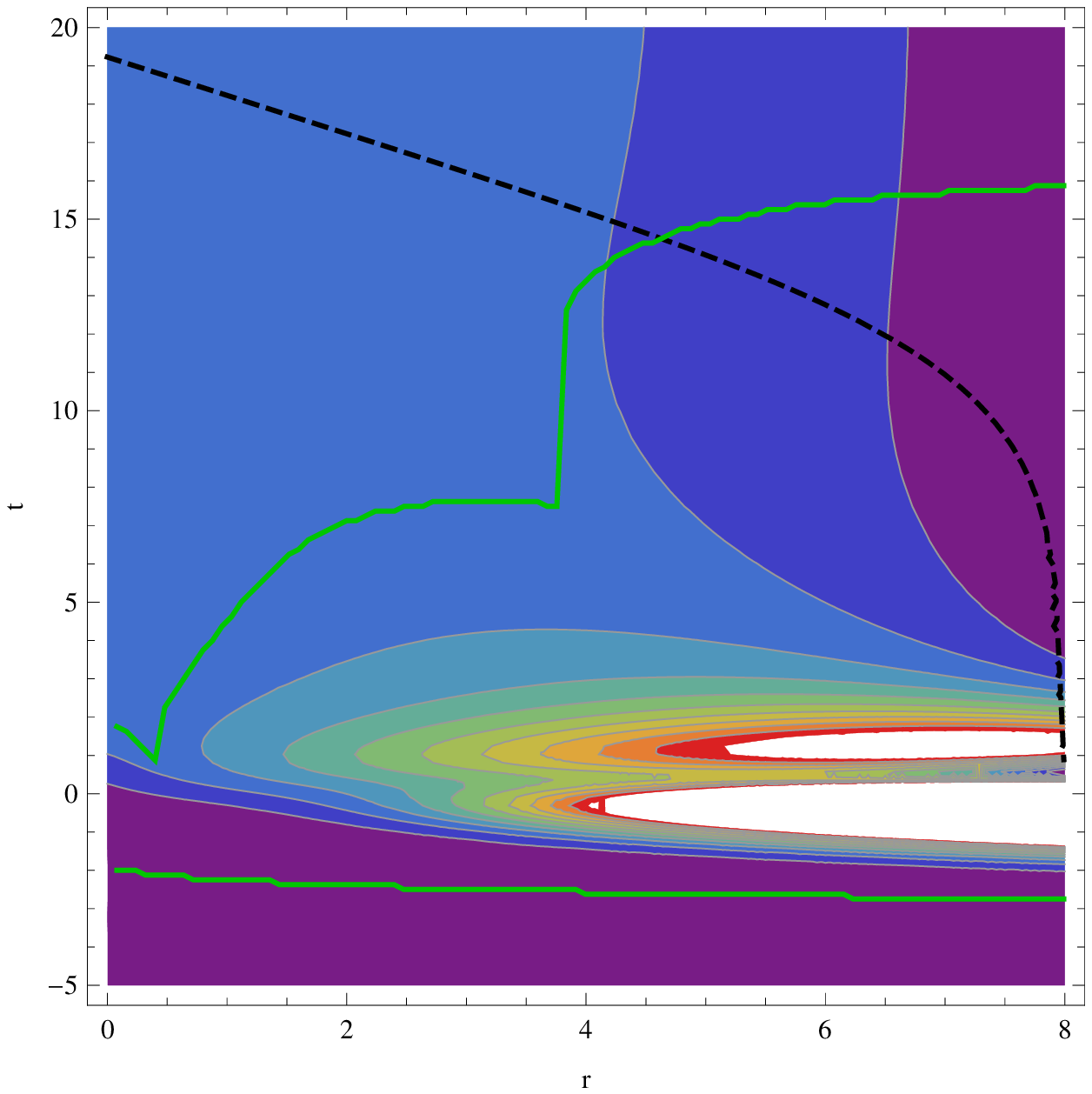}
  \raisebox{0.4\height}{\includegraphics[width=.3in]{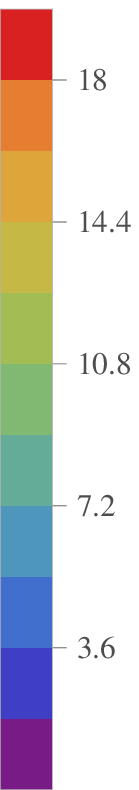}}
\end{center}
\caption{(Colour online) A contour plot of $T^{\phi}_{00}$ with $\a=\frac 18$ as a function of $\r$ and $\t_\ast$.
We add in the
excitation and equilibration curves for the tensor component in green, as we did for the scalar field in
figure \ref{phitherm}.  We also show the same contour for the thermalized entanglement entropy. \label{t00therm}}
\begin{center}
\psfrag{r}[Br][tl]{{$\scriptstyle{\r}$}}
\psfrag{t}[c]{{$\scriptstyle{\t}$}}
  \includegraphics[width=3in]{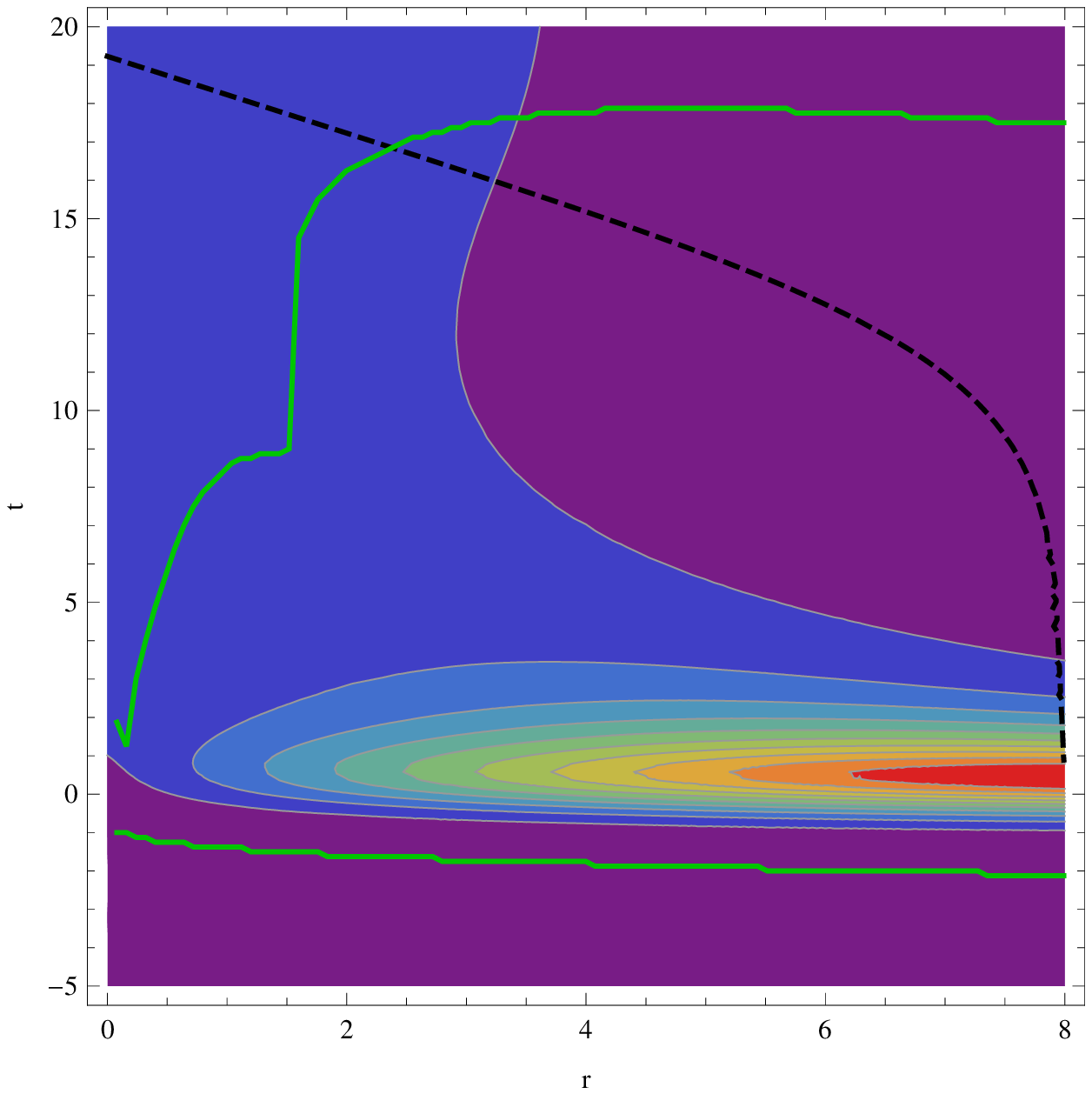}
  \raisebox{0.4\height}{\includegraphics[width=.3in]{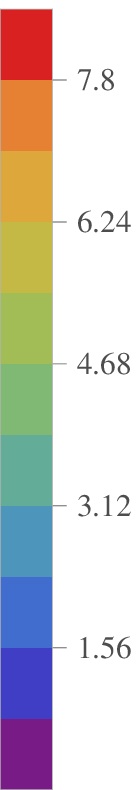}}
\end{center}
\caption{(Colour online) A contour plot of $T^{\phi}_{\r\r}$ with $\a=\frac 18$ as a function of $\r$ and $\t_\ast$.  We add in the
excitation and equilibration curves for the tensor component in green, as we did for the scalar field in
figure \ref{phitherm}.  We also show the same equilibration curve for the entanglement entropy. \label{trrtherm}}
\end{figure}

We show the contour plot with excitation and equilibration curves of the scalar $\frac{\phi}{\r}$ in figure \ref{phitherm}.  We show $\frac{\phi}{\r}$ rather than $\phi$, because $\frac{\phi}{\r}$ is
the natural quantity that was calculated in our numerical simulations.  Also note that in this figure,
as well as in figures \ref{t00therm} and \ref{trrtherm} we show a contour plot of the fields' profiles,
where its values are the contours shown in the plot, while the solid coloured regions between the
contours have intermediate values.
We remind the reader that lines of constant $\t$ in these plots are null
rays infalling into the black brane, rather than constant time slices, as also explained in section
\ref{anexcor} after equation \eqref{tpoinc}.

We will also plot two of the components
of the stress-energy of the scalar field, $T^{\phi}_{00}$ and $T^{\phi}_{\r\,\r}$, because it is the
``matter'' stress-energy which sources the backreaction of the metric in the Einstein equations.
The stress-energy is given by
\begin{eqnarray}
T^{\phi}_{\mu\nu} &=& -2\frac{\delta\,S_{\phi}}{\sqrt{-g}\delta\,g^{\mu\nu}} \nonumber\\
&=& \partial_{\mu}\phi\,\partial_{\nu}\phi-\frac{1}{2}
\left(\left(\partial\phi\right)^2+m^2\phi^2\right)g_{\mu\nu}.
\end{eqnarray}
In the first line above, $S_{\phi}$ is the part of the bulk action \eqref{action5} containing only scalar field terms,
\ie the matter action.  We show the contour plots for the these two components of the stress-energy
in figures \ref{t00therm} and \ref{trrtherm}, respectively.

\begin{figure}
\begin{center}
\psfrag{t}[Br][tl]{{$\scriptstyle{\t}$}}
\psfrag{fth}[c]{{$\scriptstyle{\phi_{(therm)}}$}}
  \includegraphics[width=3in]{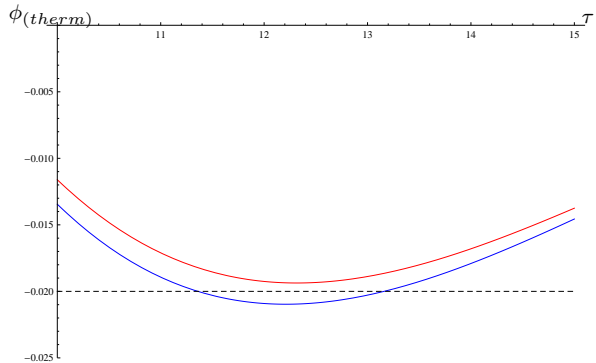}
\end{center}
\caption{(Colour online) A zoomed-in version of the thermalization curves (as defined by the $2\%$ criterion from equation \eqref{fth}) for the scalar field for particular
radial values on either side of the discontinuity seen at $\r\approx 3.8$ in fig \ref{phitherm}.  The
red curve is for $\r$ slightly smaller than $3.8$, and the blue curve for $\r$ slightly larger.  The
blue one crosses the dashed line representing the
equilibration threshold, and will therefore have a much later equilibration time than the red curve,
although the behaviour of the function is very similar at the two radii.}
\label{discont}
\end{figure}

We see several discontinuities in the equilibration curves of these quantities.
This however is not showing some
novel physics, but is rather a remnant from the strict $2\%$ cut-off, as seen in figure \ref{discont}.
That is to say that points on either side of such a discontinuity do not have very different
behaviours in time, but rather one would have a slightly higher amplitude, which allows it to cross
the $2\%$ threshold at a much later time than one with a slightly smaller amplitude, giving the
discrete jump in equilibration time.
What is interesting in each of the plots \ref{phitherm} -- \ref{trrtherm},
is that the equilibration time deep into the bulk is much later than near the boundary. 
We have also included the
$\t$ profile of a minimal surface $\gamma$, corresponding to a wide entangling region at the thermalization time of the corresponding entanglement entropy.  As can be seen in the three figures,
the profile is mostly outside of the spacetime regions where the scalar field
and the stress tensor fluctuate most.
We can therefore think of these contour profiles as indicating the level of disturbance the
EE (and correlator) experience at a certain boundary time $\t_\ast$, from how much of the $\t$-profile
extends into these regions.  The minimal surface or geodesic can be seen as being dragged through this
contour plot of $\frac{\phi}{\r}$ and $T^{(\phi)}$, exciting the entropy and two-point correlator, until most
of this profile has passed through the disturbed region and is deemed thermalized.  Because the time-profile
of the minimal surfaces/geodesic can stretch infinitely far into the past as $\r_m\to\r_h$, we can
expect that the thermalization time of the entanglement entropy or two-point correlator could be made
arbitrarily long.

\subsection{Heuristics of thermalization} \label{thermtimes}

As we have seen, the nonlocal probes that thermalized most like the non-normalizable mode 
were those that had relatively small separations, and reflect the physics closer to the AdS boundary.
Those that thermalized most slowly were those with larger separations.  The time scales here increased monotonically,
and potentially exceed the thermalization
time of the one-point function for wide-enough surfaces.

For the local quantities in sections \ref{thermprofile} and \ref{thermscalar},
we saw a wide range of equilibration times.  For the quantities close
to the boundary, we saw that they equilibrated on a time scale similar to the non-normalizable mode.
This makes sense, since
near the boundary, the dominant term in the respective asymptotic series is in fact the one containing the
source term $p_0$, or $p^2_0$.  However, deeper into the geometry the higher-order terms in the expansion
containing the response coefficient $p_2$ will have an increasing contribution, so that we can expect
longer thermalization times.  This is exactly what was observed.

The only scales that we introduced in this system are the quenching parameter $\a$
(corresponding to the non-normalizable mode $p_0$), the emergent response
$p_2$, and the temperature of the system.  The other scale that comes into play
for the geodesic or entanglement entropy is the width of
the probes.  We should expect wider separations in the entangling surface and the points of the
correlator to introduce an extra scale, since their boundaries are causally disconnected in the
boundary spacetime.  In figure \ref{lineeth} we see that the thermalization time of the
two-point correlator and entanglement entropy grow (at least, roughly) linearly as a function of the separation in
the function.  The thermal wavelength of the dual field theory is
$\lambda_T\equiv \frac{1}{T} = \pi$, given our conventions \cite{blmv}.  One might expect that
if the thermalization process occurs
quasilocally in the field theory, correlation functions or entanglement entropies on scales
larger than the thermal wavelength (\ie involving points or boundaries separated by more than $\lambda_T$) should thermalize with approximately the
same time.  Nevertheless, we see monotonic, linear growth in these probes' thermalization times
for separations $2\,y_m>\lambda_T$.  In fact, in figure \ref{lineeth}, the widest separation for our
two-point function is approximately twice the thermal wavelength.  This behaviour suggests that arbitrarily
large regions will see arbitrarily long thermalization times set by the time for these points to come
into causal contact. 
This is the behaviour observed by Calabrese and Cardy \cite{cc0} in
considering the entanglement entropy of an interval in a two-dimensional free field theory, and which
lends itself to a simple quasiparticle interpretation.
They saw a linear increase in the entropy after the quench, until such a time as the two ends
of the region would come into causal contact (as though the information was carried
by quasiparticles), after which the EE would quickly thermalize.  As in our case, they did not see an
upper bound on the thermalization time. For holographic calculations
of two-point functions and EE in a two and higher dimensional boundary theories, similar behaviour was found\footnote{In some Vaidya studies \cite{therm1}, the thermalization time at a particular length scale
is estimated as the time when the entire probe is completely contained inside the collapsing shell. With this geometric definition, it is clear that the thermalization time can become arbitrarily long since in terms of the boundary time, the shell takes infinitely long to cross the location of the black hole horizon.} in studies using a Vaidya metric in the bulk \cite{therm1,vaidya} and the precise evolution by which the entanglement entropy is saturated was extensively studied in \cite{hong}.  Hence our results are in agreement with these other holographic studies and hence it seems that the trend of longer thermalization times at larger length scales holds true in both strongly and weakly coupled field theories.

This discussion pertains only directly to perturbative quenches.  However, we may expect to see similar
behaviour for fully nonlinear quenches, since the response $p_2$ would be appropriately modified in the
nonlinear regime.

\section{Conclusion}

The standard toolkit of numerical relativity \cite{poisson,lreview} faces challenges when confronted with
typical problems in asymptotically anti-de Sitter spacetimes, motivated by the gauge theory/string
theory correspondence \cite{m1}. The main challenge is that gravitational simulations in asymptotically Minkowski spacetimes
mostly have a compact physical dependence domain; on the contrary, in AdS, control over the whole space-time,
and especially near the boundary is crucial. The latter is emphasized in problems related
to holographic quenches, where the temporal history of a quantum gauge theory coupling is encoded
as a non-normalizable component of the gravitationally dual bulk scalar field near the boundary.
In this paper we described the application of pseudo-spectral methods based on Chebyshev
polynomial expansion to problems of holographic quantum quenches. We paid special attention to
convergence and accuracy issues of the proposed spectral framework.

Our main physical application was the extension of the earlier work on holographic quenches
in strongly coupled  $\caln=4$ supersymmetric gauge theory plasma induced by a time-varying
coupling of certain dimension $\Delta=3$ operator \cite{blm}.
Here, having access to the full bulk metric (albeit only to the leading order in the
gravitational scalar backreaction), enabled us to compare {\it local}
 and {\it nonlocal} probes of the ensuing thermalization process.
Specifically, we compared the relaxation of event and apparent horizons,
the equal time two-point correlation function of operators of large conformal dimension, the
entanglement entropy of  strip-shaped regions  with the relaxation of the
one-point correlation function of the quenching  operator. The nonlocal probes of
thermalization were discussed earlier in the literature \cite{therm1,vaidya}.
In fact, our discussion of non-local
probes parallel that of \cite{periodic}. The important difference is that the authors of
the latter work considered periodically-driven holographic quenches, in which the boundary gauge theory
never reaches an equilibrium, thus making the comparison of various thermalization criteria impossible.

As a  criterion  for thermalization of a probe $f$ we considered the quantity \eqref{fth}
\begin{equation}
f_{th} (\tau)\equiv \frac{f(\t)-f(\infty)}{f(\infty)-f(-\infty)}.
\eqlabel{fthrepeat}
\end{equation}
If $f$ is a non-local probe, there is an additional dependence in $f_{th}$ on a characteristic
energy scale of $f$: the separation between the points in two-point equal-time correlation
functions; the size of the entanglement region. Note that if $f$ vanishes in the initial
state (as it did for our probes), $f_{th}(-\infty)=-1$ and $f_{th}(\infty)=0$.  If
$f= \langle\calo_\Delta\rangle$ is the expectation value of the quenching operator
of dimension $\Delta$, $\max [\ft{f(\t)}{f(\infty)}]\propto \a^{4-2\Delta}$ ($\propto \a^{-2}$ for $\Delta=3$)
in the limit of fast quenches, \ie as
$\a\ll 1$ \cite{bmv}. Probably our most dramatic finding is that for all nonlocal
probes discussed (and for wide range of probing energy scales, if applicable), $f_{th}^{non-local}$
remains finite in the limit of fast quenches. As a result the thermalization of non-local
probes, within criteria \eqref{fthrepeat}, appears to be  faster than that of $\langle\calo_\Delta\rangle$.
This effect becomes more pronounced as the quenching rate $\ft 1\a$ increases.  While we focused
in this paper on $d=4$ boundary space-time dimensions and for $\Delta=3$ of the quenching operator,
we believe that this observation would extend to general $d$ and $\Delta$ insofar as $\ft d2< \Delta <d$.

For moderate quenches, $\a=\{1,\ft 12\}$, we observe the expected characteristic energy dependence in
$f_{th}$, in comparison with $\langle\calo_3\rangle$, see top panels of figure \ref{linafig}.
A short distance two-point correlator probes the dual bulk geometry close to the boundary;
thus, its evolution would mimic that  of the square of  the  quenching coupling
--- it would thermalize earlier than $\langle \calo_3\rangle$ (see the
dark blue curve, or figure \ref{linvsp02}).  As the point separation in the two-point correlator
increases, the dual geodesic dips deeper into the bulk, probing the more infrared features ---
as a result it thermalizes similarly to  $\langle \calo_3\rangle$ (see the
red curve).  As mentioned before, we expect the two-point correlator to 
thermalize later than $\langle \calo_3\rangle$ for wide enough surfaces. The entanglement entropy behaves in a similar manner: entanglement of narrow strips
evolve as the quenching coupling squared; the entanglement of wider regions takes
a longer time to thermalize.

We now finish with open problems. First, it is important to lift the restriction of
`leading order backreaction' --- for this, one needs to extend the proposed numerical
framework to the full nonlinear evolution. We believe that this does not
pose conceptual or technical difficulties: all the numerical steps can be easily generalized,
even our spectral uplift procedure from $g_c(\t,\r)$ to $a_c(\t,\r)$ (see
appendix \ref{appA2}). Likewise, using  realistic dual scalar potentials, as in \cite{pw},
does not pose a problem either.
Another benefit of the spectral approach is that it
relatively easy allows for a generalization to spatially non-homogeneous and non-isotropic
quenches. We hope to report on the latter problem in a future work.

\section*{Acknowledgments}
We would like to thank Luis Lehner, Erik Schnetter and Johannes Walcher for valuable discussions.
We are indebted to Mukund Rangamani and Moshe Rozali for discussions about the apparent horizon 
that allowed us to track a mistake in the earlier version of the code: namely, the incorrect boundary 
condition for $g_c(\t,0)$ in \eqref{bcconst}.  
Research at Perimeter Institute is supported by the Government of Canada through Industry
Canada and by the Province of Ontario through the Ministry of Research
\& Innovation. AB and RCM gratefully acknowledge support from NSERC
Discovery grants. Research by RCM is further supported by
funding from the Canadian Institute for Advanced Research.

\appendix
\section{Appendix: Numerical solution of the dynamical metric and scalar field} \label{a1}
\subsection{Definition and solution of fields in perturbative regime}\label{appA1}

Let us introduce the functions that are involved in the numerical
recipe, namely $\hp_c$, $g_c$ and $b_c$. These fields are related to the
dimensionless warp factors and scalar field $\ha$, $b$ and $\hp$ defined
in eqs.~\eqref{hpl2}-\eqref{bl2} in
various ways.  The solution of the usual dimensionless functions can
be expressed in the rescaled coordinates in terms of the new functions as:
\begin{equation}
\begin{split}
\hp(\t,\r)=&\phi_{log}(\t,\r)+\phi_c(\t,\r),\\
b(\t,\r)=&b_{log}(\t,\r)+b_c(\t,\r),\\
\ha(\t,\r)=&-\frac 16 (p_0)^2+a_{log}(\t,\r)+a_c(\t,\r).
\end{split}
\eqlabel{ndef}
\end{equation}
The $\phi_{log}$, $a_{log}$ and $b_{log}$ terms remove (subtract) logarithms\footnote{We found that
the presence of the logarithmic terms in the Ads boundary asymptotics of various
fields renders spectral (or finite difference) numerical method unstable.}
close to the boundary ($\rho\to 0$) in the
asymptotic expansion of $\hp$ and the warp factors $a$ and $b$, while staying bounded
close to the horizon ($\r\to \frac 1\a$).
Of course, there is a choice in selecting $\phi_{log}$, $a_{log}$ and $b_{log}$.
For $\phi_{log}(\t,\r)=\phi_{log}(p_0(\t),\r)$ we choose
\begin{equation}
\phi_{log}=\log\r \sum_{i=2}^8 \frac{\r^i}{(1+\r)^{1+i}} \calf_i(p_0(\t)),
\eqlabel{deflogphi}
\end{equation}
where the coefficients $ \calf_i(p_0(\t))$ are (uniquely) adjusted in such a way that the resulting $\hp_c$
are free from $\ln\r$ up to terms $\calo(\r^9\log\r)$. Explicitly, the first few coefficients $\calf_i(p_0(\t))$
are
\begin{equation}
\begin{split}
\calf_2=\frac 12 p_0'',\qquad \calf_3=\frac 12 p_0'''+\frac 32 p_0'',\qquad \calf_4=
\frac{5}{16} p_0^{(4)}+2 p_0'''+3 p_0''.
\end{split}
\eqlabel{deffi}
\end{equation}
Note that the subtraction  $\phi_{log}$ remains bounded all the way to the horizon for fast quenches  $\a\le 1$.
Similarly, we take
\begin{equation}
\begin{split}
b_{log}=&(\a\r)^2\biggl[\log\r \sum_{i=2}^5 \frac{\r^i}{(1+\r)^{1+i}} \calb_{1,i}(p_0(\t),p_2(\t))
+\log^2\r \sum_{i=4}^5 \frac{\r^i}{(1+\r)^{1+i}} \calb_{2,i}(p_0(\t))
\biggr],\\
a_{log}=&\log\r \sum_{i=2}^5 \frac{\r^i}{(1+\r)^{1+i}} \cala_{1,i}(p_0(\t),p_2(\t))
+\log^2\r \sum_{i=4}^5 \frac{\r^i}{(1+\r)^{1+i}} \cala_{2,i}(p_0(\t)).
\end{split}
\eqlabel{defloab}
\end{equation}
Ideally, we would like to subtract as many log-terms near the boundary as possible;
this would make
spectral expansion of the functions more precise.
It is possible to expand \eqref{deflogphi} to arbitrary
order: for any $i$, $\calf_i$ depends on a source $p_0$ and its higher time derivatives,
and thus is known analytically for our quenches, where
\begin{equation}
p_0=\frac 12 \left(1+\tanh \t\right).
\eqlabel{defp0}
\end{equation}
The asymptotic expansions for $\ha$ and $b$ contain log-terms with prefactors that, in addition to the
functional source dependence, depend on response function $p_2(\t)$ and its derivatives.
Specifically, both $\calb_{1,i}$ and $\cala_{1,i}$ for $i\ge 6$ depend on the derivatives of
the response $p_2(\t)$ up to order $(i-5)$. We can extract reliably $p_2(\t)$ from the
evolution of the $\phi_c$:
\begin{equation}
p_2(\t)= \frac 12 \del^2_{\r} \phi_c(\t,0),
\eqlabel{defp2phic}
\end{equation}
 however, we find that the errors in extracting derivatives
of $p_2(\t)$ does not justify truncating the \eqref{defloab} beyond the terms employed.
Explicit expressions of the first few logarithm prefactors in \eqref{defloab} are given by
\begin{equation}
\begin{split}
&\calb_{1,2}=-\frac{1}{24}p_0p_0'',\qquad \calb_{1,3}=-\frac{1}{30}p_0p_0'''
-\frac{1}{20}p_0'p_0''-\frac 18 p_0p_0'',\\
&\cala_{1,2}=\frac 16 ((p_0')^2- p_0 p_0''),\qquad
\cala_{1,3}=-\frac{1}{12}(p_0p_0'''-p_0'p_0'')-\frac 12(p_0p_0'' -(p_0')^2),\\
&\calb_{2,4}=-\frac{1}{80} (p_0')^2,\qquad  \cala_{2,4}=-\frac{1}{40}(p_0'')^2.
\end{split}
\eqlabel{defabi}
\end{equation}

We further define the functions that occur naturally in equations
\eqref{neq1} -- \eqref{neq3}, namely $\pi$, $\b$ and $g_c$:
\begin{equation}
\begin{split}
\pi(\t,\r)=&\del_t \hp(\t,\r)+\frac {\a^4\r^4-1}{2}\ \del_\r \hp(\t,\r),\\
\b(\t,\r)=&\del_t b(\t,\r)+\frac {\a^4\r^4-1}{2}\ \del_\r b(\t,\r).
\end{split}\eqlabel{ndef2}
\end{equation}
As in \eqref{deflogphi}, we subtract the logarithmic terms of $\pi(\t,\r)$ near the boundary
\begin{equation}
\begin{split}
&\pi(\t,\r)=\pi_c(\t,\r)+\pi_{log}(\t,\r),\\
&\pi_{log}=\ln\r \sum_{i=1}^7 \frac{\r^i}{(1+\r)^i}\calp_i(p_0(\t)), \\
&\calp_1=-\frac 12 p_0'',\qquad \calp_2=-\frac 14 p_0'''-p_0'',\qquad
\calp_3=-\frac 18p_0^{(4)} -\frac 34 p_0'''-\frac 32 p_0''.
\end{split}
\eqlabel{defpic}
\end{equation}

We now present the equations which we separate into the {\it evolution}
(containing time derivatives of the functions) and the
{\it constraint} (without time derivatives of the functions) ones,
\nxt evolution equations:
\begin{equation}
\del_t \phi_c(\t,\r)=\pi_c(\t,\r)+\frac {1-\a^4\r^4}{2}\ \del_\r \phi_c(\t,\r)+k_{log}(\t,\r),
\eqlabel{evolve1}
\end{equation}
with
\begin{equation}
k_{log}(\t,\r)=\pi_{log}(\t,\r)+\frac {1-\a^4\r^4}{2}\ \del_\r \phi_{log}(\t,\r)
-\del_\t \phi_{log}(\t,\r).
\eqlabel{evolve2}
\end{equation}
\nxt constraint equations:
\begin{equation}
\begin{split}
&\del_\r \pi_c-\frac{1}{2\r}\ \pi_c=-\biggl(J_{\pi}+\del_\r \pi_{log}-\frac {1}{2\r} \pi_{log}\biggr),
\end{split}
\eqlabel{const1}
\end{equation}
\begin{equation}
\del_{\r}^2 b_c=-\biggl(J_{b}+\del^2_{\r} b_{log}\biggr),
\eqlabel{const2}
\end{equation}
with
\begin{equation}
\begin{split}
&J_{\pi}=\frac{1}{4\r} \del_\r\hp-\frac14 \a^4\r^3 \del_\r \hp+\frac 12 \a^4 \r^2 \hp,
\end{split}
\eqlabel{hconst1}
\end{equation}
\begin{equation}
\begin{split}
&J_{b}=\frac 16\a^2 \left(\hp +\r \del_\r \hp\right)^2.
\end{split}
\eqlabel{hconst2}
\end{equation}
One additional constraint equations is obtained combining \eqref{neq2} and \eqref{neq3}.
First, using the second equation in \eqref{ndef2} we rewrite the latter equation as
\begin{equation}
\begin{split}
\del_{\r}^2\ha+\frac2\r\ \del_\r \ha-\frac{6}{\r^2}\ \ha-\frac{12}{\a^2\r^3}\ \beta=&-J_{\ha},\\
\del_\r\ \beta-\frac 3\r\ \beta+\a^2\left(\frac\r2\ \ \del_\r\ \ha-\ha\right)=&-J_{\beta},
\end{split}
\eqlabel{const3}
\end{equation}
with
\begin{equation}
\begin{split}
&J_{\ha}=\left(\frac{6}{\a^2\r^3}-6\a^2\r\right) \del_\r b-\del_\t\hp \left(\del_\r\hp+\frac 1\r \hp\right)
+\frac 12 (1-\a^4\r^4) \left(\del_\r\hp+\frac 1\r \hp\right)^2+\frac{1}{2\r^2}\hp^2,
\\
&J_{\beta}=\left(\frac {3}{2\r}-\frac{3\a^4\r^3}{2}\right)\del_\r b+\frac 14\a^2 \hp^2.
\end{split}
\eqlabel{hconst3}
\end{equation}
Algebraically solving for $\beta(\t,<r)$ from the first equation in \eqref{const3}, we can represent the
remaining equation in \eqref{const3} as
\begin{equation}
\begin{split}
&\del_\r g=-J_g,\qquad g\equiv \del_{\r}^2 \ha+\frac2\r \del_\r \ha,\\
\end{split}
\eqlabel{defg}
\end{equation}
or
\begin{equation}
\begin{split}
&\del_\r g_c=-J_{g_c},\qquad g_c\equiv \del_{\r}^2 a_c+\frac2\r \del_\r a_c,\\
&J_{g_c}=J_g+\del_\r \left[\del_{\r}^2 a_{log}+\frac2\r \del_\r a_{log}\right],
\end{split}
\eqlabel{defgc}
\end{equation}
with
\begin{equation}
\begin{split}
J_g=&\left(\frac{1}{2\r^2}\hp-\del^2_{\r} \hp-\frac{3}{2\r}\del_\r \hp\right)\pi
+\left(\frac{1}{2\r}\hp-\frac 12 \a^4\r^3 \hp\right)\del^2_{\r} \hp-\left(\frac 14\a^4 \r^2 \hp
+\frac{1}{4\r^2}\hp\right)\del_\r \hp\\
&+\left(\frac14\a^4\r^3-\frac{1}{4\r}\right)(\del_\r\hp)^2
+\frac 12\a^2\left(\a^2\r\hp^2 -48 \del_\r b\right).
\end{split}
\eqlabel{defjg}
\end{equation}
Note that given $g_c(\t,\r)$, and using the definition of $\ha$ in \eqref{ndef},
 we can always reconstruct $a_c(\t,\r)$ as
\begin{equation}
a_c(\t,\r) = \int_0^\r \frac{dx}{x^2}\biggl[\int_0^x dy\ y^2 g_c(\t,y)\biggr].
\eqlabel{defac}
\end{equation}

So far, we have not used \eqref{neq5} --- this equation contains two $\t$ derivatives,
and so appears to be an evolution equation. It turns out that this is a {\it momentum constraint},
and should be imposed at a single spatial point, say $\r=0$;
the equations \eqref{neq1}-\eqref{neq4}
guarantee that \eqref{neq5} would then be true at any other point. The latter constraint
determines $a_{2,2}(\t)$ (see (\ref{al2})) in terms of the source $p_0(\t)$ and the response $p_2(\t)$:
\begin{equation}
0=a_{2,2}' +\frac13 (p_0 p_2'-p_0'p_2)+\frac{1}{18}p_0'p_0''-\frac 29 p_0 p_0'''.
\eqlabel{a2const}
\end{equation}
In practice, we find it convenient to introduce
\begin{equation}
\ha_2\equiv a_{2,2}+\frac {5}{36} \left(p_0'\right)^2-\frac 29 p_0 p_0''+\frac 13 p_0 p_2,
\eqlabel{defa2h}
\end{equation}
which allows to rewrite \eqref{a2const}  as
\begin{equation}
0=\ha_2'-\frac 23 p_0' p_2.
\eqlabel{consth}
\end{equation}

Evolution equations \eqref{evolve1} and \eqref{consth} are solved subject to appropriate initial conditions.
In our simulations we assume thermally equilibrium $\caln=4$ state in the limit $\t\to -\infty$:
\begin{equation}
\begin{split}
\phi_c(-\infty,\r)=0,\qquad \ha_2(-\infty)=0.
\end{split}
\eqlabel{inititalcond}
\end{equation}
The constraint equations \eqref{const1}, \eqref{const2}
and \eqref{defgc}
are solved subject to the boundary condition at $\r=0$, which are found using the asymptotic expansions
(\ref{hpl2})-(\ref{bl2}) and following the chain of redefinitions \eqref{ndef}, \eqref{defpic} and
\eqref{defgc}:
\begin{equation}
\begin{split}
&\pi_c(\t,0)=\frac 12 p_0',\\
&b_c(\t,0)=0,\qquad \del_\r b_c(\t,0)=0,\\
&g_c(\t,0)=6 a_{2,2}.
\end{split}
\eqlabel{bcconst}
\end{equation}
As explained in \cite{blm}, there is no need to impose the boundary condition at the horizon provided
we extend the radial integration past its location:
\begin{equation}
\r\ \in [0,L_\rho],\qquad L_\r=\frac{1.2}{\a},\qquad \r_{horizon}=\frac 1\a \biggl(1+\calo(\ell^2)\biggr).
\eqlabel{rangerho}
\end{equation}

\subsection{Numerical implementation}\label{appA2}
We use pseudo-spectral methods \cite{chebyshev} to solve numerically equations
\eqref{evolve1}, \eqref{consth} and \eqref{const1}, \eqref{const2}, \eqref{defgc}, subject to the initial
and boundary conditions \eqref{inititalcond} and \eqref{bcconst} on a domain:
\begin{equation}
\r\ \in [0,L_\r],\qquad \t \in [\t_{initial},\t_{final}].
\eqlabel{numdom}
\end{equation}
In practice we choose $\t_{initial}=-7.5$, corresponding to the source value $p_0(\t_{initial})\approx 3\times
10^{-7}\ll 1$ (see \eqref{defp0}); and $\t_{final}=12.5$ (or later for smaller $\a$). In a nutshell, any function $f(\t,\r)$
we represent as truncated sum over Chebyshev polynomials $T_j(x)$,
\begin{equation}
\begin{split}
&f(\t,\r)\sim  \sum_{j=1}^N  \calf_f^j(\t) T_{j-1} \left(-1+\frac{2\r}{L_\r}\right),\\
&T_{0}(x)= 1,\qquad T_1(x)=x,\qquad T_{j+1}(x)= 2 x T_{j}(x)-T_{j-1}(x),\qquad j\ge 1.
\end{split}
\eqlabel{expandf}
\end{equation}
All constraints equations are then reduced to linear-algebraic equations evaluated at $N$-collocation points
\cite{chebyshev}. We use fourth-order Runge-Kutta method (RK4) to evolve functions in time.

We now describe the implementation steps of our numerical package in detail:
\nxt The range of $\rho$:
\begin{equation}
\r\in \left[0,L_\r\right],
\eqlabel{num1}
\end{equation}
where we include the boundary points;
\nxt we introduce the collocation grid points:
\begin{equation}
x_i=\cos \frac{(i-1)\pi}{N-1},\qquad i=1,\cdots N\,;\qquad \r_i=\frac{L_\r}{2} (1+x_i).
\eqlabel{num2}
\end{equation}
Note: $\r_1=L_\r$ and $\r_N=0$, and
\begin{equation}
\frac {dx}{d\r}= \frac{2}{L_\r}.
\eqlabel{dxdrho}
\end{equation}
\nxt We use a recursive relation to compute Chebyshev
polynomials $T(i,j)\equiv T_{j-1}(x_i)$,
and their derivatives at the collocation points,
\ie $d_nT(i,j)\equiv P^{(n)}_{i-1}(x_i)$, for $n=1,2$ :
\begin{equation}
\begin{split}
&T(i,1)=1,\qquad T(i,2)=x_i,\qquad
T(i,j+2)=2 x_i\ T(i,j+1)- T(i,j),\\
&d_nT(i,j+2)=2 x_i\ d_nT(i,j+1)+2 n x_i\ d_{n-1}T(i,j+1)-d_nT(i,j).
\end{split}
\eqlabel{defcheb}
\end{equation}
\nxt We store data at spatial collocation point
at time $\t=\t^o$ in arrays with superscript $ ^o$,
and data at time $\t=\t^n\equiv \t^o+n\,\Delta\t$ in arrays with superscript $ ^n$.
For convergence, we choose
\begin{equation}
\Delta \t= \frac{1}{N^2} \times \min\left\{1,\a\right\}.
\eqlabel{dtchoice}
\end{equation}
\nxt RK4 is used to evolve from $\phi_c$ and $\ha_2$:
\begin{itemize}
\item \underline{\bf RK step 1}: Given,
\begin{equation}
\phi^o_i\equiv \phi_c(\t^o,\r_i),\qquad a_{2,2}^o\equiv a_{2,2}(\t^o),
\eqlabel{defhp}
\end{equation}
we compute Chebyshev coefficients $\calf_{\phi_c}^j$ solving
\begin{equation}
\phi^o_i = \sum_{j=1}^N\ \calf_{\phi_c}^j\ T(i,j),\qquad i=1,\cdots N.
\eqlabel{eomfhp}
\end{equation}
Next, we evaluate
\begin{equation}
\begin{split}
&d_1\phi^o_i\equiv \del_\r \phi_c(\t^o,\r_i)=\sum_{j=1}^N\ \calf_{\phi_c}^j\ d_1T(i,j)\ \frac{dx}{d\r},\\
&p_2^o=\frac 12 \sum_{j=1}^N\ \calf_{\phi_c}^j\ d_2T(i,j)\  \left(\frac{dx}{d\r}\right)^2,
\end{split}
\eqlabel{defdhp}
\end{equation}
\begin{equation}
\ha_2^o=a_{2,2}^o+\frac {5}{36} \left(p_0'(\t^o)\right)^2-\frac 29 p_0(\t^o) p_0''(\t^o)
+\frac 13 p_0(\t^o) p_2^o,
\eqlabel{defadhd}
\end{equation}
\begin{equation}
\begin{split}
&(\phi_{log})_i\equiv \phi_{log}\phi_{log}(\t^o,\r_i),\qquad d_1 (\phi_{log})_i\equiv \del_\r \phi_{log}(\t^o,\r_i),\\
&(\pi_{log})_i\equiv \pi_{log}(\t^o,\r_i),\qquad d_1 (\pi_{log})_i\equiv \del_\r \pi_{log}(\t^o,\r_i),\\
&(k_{log})_i\equiv k_{log}(\t^o,\r_i).
\end{split}
\eqlabel{log}
\end{equation}
We now  have all the data needed to compute (see \eqref{hconst1})
\begin{equation}
J_{\pi,i}\equiv J_{\pi}(\t^o,\r_i; \del_\r\hp= d_1\phi_i^o+d_1 (\phi_{log})_i, \hp=\phi_i^o+(\phi_{log})_i).
\eqlabel{defjpi}
\end{equation}
Next, we use \eqref{const1} and the boundary condition in \eqref{bcconst}
to compute  Chebyshev coefficients $\calf_{\pi_c}^j$:
\begin{equation}
\begin{split}
&i=1,\cdots N-1\ :\\
&\sum_{j=1}^N \left(d_1T(i,j) \frac{dx}{d\r}-\frac{1}{2\r_i} T(i,j)\right) \calf_{\pi_c}^j=
-\left(J_{\pi,i}+d_1 (\pi_{log})_i-\frac{1}{2\r_i} (\pi_{log})_i\right),
\\
& \sum_{j=1}^N T(N,j)\ \calf_{\pi_c}^j=\frac 12 p_0'(\t^o).
\end{split}
\eqlabel{eomphih}
\end{equation}
We can now determine
\begin{equation}
\pi_i\equiv \pi_c(\t^o,\r_i)= \sum_{j=1}^N \calf_{\pi_c}^j\ T(i,j),\qquad i=1,\cdots N.
\eqlabel{compphii}
\end{equation}
Finally, we complete the first RK step ($i=1,\cdots N$):
\begin{equation}
\begin{split}
&k_{1,\phi_c,i}=\Delta\t\ \left(\pi_i+\frac 12 (1-\a^4\r_i^4)\ d_1\phi_i^o+(k_{log})_i\right),\\
&k_{1,\ha_2}=\Delta\t\ \frac 23 p_0'(\t^o)\ p_2^o.\\
\end{split}
\eqlabel{k1}
\end{equation}
\item \underline{\bf RK step 2}: With the shift
\begin{equation}
\t^o\to \t^o+\frac 12\Delta\t,\qquad \phi_i^o\to \phi_i^o+\frac 12 k_{1,\phi_c,i},
\eqlabel{shiftk2}
\end{equation}
we repeat RK step 1, producing $k_{2,\phi_c,i},\, k_{2,\ha_2}$.
\item \underline{\bf RK step 3}: With the shift
\begin{equation}
\t^o\to \t^o+\frac 12\Delta\t,\qquad \phi_i^o\to \phi_i^o+\frac 12 k_{2,\phi_c,i},
\eqlabel{shiftk3}
\end{equation}
we repeat RK step 1, producing $k_{3,\phi_c,i},\, k_{3,\ha_2}$.
\item \underline{\bf RK step 4}: With the shift
\begin{equation}
\t^o\to \t^o+\Delta\t,\qquad \phi_i^o\to \phi_i^o+ k_{3,\phi_c,i},
\eqlabel{shiftk4}
\end{equation}
we repeat RK step 1, producing $k_{4,\phi_c,i},\, k_{4,\ha_2}$.
\item  We now update to a time-step $\t^n$:
\begin{equation}
\begin{split}
&\phi_i^n=\phi_i^o+\frac 16 k_{1,\phi_c,i}+\frac 13 k_{2,\phi_c,i}+\frac 13 k_{3,\phi_c,i}+\frac 16 k_{4,\phi_c,i},
\qquad i=1,\cdots N, \\
&\ha_2^n=\ha_2^o+\frac 16 k_{1,\ha_2}+\frac 13 k_{2,\ha_2}+\frac 13 k_{3,\ha_2}+\frac 16 k_{4,\ha_2}.
\end{split}
\eqlabel{update}
\end{equation}
\item At this stage we introduce dissipation \cite{dissipation}. We compute
Chebyshev coefficients $\calf_{\phi_c}^j$ solving
\begin{equation}
\phi^n_i = \sum_{j=1}^N\ \calf_{\phi_c}^j\ T(i,j),\qquad i=1,\cdots N.
\eqlabel{eomfhpn}
\end{equation}
and re-evaluate $\phi^n_i$ suppressing the higher harmonics:
\begin{equation}
\begin{split}
&\phi^n_i = \sum_{j=1}^{N-N_{diss}}\ \calf_{\phi_c}^j\ T(i,j),\qquad i=1,\cdots N,\\
&p_2^n=\frac 12 \sum_{j=1}^{N-N_{diss}}\ \calf_{\phi}^j\ d_2T(i,j)\  \left(\frac{dx}{d\r}\right)^2.
\end{split}
\eqlabel{eomfhpn2}
\end{equation}
where, in practice, we choose
\begin{equation}
N_{diss}=\left[0.2 N\right].
\eqlabel{defdiss}
\end{equation}
We use \eqref{defa2h} to compute $a_{2,2}^n$:
\begin{equation}
a_{2,2}^n=\ha_2^n-\frac {5}{36} \left(p_0'(\t^n)\right)^2+\frac 29 p_0(\t^n) p_0''(\t^n)
-\frac 13 p_0(\t^n) p_2^n.
\eqlabel{defanew}
\end{equation}
\item In preparation to computation of
\begin{equation}
b_i^n\equiv b_c(\t^n,\r_i),\qquad g_i^n\equiv g_c(\t^n,\r_i),
\eqlabel{defbn}
\end{equation}
 we evaluate $d_1 \phi_i^n\equiv \del_\r \phi_c(\t^n,\r_i)$,
$d_2 \phi_i^n\equiv \del^2_{\r} \phi_c(\t^n,\r_i)$ and  $\pi_i^n\equiv \pi_i(\phi_i^n;a_{2,2}^n)$,
following corresponding  computations in RK step 1, and
further identify
\begin{equation}
\begin{split}
&(\phi_{log})_i\equiv \phi_{log}(\t^n,\r_i),\qquad d_1 (\phi_{log})_i\equiv \del_\r \phi_{log}(\t^n,\r_i),\\
&d_2 (\phi_{log})_i\equiv \del^2_{\r} \phi_{log}(\t^n,\r_i),\\
&(b_{log})_i\equiv b_{log}(\t^n,\r_i; p^n_2),\qquad d_1 (b_{log})_i\equiv \del_\r b_{log}(\t^n,\r_i; p^n_2),\\
&d_2 (b_{log})_i\equiv \del^2_{\r} b_{log}(\t^n,\r_i; p^n_2),\\
&d_1 (a_{log})_i\equiv \del_\r a_{log}(\t^n,\r_i; p^n_2),\qquad d_2 (a_{log})_i\equiv \del^2_{\r} a_{log}(\t^n,\r_i; p^n_2),\\
&d_3 (a_{log})_i\equiv \del^3_{\r\r\r} a_{log}(\t^n,\r_i; p^n_2),\\
&d_1 (g_{log})_i\equiv d_3 (a_{log})_i +\frac {2}{\r_i} d_2 (a_{log})_i - \frac{2}{\r_i^2} d_1 (a_{log})_i.
\end{split}
\eqlabel{defbglog}
\end{equation}
\item
Note that at this stage we have all the data necessary to evaluate $J_{b,i}$ (see \eqref{hconst2}),
\begin{equation}
J_{b,i}\equiv J_b(\t^n,\r_i; \del_\r\hp=d_1\phi_i^n+d_1(\phi_{log})_i, \hp=\phi_i^n+(\phi_{log})_i).
\eqlabel{defjb}
\end{equation}
We compute Chebyshev coefficients $\calf_{b_c}^j$ solving \eqref{const2}
\begin{equation}
\sum_{j=1}^N\ \calf_{b_c}^j\ d_2T(i,j)\ \left(\frac{dx}{d\r}\right)^2=-(J_{b,i} +d_2 (b_{log})_i),\qquad i=1,\cdots N-2.
\eqlabel{eomfb1}
\end{equation}
along with the boundary conditions \eqref{bcconst}:
\begin{equation}
\sum_{j=1}^N\ \calf_{b_c}^j\ T(N,j) =0,\qquad
\sum_{j=1}^N\ \calf_{b_c}^j\ d_1T(N,j)\ \left(\frac{dx}{d\r}\right)=0.
\eqlabel{eomfb2}
\end{equation}
Given $\calf_{b_c}^j$ we evaluate
\begin{equation}
\begin{split}
&b_i^n =\sum_{j=1}^N \calf_{b_c}^j\ T(i,j),\qquad
d_1 b_i^n =\sum_{j=1}^N \calf_{b_c}^j\ d_1 T(i,j) \left(\frac{dx}{d\r}\right)
,\qquad i=1,\cdots N.
\end{split}
\eqlabel{bcidbci}
\end{equation}
\item We can now compute (see \eqref{defjg})
\begin{equation}
J_{g,i}\equiv J_g(\t^n,\r_i),
\eqlabel{defjgi}
\end{equation}
with obvious substitutions:
\begin{equation}
\begin{split}
&\del^2_{\r}\hp=d_2\phi_i^n+d_2(\phi_{log})_i,\qquad  \del_\r\hp=d_1\phi_i^n+d_1(\phi_{log})_i,\qquad  \hp=\phi_i^n+(\phi_{log})_i,\\
&\pi=\pi_i^n+(\pi_{log})_i,\qquad \del_\r b= d_1 b_i^n+ d_1 (b_{log})_i.
\end{split}
\eqlabel{substg}
\end{equation}
Next, we solve for Chebyshev coefficients following \eqref{defgc}, \eqref{defbglog}
\begin{equation}
\sum_{j=1}^N\ \calf_{g_c}^j\ d_1T(i,j)\ \left(\frac{dx}{d\r}\right)=-(J_{g,i} +d_1 (g_{log})_i),\qquad i=1,\cdots N-1,
\eqlabel{eomfg1}
\end{equation}
along with the boundary conditions \eqref{bcconst}:
\begin{equation}
\sum_{j=1}^N\ \calf_{g_c}^j\ T(N,j) =6 a_{2,2}^n +\frac 56 \left((p_0'(\t^n))^2-p_0(\t^n)p_0''(\t^n)\right).
\eqlabel{eomfg2}
\end{equation}
Given $\calf_{g_c}^j$ we evaluate
\begin{equation}
\begin{split}
&g_i^n =\sum_{j=1}^N \calf_{g_c}^j\ T(i,j),\qquad i=1,\cdots N.
\end{split}
\eqlabel{gggres}
\end{equation}
\item The next step is computation of
\begin{equation}
a_i^n\equiv a_c(\t^n,\r_i),
\eqlabel{defacha}
\end{equation}
using \eqref{defac}. Remarkably, this can be achieved analytically, given
$\calf_{g_c}^j$.
Indeed, note that
\begin{equation}
\begin{split}
a_c(\t,\r) =& \sum_{j=1}^\infty \calf_{g_c}^j(\t)\ \int_0^\r \frac{dx}{x^2}\left[\int_0^x dy y^2 T_{j-1}
\left(\frac{2y}{L_\r}-1\right)\right]\\
=&\left(\frac{L_\r}{2}\right)^2 \sum_{j=1}^{\infty} \calf_{g_c}^j(\t) \int_{-1}^{2\r/L_\r-1} \frac{dx}{(1+x)^2}
\left[\int_{-1}^x (1+y)^2 T_{j-1}(y)\right],
\end{split}
\eqlabel{helpa1}
\end{equation}
where in the second line we changed the integration variables
\begin{equation}
y\to \frac{L_\r}{2}(1+y),\qquad x\to \frac{L_\r}{2}(1+x).
\eqlabel{varchange1}
\end{equation}
Furthermore,
\begin{equation}
 \int_{-1}^{z} \frac{dx}{(1+x)^2}
\left[\int_{-1}^x (1+y)^2 T_{j-1}(y)\right]=\sum_{s=1}^{j+2}\calc_{j,s}\ T_{s-1}(z).
\eqlabel{tjm1int}
\end{equation}
where the rational coefficients $\calc_{j,s}$ can be computed using the orthonormality properties of
the Chebyshev polynomials. The first several coefficient sets are:
\begin{equation}
\begin{split}
&\calc_{1,s} = \left\{\frac 14,\, \frac 13,\, \frac {1}{12}\, \right\},\\
&\calc_{2,s} = \left\{-\frac{1}{24},\, -\frac{1}{48},\, \frac {1}{24},\,  \frac {1}{48}\, \right\},\\
&\calc_{3,s} = \left\{-\frac{7}{48},\, -\frac{13}{60},\, -\frac {1}{15},\,  \frac {1}{60},\,
\frac {1}{80}\,\right\},\\
&\qquad \cdots.
\end{split}
\eqlabel{cjsdef}
\end{equation}
Thus,
\begin{equation}
a_c(\t,\r)=\left(\frac{L_\r}{2}\right)^2 \sum_{j=1}^{\infty}  \calf_{g_c}^j(\t) \biggl[\sum_{s=1}^{j+2} \calc_{j,s}\  T_{s-1}
\left(\frac{2\r}{L_\r}-1\right)   \biggr],
\eqlabel{helpa2}
\end{equation}
and (truncating the Chebyshev modes to order $(N-2)$)
\begin{equation}
a_{i}^n= \left(\frac{L_\r}{2}\right)^2\  \sum_{j=1}^{N-2} \sum_{s=1}^{j+2}\  \calf_{g_c}^j\ \calc_{j,s}\  T(i,s).
\eqlabel{ainfinal}
\end{equation}
\item Finally, we identify
\begin{equation}
\left\{\ \t^n,\, p_2^n,\, a_{2,2}^n\,;\, \phi_i^n,\, b_i^n,\, a_i^n  \right\}\qquad \to\qquad
\left\{\ \t^o,\, p_2^o,\, a_{2,2}^o\,;\,  \phi_i^o,\, b_i^o,\, a_i^o \right\},
\eqlabel{noid}
\end{equation}
and repeat the whole process from \eqref{defhp}.
\end{itemize}

\subsection{Convergence tests}\label{appA3}
All our simulations were performed with $N=40$
collocation points. In this section we discuss the convergence of the
simulations as the number of collocation points is varied, and also the accuracy of
solving the constraint equations \eqref{const1}, \eqref{const2}, and \eqref{defgc}.

\begin{figure}
\begin{center}
\psfrag{t}[Br][tl]{{$\scriptstyle{\t}$}}
\psfrag{pp}[c]{{$\scriptstyle{e_N^{\phi_c}}$}}
\psfrag{epp}[c]{{$\scriptstyle{\rm{error} [\pi_c]}$}}
  \includegraphics[width=2.7in]{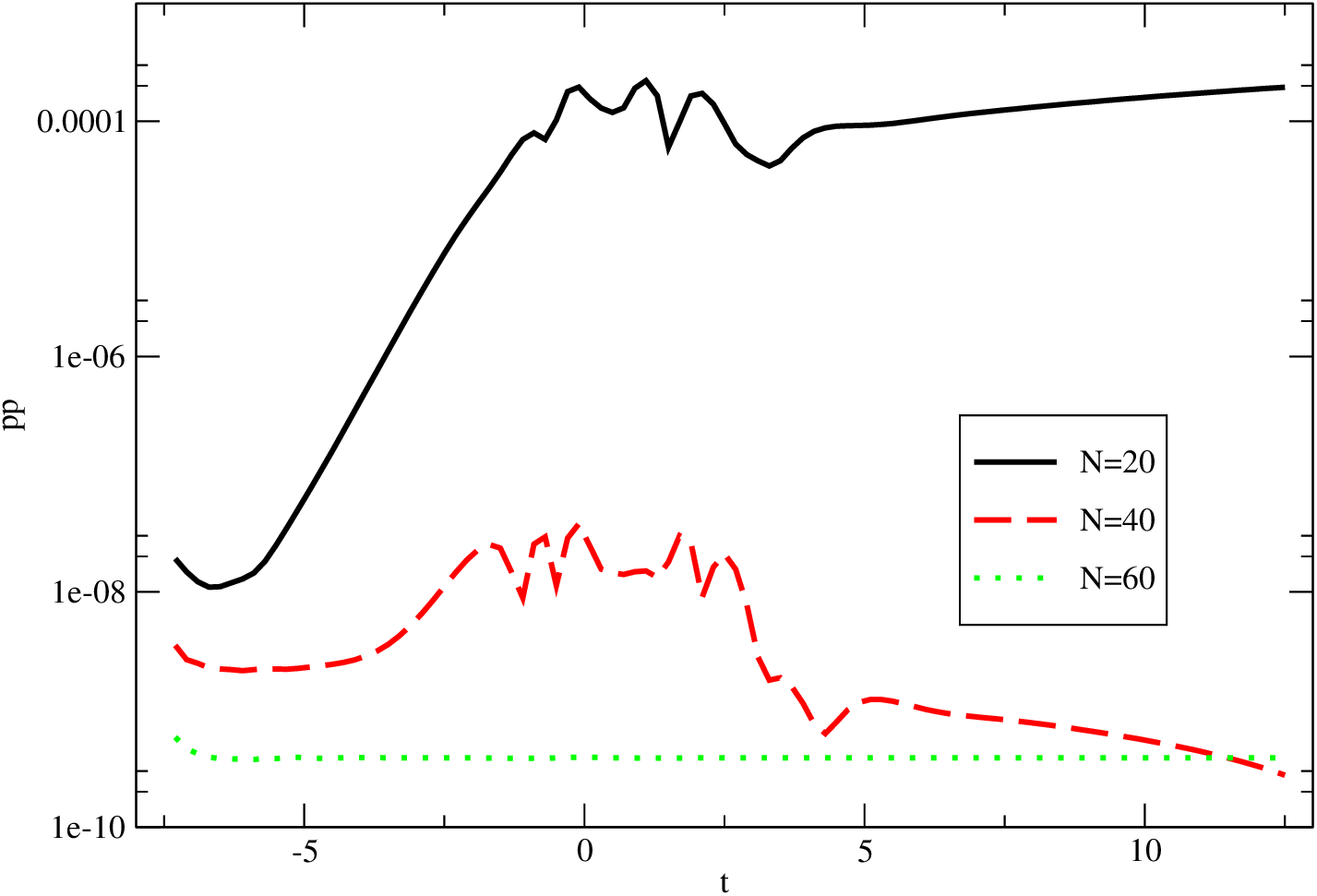}\,\,
  \includegraphics[width=2.7in]{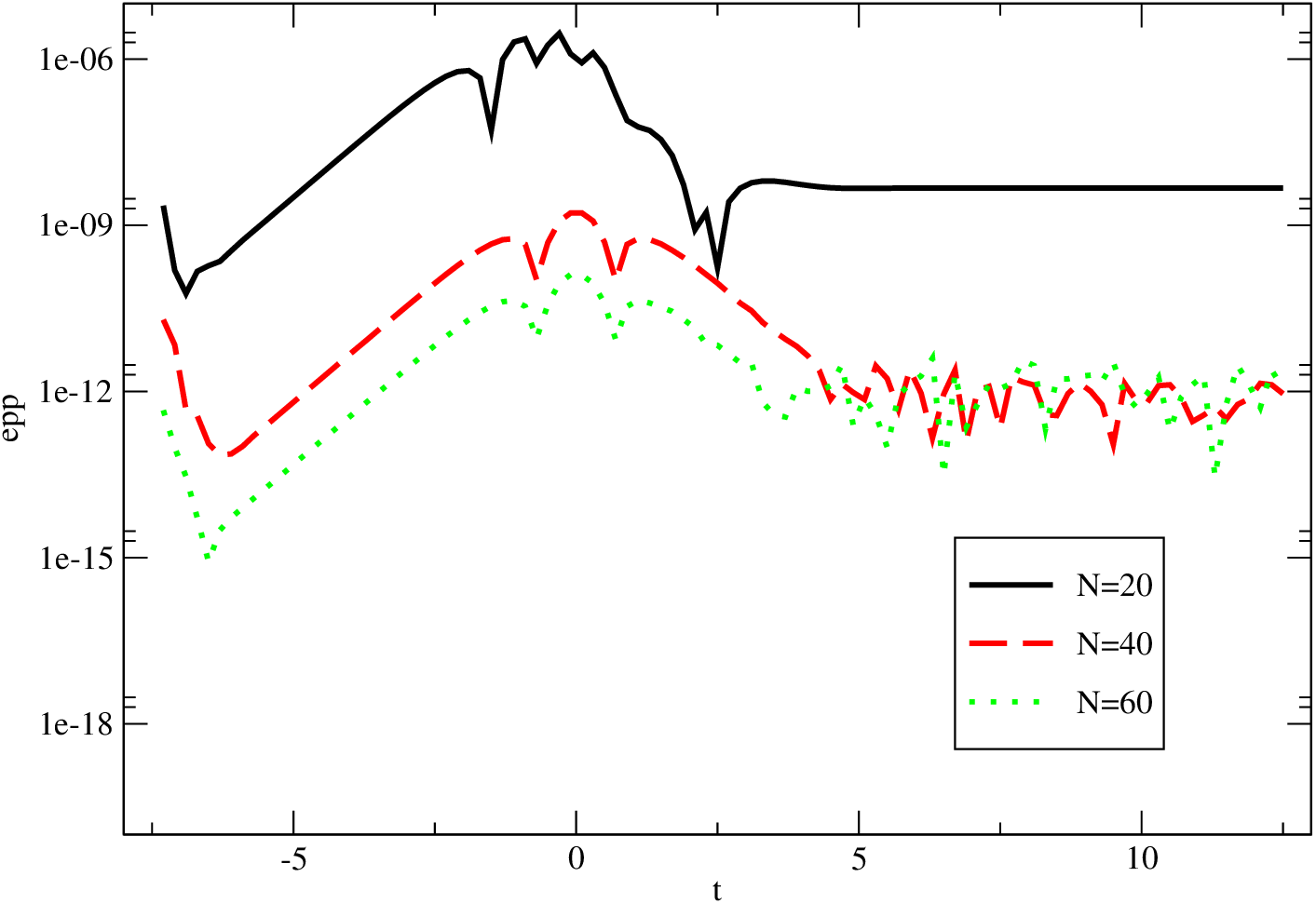}
\end{center}
  \caption{(Colour online) (Left panel) Convergence of $\phi_c$ for different number of collocation points
as a function of $\t$, see \eqref{defen}.
(Right panel) Residuals of the constraint \eqref{const1}, see \eqref{errconst1}. \bigskip }  \label{convergencetestsp}
\end{figure}

\begin{figure}
\begin{center}
\psfrag{t}[Br][tl]{{$\scriptstyle{\t}$}}
\psfrag{bb}[c]{{$\scriptstyle{e_N^{b_c}}$}}
\psfrag{ebb}[c]{{$\scriptstyle{\rm{error} [b_c]}$}}
  \includegraphics[width=2.7in]{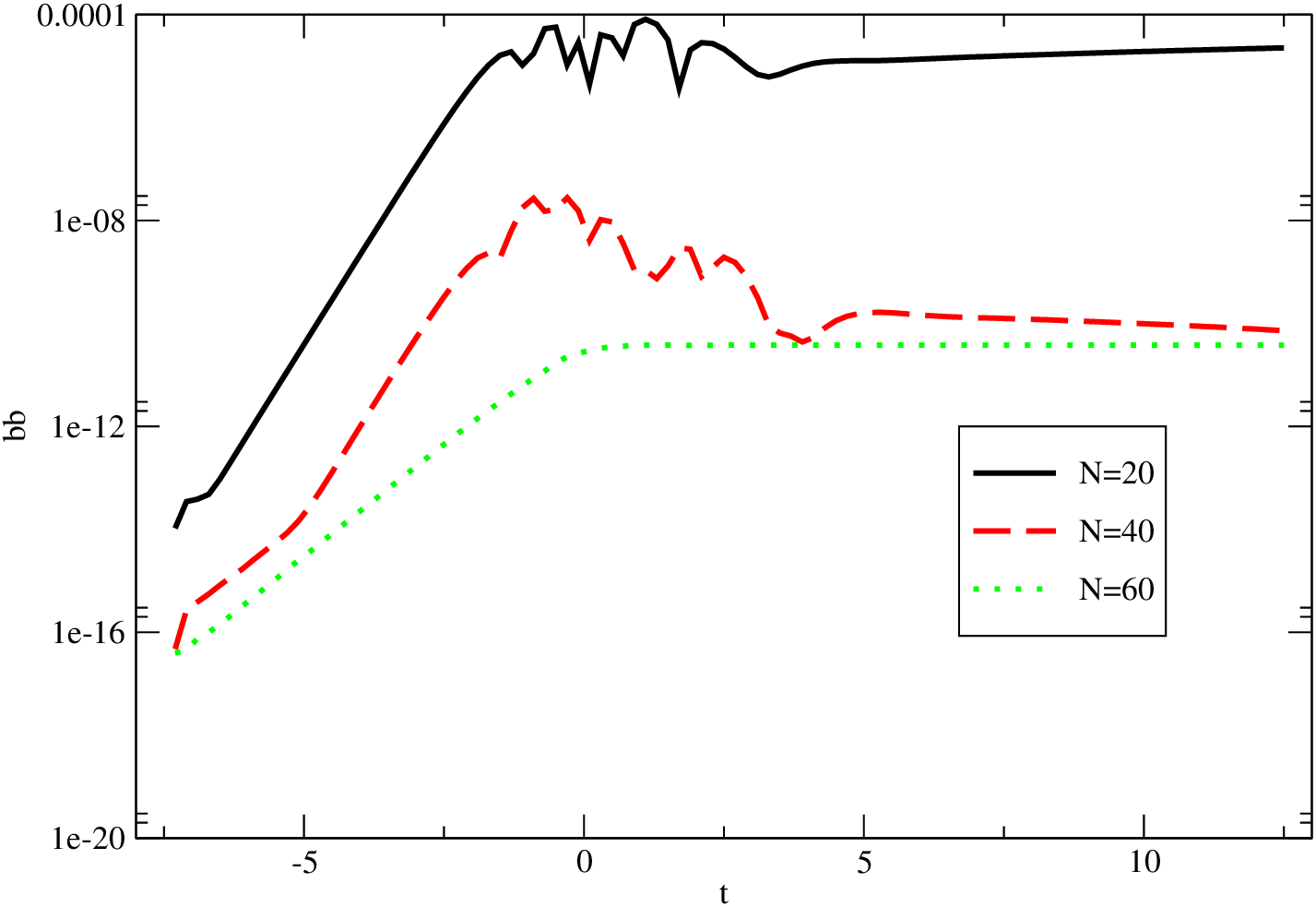}\,\,
  \includegraphics[width=2.7in]{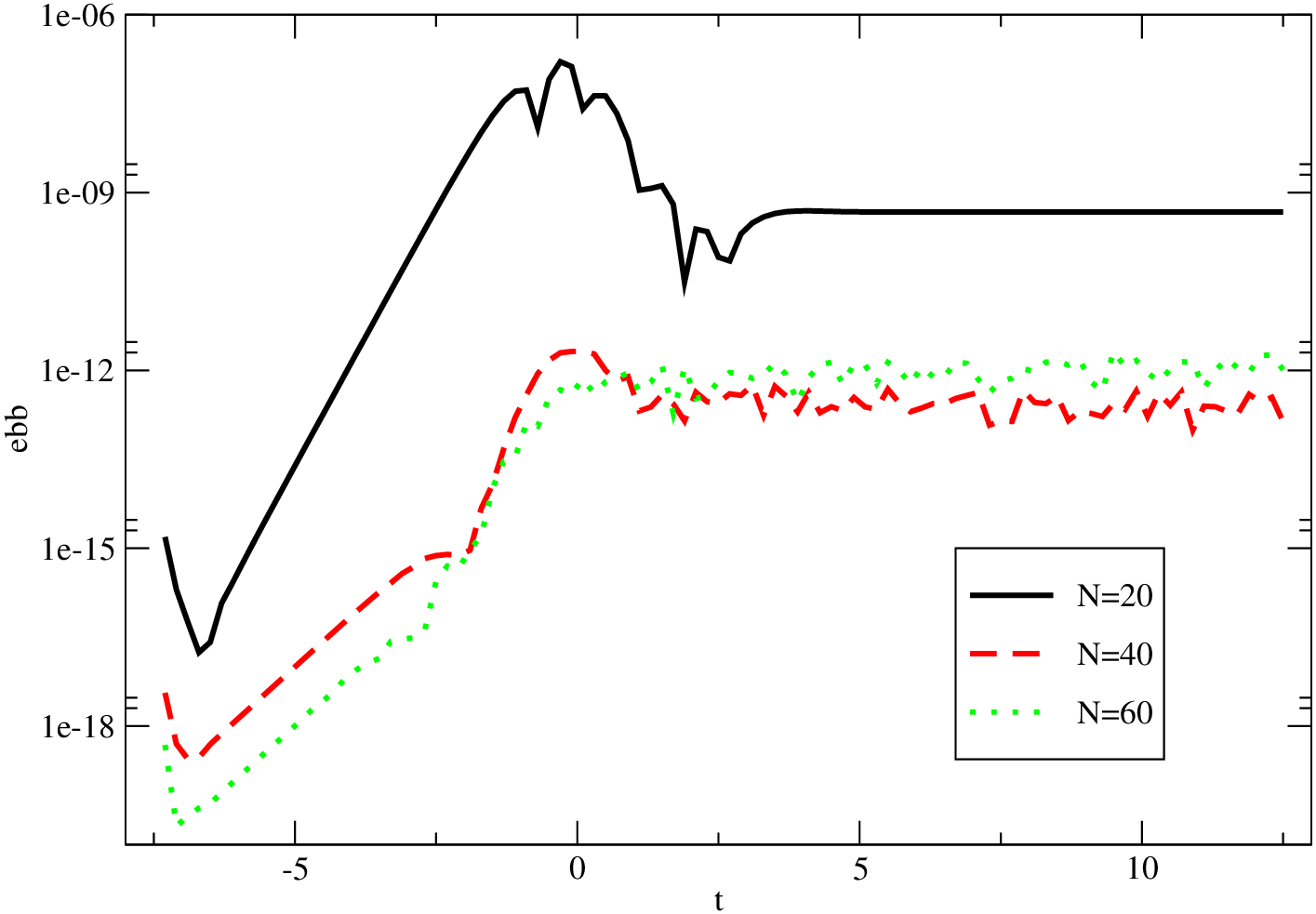}
\end{center}
  \caption{(Colour online) (Left panel) Convergence of $b_c$ for different number of collocation points
as a function of $\t$, see \eqref{defen}.
(Right panel) Residuals of the constraint \eqref{const2}, see \eqref{errconst2}. \bigskip }  \label{convergencetestsb}
\end{figure}

\begin{figure}
\begin{center}
\psfrag{t}[Br][tl]{{$\scriptstyle{\t}$}}
\psfrag{aa}[c]{{$\scriptstyle{e_N^{\ha}}$}}
\psfrag{eaa}[c]{{$\scriptstyle{\rm{error} [g_c]}$}}
  \includegraphics[width=2.7in]{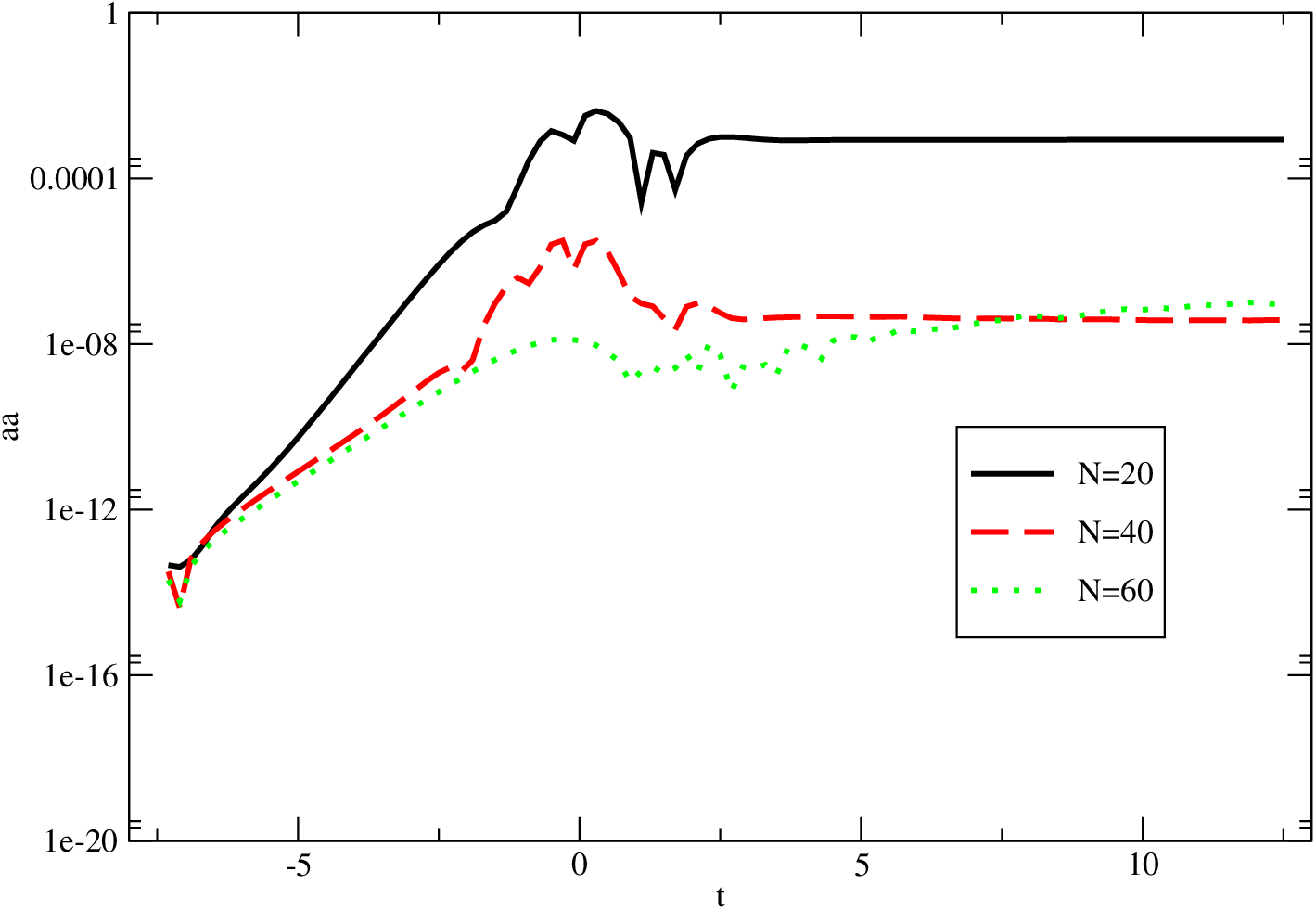}\,\,
  \includegraphics[width=2.7in]{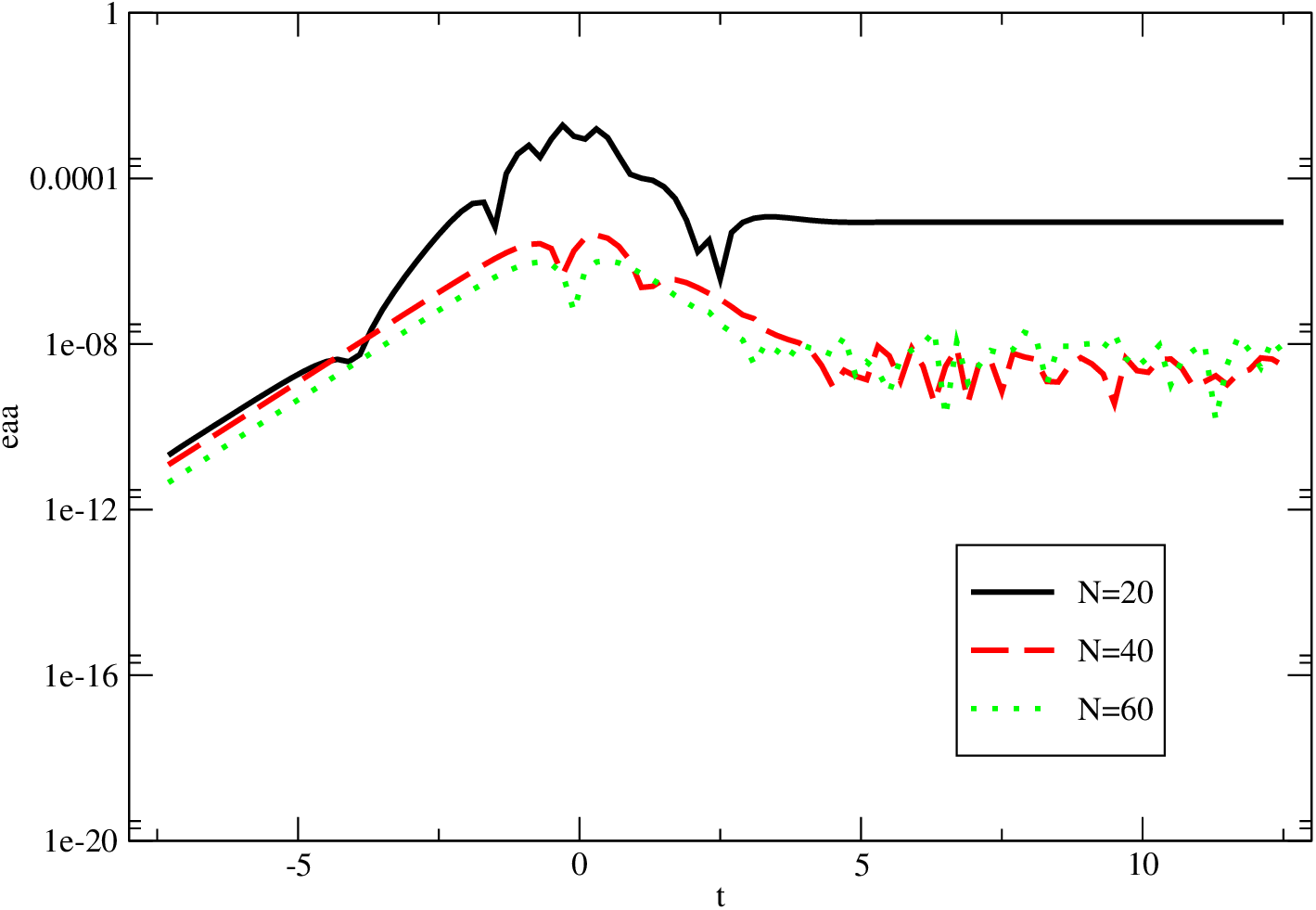}
\end{center}
  \caption{(Colour online) (Left panel) Convergence of $\ha$ for different number of collocation points
as a function of $\t$, see \eqref{defen}.
(Right panel) Residuals of the constraint \eqref{defgc}, see \eqref{errconst3}. \bigskip
}  \label{convergencetestsa}
\end{figure}

As a representation test of the code convergence behaviour, we consider $\a=1$ and different number of collocation
points: $N=N_{i=1,\cdots 4}=\{10,20,40,60\}$.
We monitor the ($L_2$ norm of the ) difference of solutions with
successive values of $N_i$, defined as
\begin{equation}
\begin{split}
e_{N_i}^{\phi_c}(\t)=&  ||\ \phi_c[N_i]- \phi_c[N_{i-1}]\ ||_2,\\
e_{N_i}^{b_c}(\t)=&  ||\  b_c[N_i]- b_c[N_{i-1}]\ ||_2,\\
e_{N_i}^{\ha}(\t)=&  ||\ \ha[N_i]- \ha[N_{i-1}]\ ||_2.
\end{split}
\eqlabel{defen}
\end{equation}
Additionally, given spectral coefficients $\calf_{\pi_c}^i$, we can verify
the accuracy of constraint \eqref{const1} defining
\begin{equation}
\begin{split}
&{\rm const}_{\pi_c}(\t,\r)\equiv J_{\pi}+\del_\r (\pi_{log})-\frac{1}{2\r} \pi_{log}+
\sum_{j=1}^N \calf_{\pi_c}^j(\t)\   \left(\del_{\r}-\frac{1}{2\r}\right) T_{j-1}\left(\frac{2\r}{L_\r}-1\right), \\
&{\rm error}[\pi_c] (\t)= || {\rm const}_{\pi_c}(\t,\r)||_2.
\end{split}
\eqlabel{errconst1}
\end{equation}
Likewise, for \eqref{const2},
\begin{equation}
\begin{split}
&{\rm const}_{b_c}(\t,\r)\equiv J_{b}+\del^2_{\r} b_{log}+
\sum_{j=1}^N \calf_{b_c}^j(\t)\   \del^2_{\r} T_{j-1}\left(\frac{2\r}{L_\r}-1\right), \\
&{\rm error}[b_c] (\t)= || {\rm const}_{b_c}(\t,\r)||_2,
\end{split}
\eqlabel{errconst2}
\end{equation}
and, for \eqref{defgc},
\begin{equation}
\begin{split}
&{\rm const}_{g_c}(\t,\r)\equiv J_{g_c}+
\sum_{j=1}^N \calf_{g_c}^j(\t)\   \del_{\r} T_{j-1}\left(\frac{2\r}{L_\r}-1\right), \\
&{\rm error}[g_c] (\t)= || {\rm const}_{g_c}(\t,\r)||_2.
\end{split}
\eqlabel{errconst3}
\end{equation}
Results are shown in figures \ref{convergencetestsp}-\ref{convergencetestsa}.
The left panels illustrate how $e_N$ decreases as the resolution is improved, while the right panels
present the accuracy of solving constraints \eqref{errconst1}-\eqref{errconst3}.

\subsection{Limit of abrupt quenches}\label{appA4}

\begin{figure}
\begin{center}
\psfrag{p0}{{$p_0$}}
\psfrag{p2}{{$p_2$}}
  \includegraphics[width=4in]{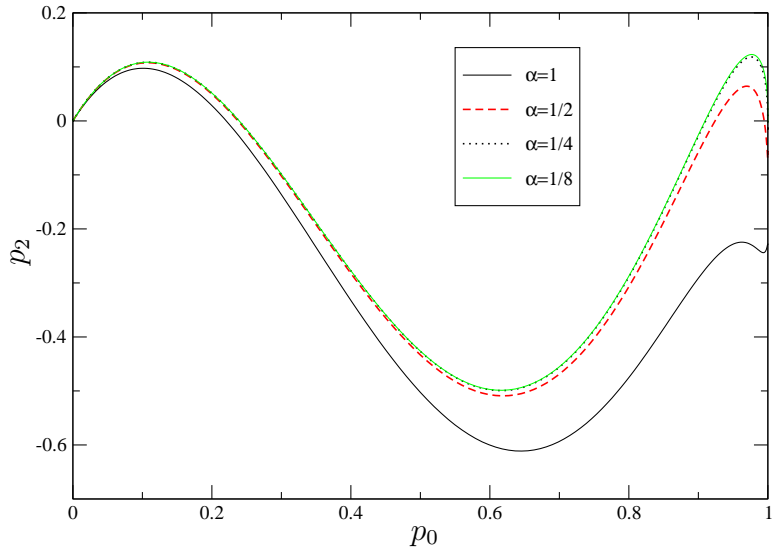}
\end{center}
  \caption{(Colour online)  Response $p_2=p_2(p_0)$ for fast quenches.
The universal regime of abrupt quenches is achieved for $\a\gtrsim \frac 14$.
\bigskip }  \label{p2alpha}
\end{figure}

Quantum quenches have two scaling regimes: the adiabatic  one ($\a\gg 1$ ),  and the regime of the
abrupt quenches ($\a \ll 1$). The former one represents an expected  slow, hydrodynamic response of the system
to external forcing \cite{blm}. It was observed in \cite{blm} (and further studied in
\cite{bmv} ) that within a holographic framework a QFT exhibits a scaling response in the limit
of abrupt quenches as well. The same scaling was observed outside of holography in a CFT deformed by a
relevant operator \cite{gm}.
Our code is ideally suited to study fast quenches since we use the
$\a$-rescaled scalar and metric variables \cite{bmv}.
Figure \ref{p2alpha} illustrates the response function $p_2$ for fast quenches, $\{\a=1,\ft 12,\ft 14,\ft 18\}$.
Note that the response becomes almost indistinguishable between $\a=\ft 14$ and $\a=\ft 18$ quenches.
We take $\a=\ft 18$  to correspond to {\it abrupt quench}.

\end{document}